\documentclass[12pt,oneside,a4paper,leqno]{article}
\usepackage{geometry} 
\usepackage{amsfonts,amssymb}
\usepackage[centertags]{amsmath}
\usepackage[all]{xy}\xyoption{all}
\usepackage{graphicx}
\RequirePackage{amsthm}[2000/10/26]
\numberwithin{equation}{section}
\makeatletter
\def\tagform@#1{\maketag@@@{\ignorespaces#1\unskip\@@italiccorr}}

\makeatother

\newtheoremstyle{par}%
     {\topsep}%
     {\topsep}%
     {\itshape}%
     {}%
     {\bfseries}%
     {}%
     {.5em}%
     {}%

\newtheoremstyle{exr}{\topsep}{\topsep}{}{}{\bfseries}{}{.5em}{}

\swapnumbers

\theoremstyle{plain}
\newtheorem{theo}[equation]{Theorem}
\newtheorem{propo}[equation]{Proposition}
\newtheorem{coro}[equation]{Corollary}
\theoremstyle{definition}
\newtheorem{defi}[equation]{Definition}
\newtheorem{remark}[equation]{Remark}
\theoremstyle{exr}
\newtheorem{example}[equation]{}

\theoremstyle{par}
\newtheorem{nr}[equation]{}
\newtheorem{lemma}[equation]{} 

\newdir{ >}{!/8pt/@{ }*@{>}}
\newdir{< }{!/-8pt/@{ }*@{<}}

\newcommand{\minus}{\smallsetminus}
\newcommand{\C}{\mathbb{C}}
\newcommand{\R}{\mathbb{R}}
\newcommand{\ze}{\mathbb{Z}}
\newcommand{\st}{\ \mathrm{|} \ }
\newcommand{\from}{\colon}

\newcommand{\X}{\mathcal{X}}
\newcommand{\bomega}{\mathbf{\omega}}
\newcommand{\T}{\mathbb{T}}
\newcommand{\I}{\mathbb{I}}
\newcommand{\action}{\mathcal{A}}
\newcommand{\h}{\mathbf{h}}
\newcommand{\n}{\mathbf{n}}
\newcommand{\kk}{\mathbf{k}}
\newcommand{\bfa}{\mathbf{a}}
\newcommand{\bfb}{\mathbf{b}}
\newcommand{\bdc}{bound to collisions}
\newcommand{\reducible}{reducible}

\newcommand{\ff}{\varphi}
\newcommand{\sphere}{\mathbb{S}}
\newcommand{\ts}{\tilde{S}}
\newcommand{\binomial}[2]{\binom{#1}{#2}}
\newcommand{\hypergeom}[8]{
{}_{#1}F_{#2} \left(
\begin{array}{ccc}
#3, &#4, & #5;\\
& #6, & #7; \\
\end{array} 
#8
\right)
}
\newcommand{\lag}{\mathcal{L}}
\newcommand{\kin}{\mathcal{K}}
\newcommand{\pot}{\mathcal{U}}

\newcommand{\matrice}[4]{%
\left[
\begin{array}{cc}
#1 & #2 \\
#3 & #4 \\
\end{array}
\right]
}

\newcommand{\Matrice}[9]{%
\left[
\begin{array}{ccc}
#1 & #2 & #3 \\
#4 & #5 & #6 \\
#7 & #8 & #9 \\
\end{array}
\right]
}

\newcommand{\rotationmatrix}[1]{%
\matrice%
{\cos \frac{2\pi}{#1}}{-\sin \frac{2\pi}{#1}} 
{\sin \frac{2\pi}{#1}}{\cos \frac{2\pi}{#1}} 
}

\newcommand{\RotationMatrix}[1]{%
\Matrice%
{\cos \frac{2\pi}{#1}} {-\sin \frac{2\pi}{#1}} { 0}
{\sin \frac{2\pi}{#1}} {\cos \frac{2\pi}{#1}} {0}
{0} {0} {1}
}
\newcommand{\mydddot}[1]{\frac{d^3{#1}}{dt^3}}

\begin{document}
\pagenumbering{arabic}

\title{%
On the Existence of  Collisionless 
Equivariant Minimizers for the Classical $n$-body 
Problem}

\author{Davide L.~Ferrario%
\footnote{%
Dipartimento di Matematica  del Politecnico di Milano,
Piazza Leonardo da Vinci, 32;  20133 Milano, Italy.
email: \emph{ferrario@mate.polimi.it}}
\ and 
Susanna Terracini%
\footnote{%
Dipartimento di Matematica e Applicazioni
Università degli Studi di Milano-Bicocca
Via Bicocca degli Arcimboldi, 8; 20126 Milano, Italy.
email: \emph{suster@matapp.unimib.it}
}
}

\date{\today}
\maketitle
\begin{abstract}
We show that 
the minimization of   the Lagrangian action functional
on suitable classes of symmetric loops
yields collisionless periodic orbits
of the $n$-body problem, provided that some simple
conditions on the symmetry group are satisfied.
More precisely, we give a fairly general condition  on
symmetry groups $G$ of the loop space $\Lambda$ 
for the $n$-body problem (with potential
of homogeneous degree ${-\alpha}$, with $\alpha>0$) which  ensures 
that the 
restriction of the Lagrangian action $\action$ 
to the space $\Lambda^G$ of 
$G$-equivariant loops is coercive and its minimizers 
are collisionless,
without any strong force assumption.
Many of the already known periodic orbits 
can be proved to exist by this result, and 
several new orbits are found with some appropriate
choice of $G$.

\vspace{12pt}
\noindent
\emph{MSC Subj. Class}: 
Primary 
70F10 (Mechanics of particles and systems: $n$-body problems);
Secondary 
70F16 (Mechanics of particles and systems: 
Collisions in celestial mechanics, regularization),
37C80  (Dynamical systems and ergodic theory: Symmetries, equivariant dynamical systems),
70G75  (Mechanics of particles and systems: Variational methods).

\vspace{12pt}
\end{abstract}

\section{Introduction}

The 
method of minimizing the Lagrangian action on a space
of loops  symmetric with respect to a well-chosen symmetry group
has been used 
in some recent papers to find new interesting periodic orbits 
for the $n$-body problem 
\cite{chenven,monchen,chen,marchalp12}.
Such a  variational approach
has been
extensively exploited in the last decades by several other authors.
We refer the reader to the following articles and references therein:
\cite{%
ambcot,
argate,
bahrab,
bessi,
chenICM,
chenciner,
dellantonio,
mate95,
montgomery,
montgomery_contmath,
moore,
riahi,
sbano,
serter1,
serter2,
andrea_thesis}.
This approach consists in seeking periodic trajectories as 
critical points
of the 
action functional  associated to a system of $n$ particles
with masses $m_i>0$, interacting through a potential of homogeneous degree 
$-\alpha$, 
\begin{equation}
\label{eq:azione}
\action(x(t)) = 
\int_0^T \left[ 
\sum_{i} \dfrac{1}{2} m_i |\dot x_i(t)|^2 +
\sum_{i<j} \dfrac{m_im_j}{|x_i(t) - x_j(t)|^\alpha} 
\right]
dt.
\end{equation}
The main difficulties reside in the fact that the action
functional is not coercive and that in principle
critical points might be trajectories with collisions.
Actually this can happen only when $\alpha<2$, since
if $\alpha\geq 2$ (strong force) the action on colliding trajectories
is not finite \cite{poincare,gordon}.
In 
the quoted papers \cite{chenven,monchen,chen}
these problems are overcame by level estimates and 
an appropriate choice of a group of symmetries
acting on the space of loops  that penalizes the global
cost of collisions.
An alternative approach consists of trying to find local variations
around a (supposed) colliding minimizer and it was 
used e.g. in \cite{serter1}. A major breakthrough in this direction
is the recent Marchal sharp contribution \cite{marchal,chenICM} (see
below remark \ref{rem:mars}).
Marchal introduced the idea of 
averaging on suitable 
sets of variations 
for the Keplerian potential $\alpha=1$ and for
discs or spheres,
so to prove that {minimizers
of the fixed-ends (Bolza) problem  are free of interior
collisions}.
The aim of the present paper is to 
go further in this direction and to give general conditions 
on the group action (the hypothesis of ~\ref{propo:coercive} and 
the \emph{rotating circle property}) 
under which 
minimizers of the action exist and are collision-free. 
Compared to the existing literature, 
our results hold for all homogeneous potentials (not only for the 
Newtonian potential with $\alpha=1$), include  many
of the known symmetric orbits and 
allow to prove the existence of new families of collisionless periodic orbits.
Also, another interesting feature of our approach is 
that these families can be found with suitable algebraic 
algorithms, which generate group actions having 
the rotating circle property.
On the other hand, our results are  not suitable 
to prove the existence of  those orbits that are found 
as minimizers in classes of paths characterized by
homotopy conditions or a mixture of symmetry 
and homotopy conditions (see \cite{montgomery_contmath,cgms}).

Let $\X$ denote the space of 
centered configurations (possibly with collisions) of 
$n>2$ point particles with masses
$m_1$, $m_2$, \dots, $m_n$ 
in the Euclidean space $\R^d$ of dimension $d\geq 2$ (that is,
$\X$ is the subspace of $\R^{nd}$ consisting of points $x=(x_1,
\dots, x_n)$ with centered center of mass $\sum_{i=1}^n m_i x_i = 0$).
Let $\T = \R/T\ze$ denote the circle of length $T=|\T|$,
embedded as $\T \subset \R^2$.

By the 
 loop space $\Lambda$ we mean  the Sobolev space $\Lambda = H^1(\T,\X)$
(see section ~\ref{section:prelimin}).
Consider a finite group $G$, a $2$-dimensional orthogonal representation
$\tau\from G \to O(2)$,  a $d$-dimensional orthogonal representation 
$\rho\from G \to O(d)$ and a homomorphism $\sigma\from G \to \Sigma_n$
to the symmetric group on $n$ elements with the property that 
\(
\forall g\in G : \left(
\sigma(g)(i) = j \implies m_i = m_j\right).
\)
Thus, if the action of $G$ on $\{1,2,\dots, n\}$ is transitive, 
we will consider the problem of $n$ equal masses; otherwise there may be
particles with different masses.
Finally, consider the subspace $\Lambda^G \subset \Lambda$ consisting 
of all loops  $x\in \Lambda$ with the property that 
\[
\forall g \in G, \forall t \in \T , \forall i=1\dots n:
\rho(g) x_{\sigma(g^{-1})(i)}(t)
=
x_i( \tau(g)t).
\]
Let $\action^G$ denote the restriction of the action functional $\action$
to $\Lambda^G$. As it is shown in section ~\ref{section:symmetry},
the homomorphisms $\rho$, $\tau$ and $\sigma$ yield an action
of $G$ on $\X$, $\T$ and $\Lambda$, and  $\Lambda^G$ is the subspace of 
$G$-equivariant loops. 

A first result is proposition ~\ref{propo:coercive}: \emph{The action
functional $\action^G$ is coercive if and only if $\X^G=0$,
where $\X^G\subset \X$ is the subspace fixed by $G$}. 
This extends a previous coercivity result by Bessi and Coti--Zelati
\cite{bessi} and 
is the first step in the variational approach to the $n$-body 
problem: if $\rho$ and $\tau$ are chosen so that $\X^G=0$,
then the minimum of $\action^G$ exists.

The next goal is 
to
exclude the occurrence of collisions  on minimizers
by performing a throughout analysis of collision--ejection trajectories.
This step 
requires the proof of asymptotic estimates 
(in the same spirit Sundman and Sperling 
estimates (see \cite{wintner,sperling}) 
for partial collisions 
that 
make possible the application of the blow-up technique.
This analysis allows us
to reduce the general case to that 
of homotetic self-similar collision trajectories.
Now a key estimate ~\ref{theo:aa}  comes into play:
\emph{it is more convenient (from the point of view of the integral
of the potential on the time line) to 
replace  one of the  point particles with 
a homogeneous circle of same mass  and fixed radius
which is moving keeping its
center in the  position of the original particle}.
This generalizes to every $\alpha>0$ 
Marchal's  result that  
\emph{minimizers
of the fixed-ends (Bolza) problem  are free of interior
collisions} (see corollary ~\ref{coro:bolza}).
We recall that the case of binary and triple
collisions was already treated in \cite{serter1}.

Finally, to prove that \emph{equivariant} minimizers are free of collisions
we need to match the averaging procedure 
with the action
of the group $G$. 
To this end, we introduce
a condition on the action of 
the group 
$G$, that we call {rotating circle property} 
(see definition ~\ref{propertyK}  and ~\ref{defi:Tisotropy}).
Under such a condition of the $G$-action we will 
prove our main result (theorem ~\ref{mt:2}):
all (local) minimizers of the action $\action^G$
on the space $\Lambda^G$ of equivariant loops are free of collision.
A major step towards the proof consists in theorem ~\ref{mt:1}:
\emph{a minimizer of the equivariant Bolza problem is free of collisions,
provided the symmetry group acts with the rotating circle property}.

Most of the symmetry groups used in the quoted literature in fact 
enjoy
the rotating circle property;
otherwise it is usually possible to find another action,
which has the same type of minimizers,
that fulfills the requirements of theorem ~\ref{mt:2}.
In particular the following actions have the rotating
circle  property:
the $n$-cyclic action of choreographies (see example ~\ref{ex:choreo}) 
and 
that of the eight-shaped orbit, for any odd
number of bodies.
A number of examples and generalizations of the known actions will be 
given in the last section ~\ref{section:examples}. 
The classification of all actions with the rotating 
circle property goes beyond of the purpose 
of this paper and is the subject of a paper in preparation.

The article is organized as follows.
\begin{enumerate}
\setlength{\itemsep}{0pt}
\setlength{\parsep}{0pt}
\setlength{\parskip}{0pt}
\setlength{\topsep}{0pt}
\item Introduction.
\item Preliminaries.
\item Symmetry constraints.
\item Coercivity and generalized solutions.
\item Isolated collisions.
\item Asymptotic estimates.
\item Blow-ups.
\item Averaging estimates.
\item The standard variation.
\item The rotating circle property and the main theorems.
\item Examples.
\end{enumerate}

This work was partially supported by the MIUR project ``Metodi
Variazionali ed Equazioni Differenziali Nonlineari''. The first 
author would like to thank the Max--Planck--Institut f\"ur
Mathematik (Bonn) where part of the work has been done.
We would like to thank all the people that helped us with 
their suggestions and criticisms: K.~Chen, A.~Chenciner, L.~Fontana, 
G.~Molteni,
R.~Montgomery, G.~Naldi and A.~Venturelli.

\section{Preliminaries}
\label{section:prelimin}
In this section we set some notation and describe 
preliminary results that will be needed later.
Let 
$V=\R^d$ denote the Euclidean space of dimension $d$  and $n\geq 2$
an integer. Let $0$ denote the origin $0\in \R^d$.
Let $m_1\dots m_n$ be $n$ positive real numbers.
The configuration space $\X$ of $n$ point particles with
masses $m_i$ respectively and center of mass in $0$  can be identified with 
the subspace of $V^n$ consisting 
of all points $x = (x_1, \dots, x_n) \in V^n$
such that $\sum_{i=1}^n m_i x_i = 0$. 
For each pair of indexes $i,j \in \{1,\dots, n\}$
let $\Delta_{i,j}$ denote the collision set of the $i$-th and $j$-th
particles $\Delta_{i,j} = \{ x\in \X \st x_i=x_j \}$.
Let $\Delta = \cup_{i,j} \Delta_{i,j}$ be the 
\emph{collision set} in $\X$. 
The space of collision-free configurations 
$\X \minus \Delta$
is denoted 
by $\hat \X$.
Let $\mathbf{n}$ denote 
the set $\{1,\dots, n\}$ of the first $n$ positive integers,
i.e.\ the set of indexes for the particles.
If $\kk\subseteq \n$ is a subset of 
the index set $\n$, let $\kk'$ denote its complement in $\n$. Given 
two subsets $\bfa,\bfb \subset \n$ with $\bfa\cap \bfb=\emptyset$, 
let $\Delta_{\bfa,\bfb}$ be the union
\(\Delta_{\bfa,\bfb} = \bigcup_{i\in \bfa, j \in \bfb} \Delta_{i,j}\).

Let $\alpha>0$ be a given positive real number.  
We consider the \emph{potential function} (the opposite of the potential
energy) defined by
\begin{equation}
\label{eq:potential}
U(x)  = \sum_{i<j} \dfrac{m_im_j}{|x_i - x_j|^\alpha}.
\end{equation}

The kinetic energy 
is defined 
(on the tangent bundle of $\X$)
by 
$K=\sum_{i=1}^n \dfrac{1}{2} m_i |\dot x_i|^2$ and 
the Lagrangian
is 
\begin{equation}
\label{eq:lagrangian}
L(x,\dot x) = L =K+U =
\sum_{i} \dfrac{1}{2} m_i |\dot x_i|^2 + 
\sum_{i<j} \dfrac{m_im_j}{|x_i - x_j|^\alpha}.
\end{equation}
When replacing the inertial frame with the uniform rotating one
the kinetic energy
needs to be changed into the corresponding form
\begin{equation}
\label{eq:kineticrotating}
K = \sum_i 
\dfrac{1}{2} m_i | \dot x_i + \Omega x_i | ^2,
\end{equation}
where $\Omega$ is a suitable linear map $V \to V$, which  
does not depend on $t$. For example, in dimension $d=3$
if 
the constant angular velocity is $\omega$ and the rotation 
axis is $\bomega$, with $|\bomega|=\omega$, we obtain 
$\Omega x_i = \bomega \times x_i$.
Accordingly, the Lagrangian for the rotating frame is
\begin{equation}
\label{eq:lagrangian_rotating}
L(x,\dot x) =
\sum_{i} \dfrac{1}{2} m_i |\dot x_i + \Omega x_i |^2 + 
\sum_{i<j} \dfrac{m_im_j}{|x_i - x_j|^\alpha}.
\end{equation}

Let $\T\subset \R^2$ denote a circle in $\R^2$ of length $T=|\T|$. 
It can be identified with $\R/T\ze$, where $T\ze$ denotes 
the lattice generated by $T\in \R$. 
Moreover, let 
$\Lambda = H^1(\T,\X)$ be the Sobolev space
of the $L^2$ loops $\T \to \X$ with $L^2$ derivative.
It is a Hilbert space with scalar product
\begin{equation}
\label{eq:scalarproduct}
x \cdot y = \int_\T ( x(t) y(t) + \dot x(t)  \dot y(t) ) dt.
\end{equation}
The corresponding norm is denoted by $\|x\|$.
An equivalent norm is  given by
\begin{equation}
\label{eq:norm_equivalent}
\|x\|' = \left( 
\sum_{i=1}^n (\int_\T \dot x_i^2 dt  
+ [x_i]^2)
\right)^{\frac{1}{2}},
\end{equation}
where $[x]$ is the average $[x] = \frac{1}{T} \int_\T x dt$.
The subspace of collision-free loops in $\Lambda$ is denoted by
$\hat\Lambda \subset \Lambda$ and is defined by
$\hat\Lambda = H^1(\T, \hat \X) \subset H^1(\T,\X)$.

Given the Lagrangian $L$ of ~\ref{eq:lagrangian} or 
~\ref{eq:lagrangian_rotating},
the positive-defined function
$\action \from \Lambda \to \R\cup \{\infty\} $
defined by
\begin{equation}
\label{eq:action}
\action(x) = \int_{\T} L(x(t),\dot x(t) ) dt
\end{equation}
for every loop $x=x(t)$ in $\Lambda$
is termed  \emph{action functional} (or 
the \emph{Lagrangian action}).

The action functional $\action$ is of class $C^1$ 
on the subspace $\hat \Lambda \subset \Lambda$ 
consisting of collisionless loops.
Hence critical points
of $\action$ in $\hat\Lambda$ are $T$-periodic classical
solutions (of class $C^2$) of the Newton equations
\begin{equation}
\label{eq:newton}
m_i \ddot x_i = \frac{\partial U}{\partial x_i}.
\end{equation}
For $x=x(t) \in \Lambda$ let $x^{-1} \Delta \subset \T$ denote
the set of \emph{collision times}. 
Our approach is to minimize $\action$ in some closed
subsets of $\Lambda$, and such minimizers are not
necessarily in $\hat\Lambda$  since they can have collisions.

Let $\kk\subset \n$ be a subset of the index set $\n$. 
The \emph{partial kinetic energy} $K_\kk$ can  be defined as
\begin{equation}
\label{eq:Kk}
K_\kk = 
\sum_{i\in \kk} 
\dfrac{m_i}{2}  |\dot x_i|^2 
\end{equation}
and in a similar way let the \emph{partial potential function} 
$U_\kk$ be defined by
\begin{equation}
\label{eq:Uk}
U_\kk = 
\sum_{i,j\in \kk, i<j} \dfrac{m_i m_j}{|x_i -x_j|^\alpha}. 
\end{equation}
For every $\kk\subset \n$ by homogeneity 
\begin{equation}
\label{eq:homogeneity}
\sum_{i\in \kk} x_i \frac{\partial U_\kk}{x_i} = -\alpha U_\kk.
\end{equation}
Moreover, let $U_{\kk,\kk'}$ be the sum
\begin{equation}
\label{eq:Ukk}
U_{\kk,\kk'} = 
U - U_\kk - U_{\kk'},
\end{equation}
which can be  defined (and it is regular) for
$x\not\in \Delta_{\kk,\kk'}$.

Given a subset $\kk \subset \n$ we define the 
\emph{partial energy}
by 
\begin{equation}
\label{eq:energy}
E_\kk =  
K_\kk  - U_\kk.  
\end{equation}
In the same way, let $L_\kk$ denote the \emph{partial Lagrangian}
$L_\kk = K_\kk + U_\kk$. 
This Lagrangian is considered as a function of $t$, once
a path $x(t)$ is chosen. The (partial) Lagrangian operator
will be denoted by $\lag_\kk$, and its value on the path
$q=(q_i)_{i\in \kk}$ simply by $\lag_\kk(q)$.

\section{Symmetry constraints}
\label{section:symmetry}
Many interesting and 
well-known periodic orbits have a non-trivial symmetry group (Lagrange
and Euler orbits, Chenciner-Montgomery eight, as well as choreographies):
this idea is at the roots of our search of 
equivariant minimizers. 
We now introduce a general method of defining transformation
groups on $\Lambda$.
Let $G$ be a finite group, acting on a space $X$. The space $X$ is
then called \emph{$G$-equivariant} space.
We recall some
standard notation on groups and equivariant spaces.
If $H\subset G$ is a subgroup of $G$, then
$G_x = \{g\in G \st gx=x \}$ is termed the \emph{isotropy} group of $x$,
or the  \emph{fixer} of $x$ in $G$.
The space $X^H\subset X$ consists of all  points
$x\in X$ which are fixed by $H$, that is,
$X^H=\{x\in X \st G_x \supset H \}$.
Given two $G$-equivariant spaces $X$ and $Y$,
an \emph{equivariant map} $f\from X \to Y$ is a map
with the property that $f(g\cdot x) = g\cdot f(x)$
for every $g\in G$ and every $x\in X$.
An equivariant map $f$ induces, by restriction to the spaces
$X^H$ fixed by subgroups $H\subset G$, maps
$f^H\from X^H \to Y^H$.
If $H\subset G$ is a subgroup of $G$, its \emph{normalizer}
$N_GH$ is defined as $N_GH = \{ g\in G \st H^g = H \}$,
for every $H\subset G$ and every $g\in G$ the subgroup 
$H^g$ is defined as $H^g= g^{-1} H g$. 
The \emph{Weyl group} $W_GH$ is defined by
$W_GH = N_GH/H$.
For every $H\subset G$ 
the Weyl group $W_GH$ acts on $X^H$, and a  $G$-equivariant map
$f\from X \to Y$ between $G$-spaces induces by restriction 
a $W_GH$-equivariant map $f^H\from X^H \to Y^H$.

Consider a finite group $G$, a $2$-dimensional orthogonal representation 
$\tau\from G \to O(2)$ 
of $G$ and a $d$-dimensional orthogonal representation
$\rho\from G \to O(d)$. 
By $\tau$ and $\rho$ we can let $G$ act on the time circle $\T \subset \R^2$
and on the Euclidean space $V$, respectively.
Moreover, 
if $\sigma\from G \to \Sigma_n$ is a given group homomorphism 
from $G$ to the symmetric group $\Sigma_n$ on $n$ elements,
we can endow the set of indexes $\n= \{ 1,\dots, n \}$ (of the masses)
with a $G$-action.
As stated in the introduction, 
we will consider 
only homomorphisms $\sigma$ with the property that 
\begin{equation}
\label{eq:property}
\forall g\in G : \left( \sigma(g)(i) = j \implies m_i = m_j \right).
\end{equation}

Given $\rho$ as above and $\sigma$ with property ~\ref{eq:property}, 
$G$ acts orthogonally on the configuration space 
$\X$ 
by
\begin{equation}
\label{eq:Gaction}
\forall g\in G : g \cdot (x_1, x_2, \dots, x_n) = 
( \rho(g) x_{\sigma(g^{-1})(1)}, \rho(g) x_{\sigma(g^{-1})(2)}, \dots,
\rho(g) x_{\sigma(g^{-1})(n)} ).
\end{equation}
In short 
\[
\forall i \in \n : \left( gx \right)_i = g x_{g^{-1}i}.  
\]
As a representation  over the reals, $\X$ is equivalent 
to the tensor product 
$ V \otimes_\R \R_0[\n]$, 
where $\R_0[\n] = \R[\n] - 1$ is equal to 
the natural representation $\R[\n]$ minus the trivial 
representation.

Furthermore, given the representation $\tau$ and the action of $G$
on $\X$ by ~\ref{eq:Gaction}
we can consider the action of 
$G$ on $\Lambda$ given by
\begin{equation}
\label{eq:Gaction2}
\forall g \in G , \forall t \in \T , \forall x\in \Lambda :
(g \cdot x ) (t)  =  g x (g^{-1}t).
\end{equation}
The loops in $\Lambda^G$, i.e.\ the loops fixed by $G$,
are the equivariant loops $\T \to \X$, 
and $\Lambda ^G \subset \Lambda$ is a closed linear
subspace. Let $\action^G$ denote the restriction of the 
action functional ~\ref{eq:action} to $\Lambda^G$
\begin{equation}
\label{eq:restricted_action}
\action^G\from \Lambda^G \subset \Lambda \to \R\cup \infty.
\end{equation}

Since the action functional ~\ref{eq:action} is $G$-invariant (that 
is, $\action(x) = \action (g \cdot x )$ for every $g\in G$ and 
every $x\in \Lambda$)  and the collision set $\Delta$ 
is $G$-invariant in $\X$,
the following proposition holds 
(the Palais principle of symmetric criticality).
\begin{lemma}
\label{lemma:palais}
A critical point of $\action^G$ in $\hat\Lambda^G$ 
is a critical point of $\action$ in $\hat\Lambda$.
\end{lemma}
\begin{proof}
See \cite{palais}.
\end{proof}

Without loss of generality we can assume that 
\begin{equation}
\label{eq:assume1}
\ker \tau \cap \ker \rho \cap \ker \sigma  = 1.
\end{equation}
Otherwise,  we can consider 
$G' = G/(\ker \tau \cap \ker \rho \cap \ker \sigma)$
instead of $G$ and obtain 
$\Lambda ^{G'} = \Lambda^G$.
Moreover, we can assume that there exists no proper linear subspace
$V' \subsetneq V$ such that 
\begin{equation}
\label{eq:assume2}
\forall i\in \n, \forall x\in \Lambda^G, \forall t\in \T :
x_i(t) \in V' \subsetneq V,
\end{equation}
and that there is no integer $k\neq \pm 1$ such that 
\begin{equation}
\label{eq:assume3}
\forall x\in \Lambda^G : \exists y\in \Lambda \st
 \forall t \in \T : x(t) = y(kt).
\end{equation}
In this case we say that the action of $G$ on $\Lambda$ 
is non \emph{\reducible}.

\begin{remark}
\label{remark:assume2}
If $\ker \tau \cap \ker \sigma \neq 1$, then 
the action is reducible.
In fact, if $g\in \ker \tau \cap \ker \sigma$ and  $g\neq 1$,
then 
for every $i\in \n$, for every 
$x \in \Lambda^G$ and every $t\in \T$
the particle $x_i(t)$ belongs to  the fixed subspace 
$V^g \subset V$. Since by ~\ref{eq:assume1}
$g\not\in \ker \rho$, $V^g$ is  a proper subspace of $V$,
and hence ~\ref{eq:assume2} holds.
\end{remark}

\begin{remark}
Also if $| \ker \rho \cap \ker \sigma | > 2$
the action is reducible.
In fact, 
consider an element $g\in \ker \rho \cap \ker \sigma $ with
$g\neq 1$. By ~\ref{eq:assume1} it needs to act non-trivially on $\T$.
Since 
$x(gt) = x(t)$
for every $t\in \T$,
if $g$ acts as a rotation, then ~\ref{eq:assume3} holds. 
Now, 
$\ker \rho \cap \ker \sigma$ can be embedded naturally in $G/\ker \tau$,
which is a finite subgroup of $O(2)$. Hence if 
$|\ker \rho \cap \ker \sigma| > 2$, there needs to exist
at least an element $g$ acting as a rotation, so the claim is true.  
On the other hand, if $|\ker \rho \cap \ker \sigma | = 2$,
the non-trivial element $g$ might act as a rotation or as a reflection
in $\T$.  In the first case again ~\ref{eq:assume3} holds,
and hence the action is reducible, 
while in the second case not.
If there is a reflection $g$ of $\T$ such that $x(gt) = x(t)$
for every $t\in \T$,
then $x(t)$ is said a \emph{brake orbit}.
\end{remark}

For some choice of $\tau$, $\rho$ and  $\sigma$, it can be that 
for every equivariant loop the set of collision
times is not empty 
\begin{equation}
\label{eq:assume4}
\forall x\in \Lambda^G :
x^{-1}\Delta \neq \emptyset,
\end{equation}
that is,
\[
\emptyset = \hat\Lambda^G  \subset \Lambda^G.
\]
If this happens, we say that the action of $G$ on 
$\Lambda$ is \emph{\bdc}.

\begin{remark}
\label{remark:bound}
If $\ker \tau \cap \ker \rho \neq 1$, then 
it is easy to see that the action of $G$ is \bdc.
Therefore, $G$ is a finite subgroup of $O(2)\times O(d)$.
If $d=2$ this implies that $G$ is metabelian, since 
it is a subgroup of the direct product of two dihedral groups.
If $d=3$, then $G$ is a finite extension of a finite metabelian
group with a platonic group. Hence the only case in which $G$
is not solvable occurs when $G$ projects onto the icosahedral
group $A_5$ in $O(3)$.

Moreover, 
by remark ~\ref{remark:assume2}, it is possible to show 
that $G$ is a finite subgroup of $O(2)\times \Sigma_n$.
\end{remark}

Consider the normal subgroup $\ker \tau \lhd G$ and 
the quotient $\bar G = G / \ker \tau$. The Weyl group of $\ker \tau$ 
acts on the space
$\X^{\ker \tau}$ by restricting the action of 
$G$ on $\X$, so that the natural inclusion 
$\X^{\ker \tau} \to \X$
induces an isomorphism of Hilbert spaces
\begin{equation}
\label{diag:1}
\begin{aligned}\xymatrix{%
H^1(\T,\X^{\ker \tau})^{\bar G} \ar@{.>}[r]^{\hspace{12pt}\cong} 
\ar@{ >->}[d]
& H^1(\T,\X)^G  \ar@{=}[r] & \Lambda^G
\ar@{ >->}[d] \\
H^1(\T,\X^{\ker \tau}) \ar@{ >->}[r]^i & H^1(\T,\X) \ar@{=}[r] &  \Lambda  \\
}\end{aligned}
\end{equation}
By definition $\bar G$ acts effectively on $\T$, hence it is 
a dihedral group or a cyclic group.

\begin{defi}
\label{defi:cyclic_dihedral}
If the group $\bar G$ acts trivially on the orientation of $\T$,
then $\bar G$ is cyclic and we say that 
the action of $G$ on $\Lambda$ is of 
\emph{cyclic type}.

If the group $\bar G$ consists of a single reflection on $\T$,
then we say that action of $G$ on $\Lambda$
is of \emph{brake type}.

Otherwise, we say that the action of $G$ on $\Lambda$
is of \emph{dihedral type} (and it is possible to show that 
$\bar G$ is a dihedral group).
\end{defi}

\begin{defi}
\label{defi:Tisotropy}
The isotropy subgroups of the action of $G$ on $\T$ 
via $\tau$ are called \emph{$\T$-isotropy subgroups} of $G$. 
\end{defi}

\begin{remark}
Let $l$ be the number of distinct isotropy subgroups
of $\bar G$, with respect to the action of $\bar G$ in $\T$,
or, equivalently, the number of distinct $\T$-isotropy subgroups
of $G$.
If $l=1$, then the action is of cyclic type.
The maximal $\T$-isotropy group is the unique isotropy group,
which coincides with $\ker \tau$.
If $l=2$, then the action is of brake type,
and the $\T$-isotropy groups are
the maximal $\T$-isotropy subgroup and $\ker \tau$.
If $l\geq 3$, then the action is of dihedral type,
and the $\T$-isotropy subgroups are either maximal or $\ker \tau$.
This shows in particular 
that definition ~\ref{defi:cyclic_dihedral} is well-posed.
\end{remark}

\begin{defi}
\label{defi:fund_dom}
Let $\I\subset \T$ be the closure of 
a \emph{fundamental domain} for the action 
of $\bar G= G/\ker\tau$ on $\T$  defined as follows.
If the action type is cyclic, let $\I$ be a closed interval
connecting  the time $t=0$ in $\T$
with its image $zt$ under a cyclic generator 
$z$ of $\bar G$. Thus in this case $\I$ can be chosen
among infinitely many intervals.
If the action type is brake or dihedral, 
then  $\I$ is a closed interval with as boundary  two distinct 
points of $\T$ with non-minimal isotropy subgroups in $G$  
and with no other points in its interior having non-minimal isotropy. 
There are $|\bar G|$ such intervals.
\end{defi}

Let $H_0$ and $H_1$  denote the isotropy subgroups of such 
consecutive points. If the action type is brake,
it is easy to see that $H_0 = H_1 = G$; if the action type is dihedral,
then $H_0 $ and $ H_1$  are distinct proper subgroups of $G$ 
(they can be conjugated or not).
Furthermore, since 
\begin{equation}
\label{eq:fund_dom}
\T = \bigcup_{\bar g \in \bar G} g \I
\end{equation} 
and the interiors of the terms in the sum are disjoint,
the  fundamental domain $\I$ 
is always an interval of length $\dfrac{T}{|\bar G|}$.

\begin{remark}
By remark ~\ref{remark:bound} ,
if $\ker \tau \cap \ker \rho \neq 1$, then 
the action is \bdc. If the action is not of cyclic type
and for some $\T$-isotropy group $H\subsetneq G$ the intersection
$H \cap \ker \rho  \neq 1$, then $\X^H\subset \Delta$, and hence
the action is \bdc, since in this case 
at the time $t\in \T^H\neq \emptyset$
the configuration $x(t)$ necessarily belongs to $\X^H$.

Furthermore, in principle it is possible that an action is 
\bdc\space
 even if $H\cap \ker \rho = 1$ for every $\T$-isotropy $H\subsetneq G$.
If $\X^{\ker \tau}\minus \Delta$ is not connected but
$(\X^{\ker \tau} \minus \Delta)/_{\bar G}$ is connected,
then the action is \bdc, since necessarily any equivariant path
pass through $\Delta$.
\end{remark}

\section{Coercivity and generalized solutions}
\label{section:coercivity}

Existence of minimizers follows from coercivity
of the functional $\action^G$: we now prove proposition ~\ref{propo:coercive},
which gives a (necessary and sufficient) criterion to 
guarantee coercivity. Then, we will describe some 
important properties of minimizers (which \emph{a priori} might 
have collisions).
Let $I=I(x) = \sum_{i} m_i |x_i|^2$ denote  the moment of inertia 
of a configuration $x\in \X$
with
respect to the its center $0$.
The action functional $\action^G\from \Lambda^G \to \R \cup \infty$ 
of ~\ref{eq:action} is called
\emph{coercive} in $\Lambda^G $ 
if $\action^G(x)$  diverges to infinity as
the $H^1$-norm $\|x\|$ goes to infinity in $\Lambda^G$.
This property is essential in the variational approach,
since it guarantees --by classical arguments-- 
the existence of minimizers (and hence
of generalized solutions -- see definition \ref{defi:generalized_solution} 
below) 
of the restricted action
functional $\action^G$.
In this perspective proposition ~\ref{propo:coercive} 
gives a complete answer to the problem of finding 
symmetry constraints that yield a coercive functional $\action^G$.

Furthermore, even if it is a simple exercise in variational 
calculus, it is worth mentioning the important fact 
that minimizers of the 
action functional $\action^G$ are classical solutions
of the Newton equations outside collision times (see remark
\ref{remark:classical}).

\begin{propo}
\label{propo:coercive}
The action functional $\action^G$ is coercive if and only if 
$\X^G = 0$.
\end{propo}
\begin{proof}
For every $g\in G$, by changing variables, 
\[
\int_\T x(t) dt = \int_\T x(gt) dt = g \int_\T x(t) dt,
\]
therefore the average 
\(
\dfrac{1}{T}\int_\T x(t) dt \)
belongs to the fixed space \(\X^G\).
So, if $\X^G = \{O\}$, then for every $i\in \n$ the average
\[
\dfrac{1}{T}\int_\T x_i(t) dt =0
\]
vanishes -- for every loop in $\Lambda^G$,
and hence in $\Lambda^G$ the norm ~\ref{eq:norm_equivalent} 
is actually equivalent to the 
norm given by the square root of the 
integral of the kinetic energy,
so that there is a constant  $c>0$ such that 
\[
 \|x\| < c  (\int_\T \sum m_i \dot x_i^2)^{1/2}.
\]
Hence 
if $\X^G=0$
the Lagrangian action $\action^G$ is coercive:
if a sequence $x_n$ is such that 
$\|x_n\|\to \infty$, then the integrals of the kinetic
energy need to diverge to infinity, and hence
$\action^G(x_n)$ diverges.

Conversely, assume  $u \in \X^G\minus 0$. 
There is a loop $x(t)\in \Lambda^G$ for $\action^G$ 
with finite action.
If $k\gg 0$, the loop $x+ku \in \Lambda^G$
has the property that $\action^G(x + ku) < \action^G(x)$.
On the other hand $|ku| \to \infty$ as $k\to \infty$, and hence
$\action^G$ is not coercive.
\end{proof}

Now consider an action of $G$ given by 
a choice of $\tau, \rho, \sigma$ with property
~\ref{eq:property}. 
Consider 
a path $x(t)$ in 
$\X^{\ker \tau}$; up to a change of variables
we can  assume that $x(t)$ is defined in $[0,1]$.
Given a $C^1$ real function $\eta(t)$ 
with support in $[0,1]$, a real $\epsilon\geq 0$ and 
a subset $\kk \subset \n$,  
consider the 
variation $\delta^\epsilon(t)$
defined by
\begin{equation}
\label{eq:varied_path}
\begin{cases}
\delta^\epsilon_{i}(t) = x_i( t + \epsilon \eta(t) ) - x_i(t) & \mbox{if $i\in \kk$} \\
\delta^\epsilon_{i}(t) = 0 & \mbox{if $i\in \kk'$} \\
\end{cases}
\end{equation}

\begin{lemma}
\label{lemma:variation_action}
For every $C^1$ real function $\eta$  with support in $[0,1]$
and for every subset $\kk\subset \n$ such that 
$\forall t\in [0,1] : x(t) \not\in \Delta_{\kk,\kk'}$, 
the following
equation holds.
\[
\action(x+\delta^\epsilon) - \action(x) =
\epsilon 
\int_0^1 
\left( E_\kk \dot \eta + 
\frac{\partial U_{\kk,\kk'}}{x_\kk} \cdot \dot x_\kk \eta
\right) dt + o(\epsilon)
\]
\end{lemma}
\begin{proof}
The Lagrangian can be decomposed as 
\[
L = K_\kk + K_{\kk'} + U_\kk + U_{\kk'} + U_{\kk,\kk'},
\]
where the variation occurs only in the terms
$K_\kk$, $U_\kk$ and $U_{\kk,\kk'}$.
Thus
\begin{equation*}
\action(x+\delta^\epsilon) - \action(x) =
\int_0^1
\Delta L_\kk dt + \int_0^1 \Delta U_{\kk,\kk'} dt.
\end{equation*}
Now, the claim follows since
\begin{equation*}
\begin{aligned}
\int _0^1 \Delta L_\kk dt  & = 
\int_0^1 \left( (K_\kk)( 1 + \epsilon \dot \eta) + \dfrac{U_\kk}{1 + \epsilon \dot \eta}\right) dt
\\
& =\epsilon \int_0^1 E_\kk \dot\eta dt + o(\epsilon)  \\
\end{aligned}
\end{equation*}
and 
\[
\int _0^1 \Delta U_{\kk,\kk'} dt = 
\epsilon \int_0^1 \sum_{i\in\kk}\frac{\partial U_{\kk,\kk'}}{\partial x_i}
\cdot \dot x_i \eta dt + o(\epsilon).
\]
The latter equation holds because $U_{\kk,\kk'}$ is a $C^2$ 
function of $x$, since $\forall t: x(t) \not\in \Delta_{\kk,\kk'}$. 
We refer the reader to \cite{dacorogna} for further details in 
classical variational techniques.
\end{proof}

\begin{remark}
\label{remark:classical}
In particular, if $\kk = \n$, lemma 
~\ref{lemma:variation_action} 
and an argument similar to the proof of ~\ref{lemma:palais}
implies that 
minimizers of 
the action functional $\action^G$ satisfy  Euler-Lagrange equations 
outside the collision set, and hence outside collision times
minimizers 
are classical solutions of the Newton equations ~\ref{eq:newton}.
\end{remark}

\begin{defi}
\label{defi:minimizer}
We say that a path $x(t)\from (T_0,T_1) \to \X^{\ker\tau}$ 
(where $T_0$ or $T_1$ may be infinite)
is a \emph{local minimizer} if there is $\epsilon>0$ such that 
for every  variation
$\delta \in H^1( \R, \X^{\ker\tau})$ with compact 
support in $[T_0,T_1]$  and $\| \delta \|<\epsilon$,
$\action(x+\delta) \leq \action(x)$.
The path $x(t)$ is called a \emph{minimizer} 
(once the domain $[T_0,T_1]$
is fixed) if for every $\delta$ with compact support 
$\action(x+\delta) \leq \action(x)$.
\end{defi}

\begin{defi}
\label{defi:generalized_solution}
A $H^1$  path $x(t)$ defined on an interval $(T_0,T_1)$ is called a 
\emph{generalized solution} of the Newton equations ~\ref{eq:newton}
if 
$x(t)$ is a $C^2$ solution of~\ref{eq:newton} 
in $(T_0,T_1) \minus x^{-1}\Delta$,
the Lagrangian action of $x(t)$ on $(a,b)$ is finite  and 
the following property holds:
\begin{quote}
For every subset $\kk \subset \n$ 
and every interval $(t_0,t_1) \subset (T_0,T_1)$ such that 
$\forall t\in [t_0,t_1] : x(t) \not\in \Delta_{\kk,\kk'} $,
the partial energy
$E_\kk$
is a $H^1$ function of the time $t$ in $[t_0,t_1]$:
\begin{equation}
\label{eq:generalized_solution}
\forall t\in [t_0,t_1] : x(t) \not\in \Delta_{\kk,\kk'} 
\implies 
E_\kk \in H^1( (t_0,t_1), \X ).
\end{equation}
\end{quote}
\end{defi}

\begin{remark}
Since $x$ is continuous, $x^{-1} \Delta$ is closed,
so that the definition is consistent. The hypothesis that 
the action is finite implies that $x^{-1}\Delta$ has measure zero.

Furthermore, the definition describes the following property
(which (local) minimizers of the Lagrangian action have -- see later 
proposition \ref{propo:gen_sol}):
if you consider a cluster $\kk\subset \n$ and an interval
of time $[t_0,t_1]$  such that there are no collisions
between particles in $\kk$ and in its complementary
cluster $\kk'$ --- that is, particles in $\kk$ collide only 
with particles in $\kk$ --- then the partial energy $E_\kk$
is $H^1$, and in particular is continuous.
In general at collision times discontinuous energy transfers can occur
inside colliding clusters, and this property describes the simple fact
that transfers cannot occur between non-colliding clusters.
\end{remark}

\begin{remark}
In \cite{ambrosetti92,amco93} 
a path $x(t)$ defined on $(T_0,T_1)$ is called a 
weak solution of  
~\ref{eq:newton}
if $(T_0,T_1) \minus x^{-1}\Delta$ is open and dense in $(T_0,T_1)$,
$x(t)$ is a $C^2$ solution 
of~\ref{eq:newton} on $(T_0,T_1) \minus x^{-1}\Delta$ and the
energy of the solution is constant, i.e.\ the function
$E$ of ~\ref{eq:energy}
does not depend on $t\in (T_0,T_1) \minus x^{-1} \Delta$.
By applying definition ~\ref{defi:generalized_solution} with
$\kk = \n$, we obtain that 
a generalized solution is in particular
a weak solution.
\end{remark}

\begin{remark}
\label{rem:rem}
Assume that 
$x(t)$ is a generalized solution of ~\ref{eq:newton} in $(t_0,t_1)$, 
and that $\kk \subset \n$ is a subset such that the particles
in $\kk$ do not collide in $(t_0,t_1)$. Then $x_\kk(t)$ can be extended to 
a $C^2$ solution of the partial problem
\[
\forall i \in \kk : m_i \ddot x_i = \frac{\partial U}{\partial x_i}.
\]
Furthermore, given $\kk \subset \n$, 
let $x_0$ be the center of mass of the particles $x_i$ with $i\in \kk$,
\[
x_0 = \frac{1}{m_0} \sum_{i\in \kk} m_i x_i,
\]
where $m_0 = \sum_{i\in \kk} m_i$.
If $x(t)$ is a generalized solution  such that  for every $t$ the 
configuration 
$x(t) \not\in \Delta_{\kk,\kk'}$,
then 
the trajectory of the center of mass
$x_0(t)$ is a  $C^2$ curve such that 
\[
m_0 \ddot x_0 = \sum_{i\in \kk} \frac{\partial U_{\kk,\kk'}}{\partial x_i},
\]
since 
$\sum_{i\in \kk} \frac{\partial U_\kk}{\partial x_i} = 0$.
\end{remark}

\begin{propo}
\label{propo:gen_sol}
If $x\from [T_0,T_1] \to \X^{\ker \tau}$ is a  local minimizer  
of the Lagrangian action $\action$, then $x$ 
is a generalized solution
of ~\ref{eq:newton}. If collisions do not occur, 
then it is a classical solution.
\end{propo}
\begin{proof}
It is a classical result that $x(t)$  is $C^2$ outside the collision times
(see remark ~\ref{remark:classical})
and that the action $\action$ is finite.
It is only left to prove property ~\ref{eq:generalized_solution}.
Let $[t_0,t_1]\subset [T_0,T_1]$ be a finite interval  and $\kk\subset \n$ 
a subset of the index set $\n$. By reparametrizing the time interval
and rescaling the problem
we can assume that $[t_0,t_1] = [0,1]$.
So, we assume that 
for every  $t\in [0,1]$ the configuration 
$x(t) \not\in \Delta_{\kk,\kk'}$. Then, 
since $x$ is a minimum
for the function $\action(x+\delta^\epsilon)$ of $\epsilon$ defined 
in ~\ref{eq:varied_path},
by lemma ~\ref{lemma:variation_action} for every choice of $\eta$, 
\[
\int_{0}^{1} 
\left( E_\kk \dot \eta + 
\sum_{i\in \kk}
\frac{\partial U_{\kk,\kk'}}{\partial x_i} \cdot \dot x_i \eta
\right) dt = 0. 
\]
Moreover, for every $t$ of 
its domain and every $\epsilon$, $\delta^\epsilon(t) \in \X^{\ker\tau}$.
Hence 
\[
E_\kk(s) - E_\kk(0)  = 
\int_0^s
\sum_{i\in\kk}
\frac{\partial U_{\kk,\kk'}}{\partial x_i(t)} \cdot \dot x_i(t) dt
\]
Since for  every $i\in \kk$
the  function $\frac{\partial U_{\kk,\kk'}}{\partial x_i}$  
is continuous in $t$, and therefore
the product  
$\frac{\partial U_{\kk,\kk'}}{ \partial x_i} \cdot \dot x_i$ 
is in $L^2( (0,1), \X^{\ker\tau} )$,
the partial energy $E_\kk$ is a continuous function of $t$.
\end{proof}

\begin{propo}
\label{propo:generalizedsolution}
If $\X^G = 0$, then 
there exists at least a minimum of the Lagrangian action $\action^G$,
which 
yields 
a generalized solution of ~\ref{eq:newton} in $\Lambda^G$.
\end{propo}
\begin{proof}
Proposition ~\ref{propo:coercive} implies that $\action^G$ is coercive 
and so
by classical results the minimum $x(t)$ of $\action^G$ in $\Lambda^G$ exists
(see e.g. \cite{dacorogna}). 
By proposition ~\ref{propo:gen_sol} it is possible to show
that if $x$ is a local minimum for $\action^G$, since then  
its restriction $x|\I$ to 
the fundamental domain (see ~\ref{defi:fund_dom})
is a local minimizer, the restriction to a fundamental domain
$x(t)|\I$ is a  generalized solution. 
Therefore $x$ is a generalized solution in the times
of non-minimal isotropy.
To complete the proof, it is necessary to show that 
$x(t)$ is a generalized solution in $\T$,
that is, that the partial energies $E_\kk$ are continuous 
when passing times with maximal $\T$-isotropy. These exist only
if the action is not of cyclic type, since otherwise
one can change the fundamental domain and apply again proposition
~\ref{propo:gen_sol}. 
Hence assume that the action is not of cyclic type and (without
loss of generality) that
$t_0=0\in \T$ is a time with maximal
$\T$-isotropy. Let $\epsilon>0$ be a small real number.
Since $x(t)$ is a generalized solution in $[0,\epsilon]$
and in $[-\epsilon,0]$ and $E_\kk(gt) = E_\kk(t)$
for every $g\in G$,
we just need to show that there exists
$g\in G$ such that $g[-\epsilon,0]=[0,\epsilon]$.
But this is a consequence  of the fact that 
$0$ has maximal $\T$-isotropy, and so there exists an element 
$h \in G$ that acts on $\T$ as a reflection around $0$.
\end{proof}

\section{Isolated collisions}
\label{section:coll}

One can always 
assume a 
collision in a minimizer 
to be isolated (more generally, it holds 
for generalized solutions: see proposition ~\ref{propo:isolated} below).
In this section we prove this property, after some
background material,
as a consequence
of the Lagrange-Jacobi equalities for colliding clusters
~\ref{coro:lagjac} (and the definig property of generalized
solutions). We first define interior collisions,
boundary collisions and locally minimal collisions (it is to be noted
that under this definition locally minimal collisions are not
necessarily collisions in trajectories that minimize the action;
actually, nowhere in this section the minimality of the action
is assumed).

\begin{defi}
\label{defi:interior_collisions}
A collision occurring at time $t\in \T$ is called
\emph{interior collision} if $t$ has principal
isotropy type in $\T$ with respect 
to the action of $G$, i.e.\ if the isotropy group of $t\in \T$
with respect to the action of $G$ is minimal in $G$.
Otherwise, it is termed \emph{boundary collision}.
It is easy to see that 
interior collisions belong to the interior
of a fundamental domain $\I\subset \T$ (see  ~\ref{defi:fund_dom}),
while boundary collisions belong to the boundary $\partial \I$
for some choice of $\I$.
\end{defi}

\begin{defi}\label{defi:loc_min}
Let $x(t)$ be a path. A collision at time $t_0$ is termed
\emph{locally minimal} if there is a subset $\kk\subset \n$ such that 
$\forall i,j \in \kk : x_i(t_0)=x_j(t_0)$, 
$x(t_0) \not\in \Delta_{\kk,\kk'}$, and there exists
a neighborhood $(t_0-\epsilon,t_0+\epsilon)$  of $t_0$  such that 
if $t \in (t_0-\epsilon,t_0+\epsilon)$ is a collision time,
then 
$\forall i,j \in \kk : x_i(t)=x_j(t)$. 
We say that the collision is a locally minimal collision 
\emph{of type $\kk$}.
Furthermore,
we say that $t_0$ is an isolated collision time 
for $x_\kk = (x_i)_{i\in \kk}$ if 
$t_0$ is an isolated point in 
$x_\kk^{-1} \Delta$.
In this case also the collision is termed \emph{isolated}.
\end{defi}

\begin{remark}
\label{rem:loc_min}
The notion of \emph{locally minimal}
refers to the number of bodies and not to the 
lagrangian action. Actually,  the reason 
of this notation is the following:
the poset $\mathcal{P}$ of partitions of $\n$ 
(with initial object $\{ \{1\},\{2\}, \dots ,\{n\} \}$ 
and final object $\{\n\}$, ordered by inclusion) 
can be given the 
topology whose closed sets are 
the intervals 
$\{ P \in \mathcal{P} \st P\geq A \}$ 
for any partition $A\in \mathcal{P}$ of $\n$,
and the map $p$, which sends the time $t$ to 
the collisions-partition $p$ given by
the configuration $x(t)$, is continuous.
Locally minimal collision times are just local minima
of the map $p$.
For example, the reader can consider the following example
of non-locally minimal collision times: collisions
13, 24, 13, 24 \dots converging to collision 1234,
or collisions 12 13 23 12 13 23 converging to the triple
collision 123.
\end{remark}
Let  $I_\kk$ denote the momentum of inertia with respect to
the center of mass of the bodies in $\kk$, 
\begin{equation}
\label{eq:Ik}
I_\kk = \sum_{i\in \kk} m_i ( x_i - x_0)^2 
= \sum_{i\in \kk} m_i q_i,
\end{equation}
where $x_0$ is 
given by 
$x_0 = \sum_{i\in \kk} m_i x_i / m_0$,
with $m_0 = \sum_{i\in \kk} m_i$.
All the bodies in $\kk$ collide in $x_0$ if and only if 
$I_\kk = 0$.

\begin{lemma}
\label{lemma:5}
Assume that 
$x(t)$  is a continuous curve, solution  of ~\ref{eq:newton}
in $[t_0,t_1]\minus x^{-1}\Delta$, and  such that 
$x(t) \not \in \Delta_{\kk,\kk'}$ for every $t$.
Then a.e. 
\[
\frac{\ddot I_\kk}{2}  =
2E_\kk(t) + (2- \alpha) U_\kk(t) + R(t),
\]
where there remainder $R(t)$ is a $C^0$ function in $[t_0,t_1]$.
\end{lemma}
\begin{proof}
We obtain the following equalities, where the last 
holds by ~\ref{eq:homogeneity}.
\begin{equation*}
\begin{aligned}
\frac{1}{2} \ddot  I_\kk = & 
\sum_{i\in \kk} m_i ( \dot x_i ^2 + x_i \ddot x_i ) -
m_0 \dot x_0^2 - m_0 x_0\ddot x_0  \\
& = 
2 K_\kk + 
\sum_{i\in \kk} x_i \frac{ \partial U}{\partial x_i} 
- m_0 \dot x_0^2 - m_0 x_0 \ddot x_0 \\
& = 
2 K_\kk  
- \alpha U_\kk
+ \sum_{i\in \kk} x_i \frac{ \partial U_{\kk,\kk'}}{\partial x_i} 
- m_0 \dot x_0^2 - m_0 x_0 \ddot x_0.
\end{aligned}
\end{equation*}
Therefore
\begin{equation} \label{eq:src_lagjac}
\frac{1}{2} \ddot  I_\kk = 
2 E_\kk  + (2-\alpha) U_\kk + R,
\end{equation}
where $R = 
\sum_{i\in \kk} x_i \frac{ \partial U_{\kk,\kk'}}{\partial x_i} 
- m_0 \dot x_0^2 - m_0 x_0 \ddot x_0$.
The claim follows by the fact that 
$x_0(t)$ is a $C^2$ function (see remark ~\ref{rem:rem}),
$x_i(t)$  and 
$ \frac{ \partial U_{\kk,\kk'}}{\partial x_i}$   are continuous,
hence $R$ is continuous.
\end{proof}

\begin{remark}
It is possible to notice, deriving the expression of $R(t)$, that 
$\dot R$
is a linear combination of terms in $\dot x_i$, with continuous
coefficients, and hence
\begin{equation}
\label{eq:est_R}
\dot R(t) < c K^{1/2} + b.
\end{equation}
for some constants $c>0$ and $b\in \R$.

Furthermore, since the derivative of the partial energy $E_\kk$
is 
\begin{equation}\label{eq:Ei}
\dot  E_\kk = \sum_{i\in \kk} \frac{\partial U_{\kk,\kk'}}{x_i} \dot x_i 
=
-\alpha 
\sum_{i\in \kk, j\in \kk'} m_i m_j \frac{(x_i - x_j) \dot x_i}{|x_i - x_j|^{2+\alpha}},
\end{equation}
there is a constant $c>0$ such that 
\begin{equation}\label{eq:est_E}
\dot E_\kk < c K_\kk^{1/2}.
\end{equation}
\end{remark}

\begin{coro}[Lagrange--Jacobi]
\label{coro:lagjac}
Assume that 
$x(t)$  is a continuous curve,  solution  of ~\ref{eq:newton}
in $[t_0,t_1]\minus x^{-1}\Delta$, such that 
$x(t) \not \in \Delta_{\kk,\kk'}$ for every $t$.
Then for some 
function $C^0$ continuous in $[t_0,t_1]$:
 \begin{equation*}
\begin{aligned}
\frac{1}{2} \ddot I_\kk  & = 2E_\kk + (2-\alpha) U_\kk + C^0\\
& = (2-\alpha) K_\kk + \alpha E_\kk + C^0.
\end{aligned}
\end{equation*}
\end{coro}

\begin{coro}
\label{coro:loc_min}
Let $x(t)$ be a generalized solution.
Then any locally minimal collision for $x(t)$
is isolated.
\end{coro}
\begin{proof}
Assume that at the time $t_0$ 
a collision of type $\kk \subset \n$ 
occurs. 
Consider a neighborhood $(t_0-\epsilon,t_0+\epsilon)$  of $t_0$.
Without loss of generality we can assume that 
for every $t$ and $\forall i \in \kk$: $x_i(t) \not \in \Delta_{\kk,\kk'}$.
Thus the function $I_\kk$  defined in ~\ref{eq:Ik} is non-negative,
$C^2$ where non-zero (since $t_0$ is a locally minimal 
collision time) and continuous.
It is zero in the set (of measure zero)
consisting of collisions of type $\kk$. 
If $t_0$ is not isolated in $x^{-1} \Delta$, 
then necessarily there is a sequence of intervals 
$(a_j, b_j)$ such that $I_\kk(a_j) = I_\kk(b_j)=0$,
$a_j < b_j$ and $a_j\to t_0$, $b_j\to t_0$. Thus
there is a sequence $t_j \to t_0$ of points where 
$\ddot I_\kk(t_j) \leq 0$  for every $j$. 
But by lemma ~\ref{lemma:5}
\(
\frac{\ddot I_\kk}{2}(t_j) - 
2E_\kk(t_j) + (\alpha-2) U_\kk(t_j)
\)
is continuous and hence bounded. Since $E_\kk$ is $H^1$, this implies that 
$\frac{1}{2} \ddot I_\kk(t_j) + (\alpha - 2) U_\kk(t_j)$ is bounded,
which cannot be. Thus $t_0$ is isolated. 
\end{proof}

\begin{propo}
\label{propo:isolated}
Let $x$ be a generalized solution  of ~\ref{eq:newton}  
such that $x^{-1}\Delta \neq \emptyset$. 
Then there exists an isolated 
collision.
\end{propo}
\begin{proof}
If in a time interval there are collisions, 
then there are locally minimal collisions.
To complete the proof it suffices to apply 
a finite number of times corollary ~\ref{coro:loc_min} 
or  to argue 
by contradiction.
\end{proof}

\begin{remark}
An alternative proof of 
proposition \ref{propo:isolated} can be found 
also as Theorem 4.1.15 of \cite{andrea_thesis}
and as well in sec. 3 of \cite{chenICM} (for $\alpha=1$).
The basic structure of all the  proofs use the same ideas: approximate 
constancy of the energy of a cluster plus a cluster 
version of the Lagrange-Jacobi identity.
In \cite{andrea_thesis} proposition \ref{propo:isolated}
is phrased as 
``if there is
no subcluster collision in a neighborhood of $t_0$ then
$t_0$ is isolated'' and the idea of the proof is
attributed there to R. Montgomery.
It is clear by definition  \ref{defi:loc_min} 
that a locally minimal collision is 
a collision without subcluster collisions in a neighborhood.
Since we consider the presentation in this section, besides being 
more general,   simpler and more direct,
we decided to include it for the reader's convenience.
\end{remark}

\section{Asymptotic estimates}
\label{sec:asy}

Different proofs of the asymptotic estimates for 
total or partial colliding clusters 
(in the case $\alpha=1$ or for total collisions)
can be found in the literature \cite{wintner,sperling,elbialy,chen_coll},
that extend and simplify the original Sundman estimates.
As remarked already by Wintner \cite{wintner},
the Tauberian lemmata used in the proofs of \cite{sperling},
essentially concerning the regularity of the variation
of the momentum function $I$,
can be traced back
to Sundman.
In this section 
we extend
and simplify Sperling asymptotic estimates 
for partial collisions 
\cite{sperling}
to every $\alpha\in(0,2)$ and we prove some 
other results necessary to 
introduce the blow-up technique.
The main lemma that implies the estimates is 
the Tauberian lemma ~\ref{lemma:tech}
(inspired by and generalizing 
\S 337 -- \S 338, page 255--257, of \cite{wintner};
lemma ~\ref{lemma:tech} replaces the various Tauberian lemmata used in the proof
of \cite{sperling}), 
which is a key step towards the more technical lemma ~\ref{lemma:bach}
(~\ref{lemma:bach} studies for every $\alpha$
the properties of a Sundman  function).
In order to obtain asymptotic estimates for colliding
clusters, it is necessary to prove first that the
partial energies $E_\kk$ are bounded in a neighborhood
of a collision time. Our approach follows Sperling's idea
\cite{sperling} of proving the boundedness as a consequence
of the asymptotic behaviour of the total kinetic energy
of the $n$ particles.
Thus we can prove proposition ~\ref{propo:asy}, 
and some of its consequences: the limit set of 
a collision of type $\kk$ is a central configuration
of $\kk$ bodies ~\ref{propo:cent_conf}; moreover, 
in ~\ref{propo:srcblowup}  it is proved that 
any converging 
sequence of normalized configurations in the collision  
trajectory yield a sequence of solutions converging to 
a blow-up solution $\bar q$.

\begin{lemma}
\label{lemma:tech}
Let $\varphi$ be an integrable real function defined on
the interval $(0,\epsilon)$, $\epsilon>0$, differentiable, 
  such that 
\begin{equation*}
\lim_{t \to 0} \frac{1}{t} \int_0^t \varphi = \eta
\end{equation*}
and 
such that there exists a continuous function $A\from \R \to \R$
with the property that 
\begin{equation*}
\forall t \in (0,\epsilon) : \ \ 
|\frac{d\varphi}{dt}| < \frac{ A(\varphi) } { t }.
\end{equation*}
Then 
\[
\lim_{t \to 0} \varphi(t) = \eta  = 
\lim_{t \to 0} \frac{1}{t} \int_0^t \varphi.
\]
\end{lemma}
\begin{proof}
Since $\liminf \varphi \leq \eta \leq \limsup \varphi$, 
we only need to show that 
the limit exists, i.e.\ that $\liminf \varphi = \limsup \varphi$.
If, on the contrary,
$\liminf \varphi < \limsup \varphi$, 
we can have two cases: either $\eta < \limsup \varphi$
or $\eta = \limsup \varphi$.
In the first case, there exist 
$l_1, l_2 \in \R$ such that 
$\liminf \varphi \leq \eta <  l_1 < l_2 < \limsup \varphi$ 
and a sequence $[a_j,b_j]$ of intervals in $(0,\epsilon)$
such that $a_j<b_j \to 0$ and 
$\varphi a_j = l_1$, $\varphi b_j = l_2$,
$t\in [a_j,b_j] \implies \varphi(t) \in [l_1,l_2]$. 
The second case can be reduced to the first by taking
$-\varphi$ in place of $\varphi$.
Since $|\dot \varphi(t)| < A(\varphi) / t$,
\[
0 < l_2 - l_1  =  
\int_{a_j}^{b_j} \dot \varphi  <   \int_{a_j}^{b_j} \dfrac{c}{t} =
c \log \frac{b_j}{a_j},
\]
where $c>0$ is a constant such that $\varphi \in [l_1,l_2] \implies 
A(\varphi) <c$. 
Therefore for every $j$
\[
\frac{b_j}{a_j} \geq \gamma > 1
\]
where $\gamma =  e^{\frac{l_2 - l_1}{c}}> 1$ is constant.
But on the other hand for every $j$
\[
l_1 ( \frac{b_j}{a_j} - 1) 
< 
\frac{1}{a_j} \int_{a_j}^{b_j} \varphi  =
\frac{1}{a_j} \int_{0}^{b_j} \varphi - \frac{1}{a_j} \int_0^{a_j} \varphi,
\]
and therefore
\[
\gamma ( l_1 - \frac{1}{b_j} \int_0^{b_j} \varphi   ) \leq 
\frac{b_j}{a_j} ( l_1 - \frac{1}{b_j} \int_0^{b_j} \varphi )
= 
\frac{b_j}{a_j}  l_1 - \frac{1}{a_j}  \int_0^{b_j} \varphi  
\leq   
l_1 - \frac{1}{a_j} \int_0^{a_j} \varphi.
\]
By taking limits, we obtain
\[
\gamma( l_1 - \eta ) \leq (l_1 - \eta) \implies l_1 \leq \eta
\]
which contradicts the assumption that  $l_1> \eta$.
\end{proof}

\begin{lemma}
\label{lemma:bach}
Let $\ff(t)$ be a function defined on $(0,\epsilon)$, $\epsilon>0$, 
with the following properties:
\begin{itemize}
\item 
There are constants $a$, $b>2$ and  $c$ such that 
$ \dot \ff^2  + a^2 \leq  b \ff \ddot \ff + c \ff$.
\item
$\forall t$: $\ff(t) \geq 0$, $\dot \ff(t) \geq 0$.
\item $\lim_{t\to 0} \ff(t) = 0$.
\item 
There is a constant $d>0$ such that 
$\forall t$: $\ddot \ff \ff^{(b-2)/b}(t) \geq d >0$.
\end{itemize} 
Then, the following properties are true:
\begin{enumerate}
\item
$a^2=0$.
\item
There exists  the positive limit
$\lim_{t\to 0}  \dot \ff^2 \ff^{-2/b} = \eta > 0$.
\item 
\label{es:f}
$\ff(t)  =
 (\sqrt{\eta} t)^{b/(b-1)} + o(t^{b/(b-1)})$.
\item 
\label{es:fp}
$\dot \ff(t) =  
\sqrt{\eta}^{b/(b-1)} t^{1/(b-1)}  + o(t^{1/(b-1)})$.
\end{enumerate}
Moreover, if  
there is a constant $e$ such that 
$|\frac{d^3 \ff}{dt^3}| < e \ddot \ff^\gamma$, 
with the exponent $\gamma = \frac{2b-3}{b-2}$,
then 
\begin{equation}
\label{es:fs}
\ddot \ff(t) 
= 
\frac{\eta^{b/2}}{b} \left( \frac{b-1}{b} \eta^{b/2} t \right)^{\frac{2-b}{b-1}}.
\end{equation}
\end{lemma}
\begin{proof}
Define $\beta = 2/b$ and 
consider the following auxiliary function
\[
Q(t) =  \frac{c\beta}{1-\beta} \ff^{1-\beta} + 
\ff^{-\beta}( \dot \ff^2 + a^2 ).
\]
Since $b>2$, $\beta \in (0,1)$ and thus $\ff^{1-\beta} \to 0$, so that  
$Q(t)$ is bounded by below.
Consider the derivative $\dot Q(t)$:
\[
\dot Q(t) = \beta \ff^{-\beta} \dot \ff 
\left [
c + \frac{2}{\beta} \ddot \ff - \frac{ \dot \ff^2 + a^2 } {\ff }.
\right]
\]
Since $\beta>0$, $\dot \ff\geq 0 $ and 
by hypothesis the term in square brackets is non-negative,
$\dot Q(t) \geq 0$ for every $t$ and hence $Q(t)$ is monotone
and thus it has a limit 
\[
\eta = \lim_{t\to 0} Q(t).
\]
This implies that $a^2=0$ and 
$\ff^{-\beta} \dot \ff^2 \to \eta$ as $t\to 0$,
which means that
\[
\lim_{t\to 0}  \dot \ff^2 \ff^{-2/b} = \eta.
\]

Since 
there is a constant $d>0$ such that 
$\ddot \ff \ff^{(b-2)/b}(t) \geq d >0$, 
we can write
\[
\dot \ff^2 = 2 \int_0^t \dot \ff \ddot \ff \geq 2 d 
\int_0^t \dot \ff \ff^{(2-b)/b} 
=
\frac{2bd}{2-b} \ff^{2/b},
\]
and hence
\[
\eta = \lim_{t\to 0} { \dot \ff^2 \ff^{-2/b}} \geq 
\frac{2bd}{2-b}> 0.
\]

By integrating the equation
\[
\dot \ff \ff^{-\beta/2} = \sqrt{\eta}  + o(1)
\]
we obtain 
\[
\ff^{1-\beta/2} = \dfrac{\sqrt{\eta}}{1-\beta/2} t + o(t) \implies
\ff(t) = (\dfrac{\sqrt{\eta}}{1-\beta/2} t)^{2/(2-\beta)} + o(t^{2/(2-\beta)})
\]
\[
\implies 
\ff(t) = (\frac{b-1}{b} \sqrt{\eta} t)^{b/(b-1)} + o(t^{b/(b-1)}),
\]
\[
\dot \ff = 
\ff^{\beta/2} ( \sqrt{\eta} + o(1) ) =
\ff^{1/b} ( \sqrt{\eta} + o(1) ) =
\sqrt{\eta} 
(\frac{b-1}{b} \sqrt{\eta} t)^{1/(b-1)} + o(t^{1/(b-1)}) 
\]
\begin{equation}\label{eq:sp}
\implies 
\dot \ff(t) = 
(\frac{b-1}{b} \sqrt{\eta}^b t)^{1/(b-1)} + o(t^{1/(b-1)}).
\end{equation}

Now, consider the non-negative function 
\[
F(t) = \dot \ff^{b-1}.
\] 
By equation ~\ref{es:fp}, $F(t)/t = \frac{b-1}{b} \eta^{b/2} + o(1)$, i.e.
\begin{equation}
\label{eq:limit}
\lim_{t \to 0} \frac{F(t)}{t} = \frac{b-1}{b} \eta^{b/2}.
\end{equation}
Its first and second derivatives are
\[
\dot F = (b-1) \dot\ff^{b-2} \ddot \ff,
\]
\[
\ddot F = (b-1)(b-2) \dot \ff^{b-3} \ddot \ff^2 + (b-1)\dot \ff^{b-2} 
\mydddot{\ff}.
\]
Therefore 
\[
F \ddot F =  c_1 \dot ( \dot \ff^{b-2} \ddot \ff)^2 + 
c_2  \dot \ff^{2b-3} \mydddot{\ff} 
\]
for some positive constants $c_1,c_2$, and hence,
since $|\mydddot{\ff} | \leq k \ddot \ff ^{\frac{2b-3}{b-2}}$, 
there are (other) constants $c_1$, $c_2$ such that 
\[
| F \ddot F | \leq   c_1 \dot F ^2 + 
c_2  \dot F^{\frac{2b-3}{b-2}},
\]
hence there is a continuous function $A\from \R \to \R$ such that 
for every $t\in (0,\epsilon)$,
\begin{equation}
\label{eq:second}
 |\ddot F(t) | \leq \frac{A(\dot F(t))}{t}.  
\end{equation}
Since $F(t)/t = \frac{1}{t} \int_0^t \dot F$, by equation
~\ref{eq:limit} one obtains
\begin{equation*}
\lim_{t \to 0} \frac{1}{t} \int_0^t \dot F =  \frac{b-1}{b} \eta^{b/2}.
\end{equation*}
Now one can apply lemma ~\ref{lemma:tech}  to the function $\dot F$,
so to get
\[
\frac{b-1}{b} \eta^{b/2} =
\lim_{t\to 0} \dot F(t) = 
\lim_{t\to 0} 
(b-1) \dot \ff^{b-2} \ddot \ff \implies 
\]
\[
\implies
\ddot \ff =
\frac{\eta^{b/2}}{b} \left( \frac{b-1}{b} \eta^{b/2} t \right)^{-\frac{b-2}{b-1}}.
\]
This completes the proof.
\end{proof}

Now we consider a solution of the Newton equations ~\ref{eq:newton} 
$x(t)$ defined in the time interval $(0,\epsilon)$, 
without collisions in $ (0,\epsilon )$  and 
with an isolated  collision in $t=0$.
A \emph{colliding cluster} of particles  is 
a subset $\kk \subset \n$ such that $x(0) \in \Delta_\kk$
and 
$x(0) \not\in \Delta_{\kk,\kk'}$. 
Let $\kk_0 \subset \n$ be the subset of non-colliding particles. 
The colliding clusters yield a partition of $\n\minus \kk_0$ into
subsets  
which we denote by $\kk_1, \kk_2, \dots, \kk_l$.
Let $\kk$  denote a generic colliding cluster.

As above, let $x_0$ denote the center of mass
$x_0 = m_0^{-1} \sum_{i\in \kk} m_i x_i$, with
$m_0 = \sum_{i\in\kk} m_i$.
Let $c_\kk$ denote the angular momentum with respect to  $x_0$
\[
c_\kk = \sum_{i\in\kk} m_i(x_i - x_0)\times (\dot x_i - \dot x_0).
\]
For a collection $\h$ of clusters $\kk$, let 
$c_\h$ denote the sum 
$c_\h = \sum_{\kk\subset  \h} c_\kk$.
Here it is necessary a word of warning about the notation:
If $\h$ is the collections of the clusters $\kk_1$, $\kk_2$, \dots
$\kk_l$, then when we write that a sum ranges over 
$\kk\subset \h$ we mean that the index $\kk$ 
assumes the values $\kk_1$, $\kk_2$, \dots, $\kk_l$.
Hence, for example,
the sum $\sum_{\kk \subset \h} c_\kk$
means the sum of the angular momenta $c_\kk$,  as the cluster
$\kk$ ranges over the collection $\h$ of clusters.

\begin{nr}
\label{nr1}
$\dot  c_k I_\kk^{-1}$ is bounded.
\end{nr}
\begin{proof}
\[
\begin{split}
\dot c_\kk = \sum_{i\in\kk} m_i (x_i-x_0)\times ( \ddot x_i - \ddot x_0 ) 
=\\
 \sum_{i\in\kk} m_i (x_i-x_0)\times  \ddot x_i  
= 
 \sum_{i\in \kk} (x_i - x_0) \times \frac{ \partial U_{\kk,\kk'}}{\partial x_i}
=\\
\sum_{i\in \kk, j \in \kk'}
(-\alpha) m_i m_j |x_i - x_j|^{-(2+\alpha)}(x_i - x_0) \times (x_0 - x_j).
\end{split}
\]
Since
$|x_i(t) - x_j(t)|^{-\alpha - 2} = |x_0(t) - x_j(t)|^{-\alpha - 2} +
k_{i,j}(t) | x_i(t) - x_0(t) |$
for a bounded function $k_{i,j}(t)$,
\[
\dot c_\kk  = -\alpha 
\sum_{i\in \kk, j \in \kk'} m_i m_j k_{i,j}|x_i - x_0| (x_i - x_0) \times 
(x_0 - x_j ),
\]
and hence $\dot c_\kk I_k^{-1}$ is bounded as claimed.
\end{proof}

For every $i\in \kk$ 
let $q_i = x_i - x_0$, $\dot q_i = \dot x_i - \dot x_0$ and
$\ddot q_i = \ddot x_i - \ddot x_0$.
We have the equalities
\begin{equation}
\begin{aligned}
I_\kk & =  \sum_{i\in\kk} m_i q_i^2 \\
\frac{1}{2} \dot I_\kk & = 
\sum_{i\in \kk}  m_i q_i \dot q_i
\end{aligned}
\end{equation}
By Cauchy-Schwartz
\begin{equation}
\begin{aligned}
(\frac{1}{2} \dot I_\kk)^2 = &  
(\sum_{i\in \kk}  m_i q_i \dot q_i )^2
\leq 
I_\kk \sum_{i\in\kk} m_i \frac{(q_i \dot q_i)}{q_i^2} \\
c_\kk^2 =  &
(\sum_{i\in \kk}  m_i q_i \times \dot q_i )^2 \leq 
I_\kk \sum_{i\in\kk} m_i \frac{(q_i \times \dot q_i)}{q_i^2}.
\end{aligned}
\end{equation}
As a consequence of the identity $(ab)^2 + (a\times b)^2 = a^2b^2$
we obtain
\begin{equation}
\label{eq:this}
\frac{\dot I_\kk^2}{4} + c_\kk^2 \leq 
I_\kk \sum_{i\in \kk} m_i \dot q_i^2 =
I_\kk( 2 K_\kk - m_0\dot x_0^2).
\end{equation}

More generally, in the same way it is possible
to show that for a  collection of clusters $\h$
\begin{equation}
\label{eq:that}
(\sum_{\kk \subset \h} \frac{\dot I_\kk}{2} )^2 + c_\h^2
\leq
(\sum_{\kk \subset \h} I_\kk) ( 2 K_\kk + b)
\end{equation}
for a constant $b$.
The sum $\sum_{\kk\subset \h} I_\kk$
is the sum of momenta $I_\kk$ (each defined, as in ~\ref{eq:Ik},
with respect to the center of mass of the cluster $\kk$),
for any cluster $\kk$  in $\h$.

\begin{nr}
\label{nr2}
If $\sum_{\kk\subset \h} \dot I_\kk \geq 0$, then
there is a constant $b\in \R$ such that
\[
|c^2_\h(t) - c^2_\h(0)| \leq b  t \sum_{\kk\subset \h} I_\kk(t).
\]
\end{nr}
\begin{proof}
By ~\ref{nr1},
$c_\kk^2(t)$ is bounded for every $\kk$
and the limit $c_\kk^2(0)$ exists.
Moreover,
\[
\begin{split}
 (\sum_{\kk\subset \h}I_\kk)^{-1} |c_\h^2(t) - c^2_\h(0)  | \leq
  \sum_{\kk\subset \h} \int_0^t 2 | c_\kk(\tau) | |\frac{\dot c_\kk(\tau )}{\sum_\kk I_\kk(t)} | d\tau
\leq
\\
  \sum_{\kk\subset \h} \int_0^t 2 |c_\kk(\tau)| \frac{|\dot c_\kk(\tau )|}{\sum_\kk I_\kk(\tau)} d\tau.
\end{split}
\]
by Cauchy-Schwarz inequality and
since $\tau < t \implies \sum_\kk I_\kk(\tau) < \sum_\kk I_\kk(t)$
($\sum_\kk I_\kk$
is monotone increasing as a consequence of the hypothesis on
$\sum_\kk \dot I_\kk$).
Thus  there is a constant $b$ such that
$|c^2_\h(t) - c^2_\h(0)| \leq bt  \sum_{\kk\subset \h} I_\kk(t)$  as claimed.
\end{proof}

\begin{nr}\label{nr:last}
If $\sum_{\kk \subset \h} E_\kk$ is bounded 
then  $\sum_{\kk\subset \h} \dot I_\kk \geq 0$.
\end{nr}
\begin{proof}
Let $\ff$ denote the sum
$\ff=\sum_{\kk\subset \h} \dot I_\kk$.
By summing up the Lagrange-Jacobi identities
~\ref{coro:lagjac}  for every $\kk$, we obtain  the equalities
\begin{equation}
\label{eq:tew}
\ddot \ff = (2-\alpha) \sum_\kk U_\kk + B_1 = (2-\alpha) K_\h + B_2 
= (2-\alpha) U_\h + B_3,
\end{equation}
where $B_1, B_2$ and $B_3$ are suitable bounded functions.
Thus  $\lim_{t\to 0} \ddot \ff = +\infty$, 
hence $\dot \ff$ is strictly increasing and cannot be $0$
in $(0,\epsilon)$ if $\epsilon$ is sufficiently small, 
that is to say, $\ff$ is monotone in $(0,\epsilon)$. Since
$\ff(0) = 0$ and $\ff(t)\geq 0$, this implies $\dot \ff \geq 0$.
\end{proof}

\begin{nr}\label{nr:est_qi}
There are positive constants $c_1$ , $c_2$ and   $c_3$ such that 
for every $i\in \kk$
\[
|\dot q_i| \leq c_1 K_\kk^{1/2}, \ \ 
| \ddot q_i| \leq c_2 U_\kk^{(\alpha+1)/\alpha}
\mbox{ \ \ {and} \ \ }
| \mydddot{q_i}| \leq c_3 K_\kk^{1/2}  U_\kk^{(2+\alpha)/\alpha}.
\]
\end{nr}
\begin{proof}
The first is trivial. For the others, 
it suffices to consider equation ~\ref{eq:newton}, and to derive
it once.
\end{proof}

\begin{nr}
\label{nr:4}
Assume that $\sum_{\kk \subset \h} E_\kk$ is bounded. 
Then there is a constant $d$ such that for every $t\in (0,\epsilon)$:
$(\sum_\kk \frac{\dot I_\kk}{2})^2 + c_\h^2(0) 
\leq 
(\sum_\kk I_\kk)(   \sum_\kk \frac{ \ddot I_\kk}{2-\alpha} + d )$.
\end{nr}
\begin{proof}
By equation ~\ref{eq:that},
$ (\sum_{\kk \subset \h} \frac{\dot I_\kk}{2} )^2 + c_\h^2 
\leq 
(\sum_{\kk \subset \h} I_\kk) ( 2 K_\kk + b)$.
As a consequence of the  Lagrange-Jacobi formula ~\ref{coro:lagjac} 
and the fact that $\sum_\kk E_\kk$ is bounded,  
there is a constant $d$ such that 
\begin{equation}\label{eq:interin}
2 K_\h  \leq \frac{ \sum_\kk \ddot I_\kk}{2-\alpha} + d.
\end{equation}
Furthermore, by ~\ref{nr:last} we can apply
~\ref{nr2}, so that  there is a constant $b$ such that 
$c_\h^2(t) \geq c_\h^2(0) + b \sum_\kk I_\kk(t)$ and hence
there is a constant $d\in \R$ such that 
\[
(\sum_\kk \frac{\dot I_\kk}{2})^2 + c_\h^2(0) 
\leq 
(\sum_\kk \frac{\dot I^2_\kk}{2})^2 + c_\h^2(t)  - b \sum_\kk I_\kk(t) \leq 
(\sum_\kk I_\kk)(   \sum_\kk \frac{ \ddot I_\kk}{2-\alpha} + d ).
\]
\end{proof}

\begin{nr}
\label{nr:next}
Let $\h$ be the union of all the colliding clusters.
Then the sum of the energies
\[
\sum_{\kk \subset \h} E_\kk < M
\]
is bounded. 
\end{nr}
\begin{proof}
The total energy $H=E_\n$ of the $n$ bodies is constant, 
and the difference $\sum_{\kk \subset \h} E_\kk  - H$
is the sum of terms of type $m_im_j|x_i - x_j|^{-\alpha}$,
where $i$ and $j$ do not collide,
and of terms of type $\frac{1}{2} m_i \dot x_i^2$,
where $i$ do not collide with any other body.
Hence it  is a $C^2$ function  and, in particular, bounded.
\end{proof}

\begin{lemma}\label{lemma:this}
Consider a union of colliding clusters $\h = \cup_{i} \kk_i$.
If the sum of the energies
\[
(\forall t)   | \sum_{\kk\subset \h} E_{\kk}(t) | < M < \infty
\]
is bounded, then the function given by the sum
\[
\ff = \sum_{\kk} I_{\kk}
\]
has the following properties:
\begin{itemize}
\item $\lim_{t\to 0} \ff(t) = 0$.
\item
$\forall t$: $\ff(t) \geq 0$, $\dot \ff(t) \geq 0$.
\item 
There are constants $a$, $b = \frac{4}{2-\alpha}>2$ and  $c$ such that 
$ \dot \ff^2  + a^2 \leq  b \ff \ddot \ff + c \ff$.
\item 
There is a constant $d>0$ such that 
$\forall t$: $\ddot \ff \ff^{(b-2)/b}(t) \geq d >0$.
\item 
There is a constant $e$ such that 
$|\frac{d^3 \ff}{dt^3}| < e \ddot \ff^\gamma$, 
with the exponent $\gamma = \frac{2b-3}{b-2} = \frac{3}{2} + \frac{1}{\alpha}$.
\end{itemize} 
\end{lemma}
\begin{proof}
By definition, $\ff\geq 0$ and $\lim_{t\to 0} {\ff(t)} = 0$.
By ~\ref{nr:last}, $\dot \ff \geq 0$.
Moreover, ~\ref{nr:4}
yields a constant $d$ such that 
\[
(\sum_\kk \frac{\dot I_\kk}{2})^2 + c_\h^2(0) 
\leq 
(\sum_\kk I_\kk)(   \sum_\kk \frac{ \ddot I_\kk}{2-\alpha} + d ).
\]
In other words, there exist constants 
$a= 4 c_\h^2(0)$, 
$b=\frac{4}{2-\alpha}>2$ and  $c=4d$ such that 
$ \dot \ff^2  + a^2 \leq  b \ff \ddot \ff + c \ff$.

Now we prove that there is a constant $d>0$ such that 
$\forall t$: $\ddot \ff \ff^{(b-2)/b}(t) \geq d >0$.
By equation ~\ref{eq:tew}, 
$\ddot \ff \geq k_1 \sum_\kk U_\kk + k_2$
for some constant $k_1>0$ and $k_2$.
But since 
$(2-b)/b = -\alpha/2$, it follows that 
$\ddot \ff \geq k_1 \sum_\kk U_\kk + k_2 \geq d (\sum_\kk I_\kk)^{-\alpha/2}$ 
for a constant $d>0$.

It is left to show that 
there is a constant $e$ such that 
$|\frac{d^3 \ff}{dt^3}| < e \ddot \ff^\gamma$, 
with the exponent $\gamma = \frac{2b-3}{b-2} = \frac{3}{2} + \frac{1}{\alpha}$.
By deriving equation ~\ref{eq:src_lagjac}
\[
\frac{1}{2} \mydddot{I_\kk} = 
2 \dot E_\kk  + (2-\alpha) \dot U_\kk + \dot R,
\]
where by ~\ref{eq:est_R} and ~\ref{eq:est_E} 
$2\dot E_\kk + \dot R < cK^{1/2} +b$
for some constants $c>0$ and $b$. 
Furthermore, 
$\dot U_\kk$ is a combination of terms
of type $(\dot x_i - \dot x_j)(x_i - x_j)|x_i - x_j|^{-\alpha-2}$,
with $i,j\in \kk$, 
and hence there is a constant $c_2>0$ such that 
\[
\dot U_\kk < c_2 K_\kk^{1/2} U_\kk^{(\alpha+1)/\alpha}.
\]
Thus, by ~\ref{eq:tew}, there exists $c>0$ such that 
\[
| \mydddot{I_\kk}| \leq 
c K^{1/2} U_\kk ^{(\alpha+1)/\alpha}.
\]
Summing up for every $\kk \subset \h$, we obtain
\[
|\mydddot{\ff}| = 
| \sum_{\kk \subset \h} \mydddot{I_\kk} |
< c K^{1/2} \sum_{\kk \subset \h} 
U_\kk ^{(\alpha+1)/\alpha} \leq
c K^{1/2} \left( \sum_{\kk \subset \h} 
U_\kk \right) ^{(\alpha+1)/\alpha}.
\]
Hence, by ~\ref{eq:tew},
\begin{equation}\label{eq:todo}
|\mydddot{\ff}|   < c K^{1/2} \left( \ddot \ff \right) ^{(\alpha+1)/\alpha}.
\end{equation}
Now, the wanted inequality follows if we can prove that 
$K \leq c \ddot \ff$ for some $c>0$.
If $\h = \n$, then $K = K_\n = K_\h$,
and hence by, ~\ref{eq:tew},
$K \leq c \ddot \ff$ for some $c>0$. 
Thus if $\h = \n$ the conclusion of this lemma ~\ref{lemma:this} is true,
provided that the sum of the energies
\[
| \sum_{\kk \subset \n} E_\kk | < M
\]
is bounded.
But  by
~\ref{nr:next}
the sum of the all the energies
is bounded, and hence we can apply to such $\ff$ lemma ~\ref{lemma:bach},
which implies that there exists the limit
\[
\lim_{t \to 0}
K \left( \sum_{\kk \subset \n} I_\kk \right) ^{\alpha/2}  =
\lim_{t \to 0}
\ff^{\alpha/2} \ddot \ff, 
\]
and hence that  there is a constant $c_2>0$ such that 
\begin{equation}\label{eq:maybe}
\left( \sum_{\kk \subset \n} I_\kk \right) ^{-\alpha/2}  
> c_2 K.
\end{equation}

Now, 
consider again the general case of a union $\h$ of colliding clusters
$\kk$.
We have seen that $\ddot \ff \ff^{\alpha/2} \geq d >0$ 
for a constant $d$,
and thus,
\begin{equation}\label{eq:nice}
K_\h 
\geq d 
\left ( \sum_{\kk \subset \n} I_\kk \right)^{\alpha/2} 
> dc_2 K.
\end{equation}
This implies that  $K < c K_\h < d \ddot \ff$ 
for some constants $c,d>0$, 
and hence by ~\ref{eq:todo}
that 
$|\mydddot{\ff}| < c \ddot \ff^{\frac{3}{2} + \frac{1}{\alpha}}$
as claimed.
\end{proof}

\begin{lemma}
For every colliding cluster $\kk\subset \n$ there is a constant $c>0$ 
such that $K_\kk < c t^{-2\alpha/(2+\alpha)}$.
\end{lemma}
\begin{proof}
By equation ~\ref{eq:nice} it is enough to show that 
the total kinetic energy $K$ is bounded by
$K < c t^{-2\alpha/(2+\alpha)}$.
By ~\ref{nr:next}, ~\ref{lemma:this}  and ~\ref{lemma:bach}, the second 
derivative of the function
$\ff = \sum_{\kk \subset \n}$ is asymptotically equal to
\[
\ddot \ff \sim t^{-2\alpha/(2+\alpha)},
\]
while 
$K < c \ff$ for a positive constant $c$,
because of ~\ref{eq:interin}.
\end{proof}

\begin{lemma}\label{lemma:bounded_energy}
For every colliding cluster $\kk\subset \n$
the partial energy  $E_\kk$  is bounded.
\end{lemma}
\begin{proof}
By ~\ref{eq:est_E}
the derivative $\dot E_\kk$  is bounded by
\[
\dot E_\kk < 
\ddot \ff \sim t^{-\alpha/(2+\alpha)},
\]
which is integrable. 
\end{proof}

\begin{propo}
\label{propo:asy}
Let $\kk$ be a colliding cluster. 
Then there is $\kappa>0$ such that 
the following asymptotic estimates hold:
\begin{equation*}
\begin{aligned}
I_\kk &\sim  (\kappa t)^{\frac{4}{2+\alpha}} \\
\dot I_\kk &\sim  
\frac{4}{2+\alpha} \kappa (\kappa t)^{\frac{2-\alpha}{2+\alpha}}\\
\ddot I_\kk &\sim   
4 \frac{2-\alpha}{(2+\alpha)^2} \kappa^2 (\kappa t ) ^ { \frac{-2\alpha}{2+\alpha}},
\end{aligned}
\end{equation*}
and therefore 
\[
K_\kk \sim U_\kk \sim 
\frac{1}{4-2\alpha}  \ddot I_\kk \sim
\frac{2}{(2+\alpha)^2} \kappa^2 (\kappa t ) ^ { \frac{-2\alpha}{2+\alpha}}.
\]
\end{propo}
\begin{proof}
By ~\ref{lemma:bounded_energy}
the energy of a colliding cluster is bounded,
hence we can apply lemma ~\ref{lemma:this} and  consequently 
~\ref{lemma:bach}.
\end{proof}

Let $\kk\subset \n$ a colliding cluster.
Define the \emph{normalized configuration}  $s$ by 
\begin{equation}\label{eq:def_s}
s_i = 
I_\kk^{-1/2} 
(x_i-x_0) 
= 
I_\kk^{-1/2} 
q_i 
\end{equation}
for every $i\in \kk$.
Then
\begin{equation}\label{eq:dotqi}
\dot q_i =  \frac{1}{2} I_\kk^{-1/2} \dot I_\kk s_i + I_\kk^{1/2} \dot s_i
\end{equation}
and 
\begin{equation}\label{eq:ddotqi}
\ddot q_i = 
(-\frac{1}{4} I_\kk^{-3/2} \dot I_\kk^2 + \frac{1}{2} I_\kk^{-1/2} \ddot I_\kk) s_i +
I_\kk^{-1/2} \dot I_\kk \dot s_i +
I_\kk^{1/2} \ddot s_i
\end{equation}
\begin{equation}\label{eq:dddotqi}
\begin{split}
\mydddot{q_i} = 
\left( \frac{3}{8}  \dot I_\kk^3 I_\kk ^{-5/2} - 
\frac{3}{4} \dot I_\kk I_\kk^{-3/2} \ddot I_\kk +
\frac{1}{2} I_\kk^{-1/2} \mydddot{I_\kk} \right) s_i + 
\\
\left( -\frac{3}{4} \dot I_\kk ^2 I_\kk^{-3/2} + \frac{3}{2} I_\kk^{-1/2} \ddot I_\kk \right) \dot s_i +
\frac{3}{2} \dot I_\kk I_\kk^{-1/2} \ddot s_i +
I_\kk^{1/2} \mydddot{s_i}.
\end{split}
\end{equation}

\begin{nr}
\label{as:lim1}
For every $i\in \kk$:  $\displaystyle \lim_{t\to 0} t\dot s_i  = 0 $.
\end{nr}
\begin{proof}
By equation ~\ref{eq:dotqi}
\begin{equation*}
\begin{split}
2 K_\kk = \sum_{i\in \kk}  m_i \dot x_i^2 = 
\sum_{i\in \kk} m_i \dot q_i^2 
- m_0 \dot x_0^2 = \\
\sum_{i\in \kk} m_i \left( 
(\frac{1}{2} I_\kk^{-1/2} \dot I_\kk s_i)^2 + (I_\kk^{1/2} \dot s_i)^2
\right) 
- m_0 \dot x_0^2 =  \\
\frac{1}{4} \dot I_\kk^2 I_\kk^{-1}  +
I_\kk \sum_{i\in \kk} m_i  \dot s_i^2 
- m_0 \dot x_0^2. 
\end{split}
\end{equation*}
Now,
by applying proposition ~\ref{propo:asy} one can 
multiply both sides by $(\kappa t)^{2\alpha/(2+\alpha)}$ 
and take the limit as $t\to 0$:
\[
\frac{4}{(2+\alpha)^2} \kappa ^2 = 
\frac{4}{(2+\alpha)^2} \kappa ^2   +
\lim_{t\to 0} \left[ (\kappa t) ^{2\alpha/(2+\alpha)} 
I_\kk \sum_{i\in \kk} m_i  \dot s_i^2 \right] 
\]
\[
\implies 
\lim_{t\to 0} \left[ (\kappa t)^2 
\sum_{i\in \kk} m_i  \dot s_i^2\right]   = 0,
\]
and hence
\[
\lim_{t\to 0} t s_i = 0.
\]
\end{proof}

\begin{nr}
\label{as:lim2}
For every $i\in \kk$:  
$\displaystyle \lim_{t\to 0} t^2 \ddot s_i = 0$.
\end{nr}
\begin{proof}
By considering the 
form of $\frac{\partial U_\kk}{\partial x_i}$ and 
its derivative
$\frac{d}{dt}\frac{\partial U_\kk}{\partial x_i}$  
it is not difficult to show that there 
are positive constants $c_1$ and $c_2$ such that 
\[
|\ddot q_i| < c_1 I_\kk^{-(\alpha+1)/2} \sim 
c_1 t^{-2(\alpha+1)/(2+\alpha)},
\]
\[
|\mydddot{q_i}| < c_2 I_\kk^{-(2+\alpha)/2} K_\kk^{1/2}.
c_2 t^{-(3\alpha+4)/(2+\alpha)}.
\]
Therefore, 
multiplying both sides of 
~\ref{eq:ddotqi} by 
$t^{2(\alpha+1)/(2+\alpha)}$
one can see that 
$t^2\ddot s_i$ is bounded;
multiplying both sides of 
~\ref{eq:dddotqi} by
$t^{(3\alpha+4)/(2+\alpha)}$
one can see that $t^3\mydddot{s_i}$ 
is bounded.

So, for every $i \in \kk$ 
consider the function $\ff(t) = t^2 \ddot s_i(t)$.
Since
\[
\int_0^t \ff(\tau) d\tau = t^2 \dot s_i(t) 
-2 \int_0^t \tau \dot s(\tau) d\tau, 
\]
\[
\lim_{t\to 0} \frac{1}{t} \int_0^t \ff(t) =
\lim_{t\to 0} 
\left[ t\dot s_i(t) - \frac{2}{t} \int_0^t \tau \dot s_i(\tau) d\tau
\right] = 0.
\]
Moreover, since $t^2 \ddot s_i$ and $t^3 \mydddot{s_i}$ are bounded,
\[
|t \dot \ff| < 2t^2 |\ddot s_i| + t^3 |\mydddot{s_i}| < c
\]
for some constant $c>0$.
Thus, by lemma ~\ref{lemma:tech}
\[
\lim_{t\to 0} t^2 \ddot s_i = 
\lim_{t\to 0} \ff(t)  
\lim_{t\to 0} 
\frac{1}{t} \int_0^t \ff = 0.
\]
\end{proof}

\begin{propo}
\label{propo:cent_conf}
For every converging sequence
$s(t_j)$ of normalized configurations, the limit
$\lim_{j\to \infty} s(t_j)$ is a central configuration.
\end{propo}
\begin{proof}
Consider again equation ~\ref{eq:ddotqi} multiplied by
$(\kappa t)^{2(\alpha+1)/(2+\alpha)}$: 
\begin{equation}\label{eq:alkazar}
(\kappa t)^{2(\alpha+1)/(2+\alpha)} \ddot q_i = 
b(t) s_i + c_1 t \dot s_i + c_2 t^2 \ddot s_i + o(1) =
b(t) s_i + o(1),
\end{equation}
where $b(t)$ is a function with the  finite limit 
$b(0) = -\alpha \frac{2\kappa^2}{(2+\alpha)^2}$
 and $c_1$, $c_2$ 
are bounded.
A central configuration $q=(q_1,\dots, q_k)$ is a critical
point of the potential $U_\kk$ restricted to the ellipsoid
$I_\kk(q)=c$ with $c>0$ constant.  That is, there is $\lambda \in \R$
such that $\frac{\partial U_\kk}{\partial q_i} = \lambda m_i q_i$
for every $i\in \kk$.
By homogeneity
\[
-\alpha U_\kk = 
\sum_{i\in \kk}q_i \cdot \frac{\partial U_\kk}{\partial q_i} = \lambda I_\kk, 
\]
hence 
$\lambda = -\alpha \dfrac{U_\kk}{I_\kk}$ and thus 
central configurations solve the equations
\begin{equation}\label{eq:cent_conf}
\dfrac{\partial U_\kk}{\partial q_i} U_\kk^{-1} I_\kk^{1/2}= 
-\alpha m_i s_i
\end{equation}
for $i\in \kk$.
Now, since the terms $\dfrac{\partial U_{\kk,\kk'}}{\partial x_i}$
of $\dfrac{\partial U}{\partial x_i}$ are bounded,
the limit $\bar s= \lim_{j\to \infty} s(t_j)$
is a central configuration if and only if 
\[
m_i \ddot q_i(t_j) U_\kk^{-1}(t_j) I_\kk^{1/2}(t_j)
+\alpha m_i s_i(t_j)
\]
converges to $0$ as $j\to \infty$.
By ~\ref{propo:asy}, this holds if and only if 
\[
\dfrac{(2+\alpha)^2}{2 k^2} 
(\kappa t_j)^{2(\alpha+1)/(2+\alpha)} \ddot q_i(t_j)
 \to -\alpha \bar s_i
\]
as $j\to \infty$, 
and this follows by taking limits in  
~\ref{eq:alkazar}.
Thus $s(t_j)$ tends to central configurations $\bar s$
as stated.
\end{proof}

\begin{propo}
\label{propo:srcblowup}
Let $\kk\subset \n$ be a colliding cluster
and $s(t)=(s_1(t),\dots,s_k(t))$ 
its 
normalized configuration (defined  in ~\ref{eq:def_s}). 
If $\{\lambda_n\}_n$ is a sequence of  positive real numbers
such that  $s(\lambda_n)$ converges to a normalized configuration 
$\bar s$, then
\[
\forall t\in (0,1):
\lim_{n\to \infty} s(\lambda_n t) = 
\lim_{n\to \infty} s(\lambda_n) =
\bar s.
\]
\end{propo}
\begin{proof}
For every $n$ 
\begin{equation*}
| s_i(\lambda_n t) - s_i(\lambda_n) |
\leq
 \max_{u\in (t,1)} 
| \dot s_i( \lambda_n t) | \lambda_n(1-t).
\end{equation*}
Moreover, for every $\epsilon>0$ there exists $N>0$ such that 
$u<1/N \implies u \dot |s_i(u)|  \leq \epsilon$ for every $i=1\dots k$
(see  ~\ref{as:lim1}).
Thus 
for every $\epsilon>0$ and 
for every $i=1\dots k$ the following inequalities
hold:
\begin{equation*}
\begin{split}
\lim_{n\to \infty} |s_i(\lambda_n t) - s_i(\lambda_n)| 
\leq 
\lim_{n\to \infty} 
 \max_{u\in (t,1)} 
| \dot s_i( \lambda_n t) | \lambda_n(1-t).
\leq
\frac{\epsilon}{\lambda_n t} \lambda_n(1-t) \leq \epsilon\dfrac{1-t}{t},
\end{split}
\end{equation*}
hence the claim.
\end{proof}

\begin{remark}
A normalized configuration $s$ (relative to $\kk \subset \n$)
is a central configuration if and only if equations
~\ref{eq:cent_conf} hold, and hence if and only if 
for every $i\in \kk$
\begin{equation}\label{eq:cent_conf2}
\sum_{j\neq i} m_j\frac{s_i - s_j}{|s_i - s_j|^{2+\alpha}} U_\kk(s)^{-1}
- s_i   = 0
\end{equation}
for every $i\in \kk$.
Let $F_i(s)$ denote the left hand side in ~\ref{eq:cent_conf2}
and, for every subset $\h\subset \kk$, let $F_\h(s)$ denote
the sum $F_\h(s) = \sum_{i\in \h} m_i s_i F_i(s)$;
then
\begin{equation}\label{eq:F_h}
F_\h(s) = 
U_\kk(s)^{-1}
\left[
U_\h(s) + 
\sum_{\substack{i\in \h\\ j \in \h'}} 
m_im_j\frac{s_i - s_j}{|s_i - s_j|^{2+\alpha}} 
\right] - \sum_{i\in \h} m_i s_i^2,
\end{equation}
where $\h' = \kk \minus \h$.
It is easy to see that if  $s$ is a central configuration then 
$F_\h(s) = 0$. On the other hand, 
as  $s$ converges to a configuration $\bar s$ with 
a collision of type $\h$ (that is, there is $s_0 \in V$ such that 
 $\bar s_i = s_0$ for every $i\in\h$), then the function 
$F_\h(s)$ converges
to $1-(\sum_{i\in \h} m_i)s_0^2$.
Now, since $I_\kk(\bar s)=1$, there is a positive constant 
$c>0$ such that $1-(\sum_{i\in \h} m_i) s_0^2\geq c$ for every $s_0$. 
Thus there is a constant $c>0$ 
such that on all points in $s$ containing a collision
of type $\h$ the norm $|F_\h(s)|\geq c>0$. 
Now we proceed iteratively.
It is easy to see that $F_\h$ is continuous 
on the 
space of configurations not in $\Delta_{\h,\h'}$
and vanishes on central configurations;
this implies that there is a constant $c_\h>0$ such that 
if $s$ is a central configuration and $y$ a collision
of type $\bfa \supset \h$ then 
$|s - y|\geq c_\h$. 
It is true when $\h$ is maximal (that is, given by 
a collision of $k-1$ particles), and by induction 
one can take the minimum of the $c_\bfa$ with $\bfa\supset \h$.
Hence one can conclude that there is $c>0$ such that 
$|s-y|\geq c>0$
for every (normalized) central configuration $s$ and every 
(normalized) collision $y$.  In particular, the set
of normalized central configurations is compact and central 
configurations are collisionless.
\end{remark}

\begin{remark}
One of the difficulties in proving the asymptotic estimates
of  proposition ~\ref{propo:asy} is that it is necessary to prove that
the  energies $E_\kk$ of colliding clusters are 
bounded (Lemma ~\ref{lemma:bounded_energy}). 
This is easy to prove if the collision is total.
In \cite{elbialy}, proposition 2.9, it is proved 
by direct integration of 
the derivatives
that partial energies are bounded,
provided that asymptotic   estimates of the kinetic energy
are known (as a consequence of \cite{sperling}). Thus
McGehee coordinates yield asymptotic estimates only
in the total collision case, in which the energy is constant and hence 
bounded (see also \cite{chen_coll} for a conceptual proof without
Tauberian theorems). 
For partial collisions, it is 
necessary to estimate partial energies as in~\ref{lemma:bounded_energy}.
\end{remark}

\section{Blow-ups}
\label{section:bl}
Aim of this section is to prove some auxiliary results
about the convergence and continuity of integrals
of families of rescaled solutions. 
The main goal is proposition \ref{propo:blowup} which is 
a key step towards the standard variation method of 
section \ref{section:st}.
A similar technique was used in \cite{andrea_thesis} and 
ascribed there to the second author.

Assume as in section ~\ref{sec:asy}
that $x(t)$ is 
a solution of Newton equations ~\ref{eq:newton} defined in the time
interval $(0,\epsilon)$ with an isolated collision at $t=0$ 
and without other collisions in $(0,\epsilon)$.
Let $\kk\subset \n$ be a colliding cluster and define
the normalized configuration $s$ and 
the centered configuration $q$ as in ~\ref{eq:def_s}. 
If $\lambda_n$ is a sequence of real numbers such that $s(\lambda_n)$
converges (and thus necessarily to a central configuration $\bar s$,
by ~\ref{propo:cent_conf}),
let $\bar q$ be the path defined for $0\leq t  < +\infty$ as 
\begin{equation}\label{eq:barq}
\bar q_i(t) = (\kappa t)^{2/(2+\alpha)} \bar s_i
\end{equation}
for every $i\in \kk$, where $\kappa >0$ is the constant of 
proposition ~\ref{propo:asy}.
The path $\bar q$ is  termed
a (right) \emph{blow-up} of the solution $x(t)$ in $0$.
Its definition depends on the choice of the limit configuration
$\bar s$. 

For every $\lambda>0$ consider the path $q^\lambda$ defined by
\begin{equation}
\label{def:qlambda}
q^{\lambda}(t) =  \lambda^{-2/(2+\alpha)} q(\lambda t)
\end{equation}
for every $t\in [0,\lambda^{-1}\epsilon]$.

\begin{nr}
\label{bl:0}
The energy of the collision solution $\bar q$ is zero, i.e.\ the blow-up
$\bar q$  is parabolic.
\end{nr}
\begin{proof}
For such a trajectory the asymptotic estimates of ~\ref{propo:asy}
are all equalities for every $t$. Therefore,
by applying the Lagrange-Jacobi identity, the 
energy $E_\kk$ is zero. 
\end{proof}

\begin{nr}
\label{bl:1}
If $s(\lambda_n)$ converges to the normalized configuration
$\bar s$, then 
for every $T>0$ 
the sequences $q^{\lambda_n}$ and $\dfrac{d{q}^{\lambda_n}}{dt}$   converge
to the blow-up $\bar q$ and its derivative $\dot {\bar q}$
respectively, 
uniformly on $[0,T]$ and on the compact subsets of $(0,T]$
respectively.
\end{nr}
\begin{proof}
By definition 
\begin{equation*}
q^{\lambda_n}(t) = 
(\kappa t)^{2/(2+\alpha)} 
\dfrac{I_\kk^{1/2}(\lambda_nt)}{%
\lambda_n^{2/(2+\alpha)} (\kappa t)^{2/(2+\alpha)} } 
s(\lambda_n t).
\end{equation*}
Thus,
since by ~\ref{propo:srcblowup} 
$\lim_{n\to +\infty} s(\lambda_n t) = \bar s$
and by ~\ref{propo:asy} 
$I_\kk(\lambda_n t) \sim (\kappa \lambda_n t)^{4/(2+\alpha)}$,
for  every $t>0$ the limit
\begin{equation*}
\lim_{n\to +\infty} 
q^{\lambda_n}(t) = 
(\kappa t)^{2/(2+\alpha)} \bar s.
\end{equation*}
Hence $q^{\lambda_n}$ converges to $\bar q$ pointwise.
The convergence is uniform in 
$[0,T]$, since $q^{\lambda_n}$ is an $H^1$-bounded sequence.

Now we need to show that for every $\epsilon\in (0,T)$ its derivative 
$\frac{dq^{\lambda_n}}{dt}$ 
converges uniformly on $[\epsilon,T]$ to 
the derivative of $\bar q$.
By ~\ref{eq:dotqi}, ~\ref{as:lim1} and ~\ref{propo:asy}, for every $t>0$
\[
\lim_{n\to \infty} 
\dfrac{dq^{\lambda_n}}{dt}(t)  =
\dfrac{2\kappa}{2+\alpha} (\kappa t)^{-\alpha/(2+\alpha)} =
\dfrac{d\bar q}{dt}(t).
\]
Hence 
$\frac{dq^{\lambda_n}}{dt}$ 
converges pointwise to 
$\dot {\bar q}$.
By ~\ref{as:lim1}, ~\ref{as:lim2} and ~\ref{eq:dotqi}, 
the convergence is uniform in $[\epsilon,T]$ for
every $\epsilon\in (0,T)$.
\end{proof}

We say that a path $x(t)$ is \emph{centered} if its center
of mass is  zero. If a path is defined only for a cluster
$\kk \subset \n$ of particles, then it is termed centered
if the center of mass of the cluster is zero.
By definition $q(t) = x(t) - x_0(t)$, hence
$q(t)$ is a centered path.

\begin{nr}
\label{bl:2}
There exists a sequence $\psi_n\in H^1([0,T],\R^d)^k$ 
of centered paths
converging uniformly to $0$ in $[0,T]$ and with support in $[0,T]$,
with $T>0$,  such that 
for every path   $\ff(t)=(\ff_i(t))_{i\in \kk}$ 
in $H^1([0,T],\R^d)^k$,
with support in $[0,T]$ and $C^1$ in a neighborhood of $T$
\[ 
\lim_{n\to \infty} 
\int_0^T 
\lag_\kk( q^{\lambda_n} + \ff + \psi_n )
 dt  =
\int_0^T
\lag_\kk( \bar q + \ff)
dt.
\]
Furthermore, if for every $t$ the centered configuration
$q(t) \in \X^H$ for a subgroup
$H\subset G$, then $\psi_n(t) \in \X^H$ for every $t$.
\end{nr}
\begin{proof}
Consider a sequence $N_n$ (which will be chosen
properly later) going to $+\infty$ as $n\to \infty$,
and  for all $n\gg 0$
the function $\psi_n$ defined  as
\begin{equation*}
\psi_n(t) = 
\begin{cases}
\bar q(t) - q^{\lambda_n}(t) & \mbox{ if } 0\leq t\leq T-\frac{1}{n}, \\
N_n(T-t)( \bar q (t) - q^{\lambda_n}(t) ) 
& \mbox{ if } T-\frac{1}{N_n}\leq t < T.
\end{cases}
\end{equation*}
It is clear that $\psi_n(T)=0$ and that $\psi_n$ converges
to $0$ uniformly in $[0,T]$, by ~\ref{bl:1}.
For every $t$ the configuration $\psi_n(t)$ is centered, since 
$\bar q(t)$ and $q^{\lambda_n}(t)$ are centered.
Furthermore, 
if $q(t) \in \X^H$ for every $t$, then 
$q^{\lambda_n}(t) \in \X^H$ and by
taking the limit $\bar q(t) \in \X^H$, so that  
$\psi_n(t) \in \X^H$.

Since $\ff$ is continuous and $\ff(T)=0$, $\bar q(t) + \ff(t)$ is not 
a collision for every $t\in [T-1/N_n,T]$ and $n\gg  0$. 
Moreover, since $\ff$ is $C^1$, $\dot \ff$ is bounded 
in $[T-1/N_n,T]$ and $n\gg 0$. Moreover, 
by an appropriate choice of the sequence $N_n$ one can 
assume that 
there is a constant $C>0$ such that the 
 partial Lagrangians are bounded by $C$:
\(
\lag_\kk(\bar q + \ff)  < C
\) 
and 
\(
\lag_\kk(q^{\lambda_n} + \ff + \psi_n) <C
\)
for every $n>n_0$, in the interval $[T-1/N_{n_0},T]$.
Therefore,  since for every $t$ in $[0,T-1/N_n]$ 
\(
q^{\lambda_n}(t) + \psi_n(t) = \bar q(t),
\)
\[\begin{split}
\int_0^T 
\lag_\kk( q^{\lambda_n} + \ff + \psi_n )
dt  =
\int_0^{T-1/N_n}
\lag_\kk( \bar q + \ff) dt 
+
\int_{T-1/N_n}^T
\lag_\kk( q^{\lambda_n} + \ff + \psi_n )
dt & =\\
\int_{0}^T  
\lag_\kk( \bar q + \ff) dt 
-
\int_{T-1/N_n}^T  
\lag_\kk( \bar q + \ff) dt 
+
\int_{T-1/N_n}^T
\lag_\kk( q^{\lambda_n} + \ff + \psi_n ),
\end{split}
\]
which converges to 
\(
\int_{0}^T  
\lag_\kk( \bar q + \ff) dt 
\) as claimed.
\end{proof}

\begin{nr}
\label{bl:3}
For every $\lambda >0$ 
let $x^\lambda$ be as above the path $\lambda^{-2/(2+\alpha)} x(\lambda t)$.
Then for every $T>0$ and every $\lambda>0$
\[
\int_0^T
\lag(x)(t) dt =
\lambda^{\frac{2-\alpha}{2+\alpha}} 
\int_0^{T/\lambda} \lag(x^\lambda)(t) dt.
\]
\end{nr}
\begin{proof}
For every $\lambda>0$ by 
\[
\lag(x^\lambda)(t) = \lambda^{2\alpha/(2+\alpha)} \lag(x)(\lambda t),
\]
hence by integrating  and changing variables
\[
\int_0^T \lag(x^\lambda)(t) dt = 
\lambda^{\frac{\alpha-2}{2+\alpha}} 
\int_0^{\lambda T} 
\lag(x)(t) dt.
\]
\end{proof}

\begin{nr}
\label{bl:4}
Let $\kk \subset \n$ be a cluster, 
$x(t)$ a path and $q(t)=x(t) - x_0(t)$ the corresponding 
centered path as in ~\ref{eq:def_s}.
If $\ff$ is a centered path of the cluster $\kk$, 
then for every $\lambda>0$
\[
\lag_\kk( q^{\lambda} + \ff ) - \lag_\kk( q^\lambda) =
\lag_\kk( x^\lambda + \ff) - \lag_\kk(x^\lambda).
\]
\end{nr}
\begin{proof}
By the arbitrariness of $x$ and $q$ we can assume that $\lambda=1$.
Since for every $x$
\[
\lag_\kk(x) = \lag_\kk(q) + \frac{1}{2}m_0\dot x_0^2,
\]
where as above $x_0(t)=x(t) - q(t)$ 
is the center of mass of the cluster $\kk$
at time $t$,
it is also true that 
\[
\lag_\kk(x+\ff) = \lag_\kk(q+\ff) + \frac{1}{2}m_0\dot x_0^2,
\]
if $\ff$ is centered. 
\end{proof}

\begin{nr}
\label{bl:5}
Let $\ff$ be a path such that 
$\ff_i(t)\neq 0 \implies  i \in \kk$,
$T>0$ a real number 
 and $\psi_n$ the sequence 
of ~\ref{bl:2}, extended with the zero path for $i\in \kk'$.
Then  
\[
\begin{split}
\lim_{n\to \infty} 
\int_0^T 
\left[ 
\lag(x^{\lambda_n} + \ff +\psi_n) - \lag(x^{\lambda_n}) 
\right] dt 
= \\
\lim_{n\to \infty}
\int_0^T 
\left[
\lag_\kk(x^{\lambda_n} + \ff +\psi_n) - \lag_\kk(x^{\lambda_n}) 
\right] 
dt.
\end{split}
\]
\end{nr}
\begin{proof}
Since  for every $\lambda>0$, every $\ff$ and every $x$
\[
\lag(x + \ff) - \lag(x) = 
\lag_\kk(x + \ff) - \lag_\kk(x) +
U_{\kk,\kk'}(x+\ff) - U_{\kk,\kk'}(x),
\]
the 
conclusion follows once we prove that 
$\int_0^T U_{\kk,\kk'}(x^{\lambda_n}+ \ff +  \psi_n) dt$ 
and 
$\int_0^T U_{\kk,\kk'}(x^{\lambda_n}) dt$ 
converge to  zero as $n\to \infty$,
for every $\ff$.
Since
\[
U_{\kk,\kk'}(x^{\lambda}+ \ff
+ \psi_n) = 
\lambda^{2\alpha/(2+\alpha)} U_{\kk,\kk'}(x + \ff^{1/\lambda} 
+ \psi_n^{1/\lambda}),
\]
and as $\lambda \to 0$ the term 
$U_{\kk,\kk'}(x + \ff^{1/\lambda} + \psi_n^{1/\lambda})$ 
is uniformly bounded,
$\int_0^T U_{\kk,\kk'}(x^{\lambda_n}+ \ff + \psi_n^{1/\lambda}) dt$ 
converges to  zero as $n\to \infty$.
The same argument works for 
$\int_0^T U_{\kk,\kk'}(x^{\lambda_n}) dt$.
\end{proof}

\begin{propo}
\label{propo:blowup}
Let $x(t)$ be a solution in $(0,\epsilon)$, with 
an isolated collision in $t=0$.
Let $\kk\subset \n$ be a colliding
cluster
and $\bar q$ a (right) blow-up of $x(t)$ with respect to $\kk$ in $0$.
Let 
$T\in (0,\epsilon)$ and   let 
$\ff$ be a variation  of the particles in $\kk$ which 
is $C^1$ in a neighborhood of $T$, defined and 
centered for every $t \in [0,T]$.
Then the sequence 
$\psi_n$ of ~\ref{bl:2} has the following property:
\[
\lim_{n\to \infty}
\int_0^T \left[
\lag(x^{\lambda_n} + \ff +\psi_n) - \lag(x^{\lambda_n})
\right] dt
=
\int_0^T 
\left[
\lag(\bar q + \ff) - \lag(\bar q)
\right] dt.
\]
\end{propo}
\begin{proof}
As a consequence of the asymptotic estimates 
~\ref{propo:asy}, the Lagrangians $\lag(q^{\lambda_n})$ are uniformly 
bounded by an integrable function, hence by the Lebesgue theorem
\[
\lim_{n\to \infty} \int_0^T \lag(q^{\lambda_n}) dt =
\int_0^T \lag(\bar q) dt,
\]
and therefore by ~\ref{bl:2}
\[
\int_0^T 
\left[
\lag_\kk(\bar q + \ff) - \lag_\kk(\bar q)
\right] dt
=
\lim_{n\to \infty} 
\left[
\lag_\kk(q^{\lambda_n} + \ff + \psi_n) - \lag_\kk(q^{\lambda_n})
\right]
dt.
\]
Since $\psi_n$ and $\ff$ are centered, $\ff+\psi_n$ is centered
 for every $n$, and therefore by ~\ref{bl:4}
\[
\lag_\kk(q^{\lambda_n} + \ff + \psi_n) - \lag_\kk(q^{\lambda_n})
=
\lag_\kk(x^{\lambda_n} + \ff + \psi_n) - \lag_\kk(x^{\lambda_n}),
\]
so that 
\[\begin{split}
\lim_{n\to \infty} 
\left[
\lag_\kk(q^{\lambda_n} + \ff + \psi_n) - \lag_\kk(q^{\lambda_n})
\right]
dt
= \\
\lim_{n\to \infty} 
\int_0^T
\left[
\lag_\kk(x^{\lambda_n} + \ff + \psi_n) - \lag_\kk(x^{\lambda_n})
\right]
dt
\end{split}\]
But as a consequence of ~\ref{bl:5} the latter limit is equal to 
\[
\lim_{n\to \infty}
\int_0^T \left[
\lag(x^{\lambda_n} + \ff +\psi_n) - \lag(x^{\lambda_n})
\right] dt,
\]
and this concludes the proof.
\end{proof}

\section{Averaging estimates}
\label{section:aa}

Now we come to the averaging estimates ~\ref{theo:aa}.
The main idea, inspired by Marchal's argument 
as explained in  \cite{chenICM}
(see remark \ref{rem:mars} below), is
to consider a parabolic collision-ejection trajectory
$\bar q$ ~\ref{eq:barq} and 
to replace one (or more, if it is necessary -- we 
will explain how in the proof of ~\ref{mt:1})
of the point particles with a homogeneous circle
having the same mass, constant radius, and center
moving following the same trajectory of the original
particle. The key point is that this procedure
decreases the integral of the potential $U$ on the 
time line, and hence it will be possible in the next
section to define a \emph{standard variation} $v^\delta$,
following this principle, to show that 
such a homotetic  solution $\bar q$ cannot be a minimizer.

{F}or all 
$\xi,\delta\in \R^3\minus \{0\}$ let $S(\xi,\delta)$ denote the following 
integral (where $\alpha\in (0,2)$)
\begin{equation}\label{eq:S}
S(\xi,\delta) = 
\int_0^{+\infty}
\left[
\dfrac{1}{|  t^{2/(2+\alpha)}\xi  + \delta|^\alpha} - 
\dfrac{1}{|t^{2/(2+\alpha)}\xi|^{\alpha}}
\right] dt
\end{equation}
If $\lambda>0$ is a real number, then
\begin{align*}
S(\lambda \xi, \delta) &= 
\lambda^{-1-\alpha/2} S(\xi,\delta) \\
S(\xi,\lambda \delta) &=   \lambda^{-\alpha} S(\lambda^{-1}\xi, \delta) =
\lambda^{1-\alpha/2}  S(\xi,\delta)
\end{align*}
and hence 
\begin{equation}
\label{aa:homogen}
S(\xi,\delta) = 
|\xi|^{-1-\alpha/2} |\delta|^{1-\alpha/2} S( \frac{\xi}{|\xi|},
\frac{\delta}{|\delta|}).
\end{equation}

Consider a circle $\sphere\subset \R^3$ with center in $0$ (its  radius
is equal to $\frac{|\sphere|}{2\pi}$). 
For any vector $\xi \in \R^3\minus \{0\}$ let 
$\ts(\xi,\sphere)$ be defined as the average of $S(\xi,-)$
on the circle $\sphere$
\begin{equation}\label{defi:ts}
\ts(\xi,\sphere) = 
\frac{1}{|\sphere|} \int_{\sphere} S(\xi,\delta) d\delta.
\end{equation}

The purpose of this section is to prove the following theorem.
\begin{theo}
\label{theo:aa}
For every $\xi\in \R^3\minus \{0\}$ and for every 
circle $\sphere\subset \R^3$ with center in $0$,
\[
\ts(\xi,\sphere) = 
\frac{1}{|\sphere|} \int_{\sphere} S(\xi,\delta) d\delta < 0.
\]
\end{theo}
\begin{remark} \label{rem:mars}
A comparison with the argument of Marchal is in order. 
In the quoted paper \cite{chenICM} the two-dimensional
and the three-dimensional cases are considered separately
both in the case of exponent $\alpha=1$.
The three-dimensional case is the simplest: from our
point of view
the key point consists in proving an inequality like
the one in \ref{theo:aa}, with the integral taken 
over a two-dimensional sphere $\sphere = S^2$; which directly
follows from the harmonicity of the Kepler potential in $\R^3$.
To treat the planar case, Marchal used the known potential 
generated by a special radially symmetric mass distribution
on the disc $B^2$. Let us observe that, due to the homogeneity
property \ref{aa:homogen}, any mass distribution $\mu(r)$ 
on the disc would fit our purpose: indeed, integrating in
polar coordinates 
\begin{equation*}
\begin{aligned}
\dfrac{1}{|B^2|} \int_{B^2} S(\xi,x) \mu(|x|) dx  & =
\dfrac{1}{|S^1|}
\int_0^1 \int_{S^1} S(\xi,r\delta) r \mu(r) \ dr \ d\delta
\\ 
& =  
\dfrac{1}{|S^1|} \left(
\int_{S^1} S(\xi,\delta) d\delta
\right)
\left(
\int_0^1 r^{2-\alpha/2} \mu(r) dr
\right), \\
\end{aligned}
\end{equation*}
whose sign does not depend on the choice of the function $\mu$.
\end{remark}

\begin{nr}
\label{aa:nr1}
Consider 
a circle
$\sphere$ as above, and a family
of unit vectors $\xi_\gamma \in \R^3\minus \{0\}$, 
with $\gamma\in [0,\pi/2]$,
such that $\xi_\gamma$ and  the plane generated by $\sphere$
meet at an angle $\gamma$.
Then the function 
\[
\ff(\gamma) = \ts(\xi_\gamma,\sphere)
\]
is monotonically decreasing with $\gamma$ and for 
every $\gamma$
\[
\ts(\xi_\gamma,\sphere) \leq (\cos\gamma)^{1-\alpha/2}\ts( \xi_0,\sphere).
\]
\end{nr}
\begin{proof}[of \ref{aa:nr1}]
Consider the function
\[
f(\xi,\delta,t) = \frac{1}{2} \left[ 
\frac{1}{|t^{2/(2+\alpha)}\xi + \delta|^\alpha}
+
\frac{1}{|t^{2/(2+\alpha)}\xi - \delta|^\alpha}
\right]
- \frac{1}{|t^{2/(2+\alpha)}\xi|^\alpha}. 
\]
Since 
\[
\ts(\xi,\sphere) =
\frac{1}{|\sphere|} \int_\sphere \int_{0}^{+\infty} f(\xi,\delta,t) dt d\delta,
\]
the result follows from the fact that if $\theta$ denotes 
the angle between the plane generated by $\sphere$ and $\delta$, 
the following equation  holds:
\begin{equation*}
\begin{split}
f(\xi,\delta,t) = 
\frac{1}{2} \left[ 
\frac{1}{(
t^{4/(2+\alpha)} +
2 \cos\gamma \cos \theta t^{2/(2+\alpha)} +
1 
)^{\alpha/2}} +  \right. \\ + \left.
\frac{1}{(
t^{4/(2+\alpha)}- 
2 \cos\gamma \cos \theta t^{2/(2+\alpha)} +
1 
)^{\alpha/2}}
\right]  - \\
- \frac{1}{|t^{2/(2+\alpha)}\xi|^\alpha}, 
\end{split}
\end{equation*}
and the right hand side is monotonically decreasing in $\gamma$ 
in $[0,\pi/2]$ (an easy computation shows that 
the derivative is negative, whatever be the sign of $\cos\theta$).

Furthermore, since  if $|\xi_\gamma| = |\delta| = 1$
\[
\begin{split}
S(\xi_\gamma,\delta) = 
\int_0^{+\infty}
\left[ 
\frac{1}{%
|t^{4/(2+\alpha)} + 2\cos\theta\cos\gamma t^{2/(2+\alpha)} + 1|^{\alpha/2}}
-
t^{-2\alpha/(2+\alpha)} 
\right] dt 
\leq  \\
\int_0^{+\infty}
\left[ 
\frac{1}{%
|t^{4/(2+\alpha)} + 2\cos\theta\cos\gamma t^{2/(2+\alpha)} + \cos^2\gamma|^{\alpha/2}}
-
t^{-2\alpha/(2+\alpha)} 
\right] dt 
=\\
S(\xi_0, \cos\gamma \delta) = 
(\cos\gamma)^{1-\alpha/2} S(\xi_0,\delta),
\end{split}
\]
for every $\gamma\in [0,\pi/2]$
\[
\ts(\xi_\gamma,\sphere) \leq (\cos\gamma)^{1-\alpha/2} \ts( \xi_0,\sphere).
\]

\end{proof}

\begin{nr}
\label{aa:series}
If $x\in (0,1)$,
then
\[
\frac{1}{2\pi} \int_{0}^{2\pi} |1 + x e^{i\theta}|^{-\alpha} =
\sum_{k=0}^{+\infty} 
\binomial{-\alpha/2}{k}^2 x^{2k}
\]
\end{nr}
\begin{proof}
Since for $|z|<1$ and $a\in \R$
\[
(1+z)^{a} = 
\sum_{k=0}^{\infty} \binomial{a}{k} z^k,
\]
if $z=xe^{i\theta}$ with $0\leq x<1$ then
\[
\begin{split}
|1+z|^{-\alpha} = 
(1+z)^{-\alpha/2}(1+\bar z)^{-\alpha/2} = 
\\
= \sum_{k=0}^{\infty} 
\sum_{k_1+k_2=k} 
\left(
\binomial{-\alpha/2}{k_1}
\binomial{-\alpha/2}{k_2}
e^{i(k_1 - k_2)\theta}
\right)
x^{k}.
\end{split}
\]
Therefore
\[
\frac{1}{2\pi} \int_{0}^{2\pi} | 1 + xe^{i\theta}|^{-\alpha} d\theta =
\sum_{k=0}^{\infty} \binomial{-\alpha/2}{k}^2 x^{2k},
\]
since 
\[
\int_{0}^{2\pi}
\binomial{-\alpha/2}{k_1}
\binomial{-\alpha/2}{k_2}
e^{i(k_1 - k_2)\theta}
d\theta = 
\begin{cases}
2\pi & \mbox{ if $k_1 = k_2$ }\\
0 & \mbox{ if $k_1 \neq k_2$}
\end{cases}
\]

\end{proof}

\begin{nr}
\label{aa:series2}
If $x\in (1,+\infty)$,
then
\[
\frac{1}{2\pi} \int_{0}^{2\pi} |1 + x e^{i\theta}|^{-\alpha} =
x^{-\alpha}\sum_{k=0}^{+\infty} 
\binomial{-\alpha/2}{k}^2 x^{-2k}
\]
\end{nr}
\begin{proof}
\[
\begin{split}
\frac{1}{2\pi}
\int_0^{2\pi} | 1 + x e^{i\theta}|^{-\alpha} d\theta =
x^{-\alpha} 
\frac{1}{2\pi} \int_0^{2\pi} | 1 + \frac{e^{-i\theta}}{x}|^{-\alpha}
=
\\
x^{-\alpha} 
\sum_{k=0}^{\infty} \binomial{-\alpha/2}{k}^2 (\frac{1}{x})^{2k}.
\end{split}
\] 
\end{proof}

\begin{nr}
\label{aa:cases}
Consider in the complex plane  $\xi=1$. 
Then
\[
\frac{1}{2\pi} \int_{0}^{2\pi} |t^{2/(2+\alpha)} \xi + e^{i\theta}|^{-\alpha} 
d\theta =
\begin{cases}
\sum_{k=0}^{\infty} \binomial{-\alpha/2}{k}^2 t^{4k/(2+\alpha)}  &
\mbox{ if $t\in (0,1)$} \\
\sum_{k=0}^{\infty} \binomial{-\alpha/2}{k}^2 t^{-(4k+2\alpha)/(2+\alpha)}  &
\mbox{ if $t\in (1,\infty)$} 
\end{cases}
\]
\end{nr}
\begin{proof}
It is a simple consequence of ~\ref{aa:series} and 
~\ref{aa:series2}.
\end{proof}

\begin{nr} \label{aa:sum1}
\begin{equation*}
\begin{split}
\frac{1}{2\pi} \int_0^1 \int_0^{2\pi} \left[
|t^{2/(2+\alpha)} + e^{i\theta}|^{-\alpha} - t^{-2\alpha/(2+\alpha)}
\right] d\theta dt = \\
\frac{2+\alpha}{4} 
\sum_{k=0}^{\infty}
\left[ 
 \binomial{-\alpha/2}{k}^2 \frac{1}{k + \frac{2+\alpha}{4}}
\right]
- 
\frac{2+\alpha}{2-\alpha}.
\end{split}
\end{equation*}
\end{nr}
\begin{proof}
It follows by integrating  the first series 
in ~\ref{aa:cases}.
\end{proof}

\begin{nr}\label{aa:sum2}
\begin{equation*}\begin{split}
\frac{1}{2\pi} \int_1^{+\infty} \int_0^{2\pi} \left[
|t^{2/(2+\alpha)} + e^{i\theta}|^{-\alpha} - t^{-2\alpha/(2+\alpha)}
\right] d\theta dt = \\
\frac{2+\alpha}{4}
\sum_{k=1}^{\infty}
\left[
\binomial{-\alpha/2}{k}^2 \frac{1}{k + \frac{\alpha-2}{4}}
\right].
\end{split}
\end{equation*}
\end{nr}
\begin{proof}
It follows by integrating  the second series 
in ~\ref{aa:cases}, 
in which the term for $k=0$ is equal to 
$t^{-2\alpha}{2+\alpha}$.
\end{proof}

Now we can sum the two latter equations changing 
the summation index of the second, hence obtaining
\begin{equation}
\label{aa:hence}
\begin{split}
\frac{1}{2\pi} \int_0^{+\infty} \int_0^{2\pi} \left[
|t^{2/(2+\alpha)} + e^{i\theta}|^{-\alpha} - t^{-2\alpha/(2+\alpha)}
\right] d\theta dt  =  \\
\frac{2+\alpha}{4}
\sum_{k=0}^{+\infty}
\left[
\binomial{-\alpha/2}{k+1}^2 \frac{1}{k + \frac{2+\alpha}{4}}
\right] + 
\frac{2+\alpha}{4} 
\sum_{k=0}^{+\infty}
\left[ 
 \binomial{-\alpha/2}{k}^2 \frac{1}{k + \frac{2+\alpha}{4}}
\right]
- 
\frac{2+\alpha}{2-\alpha} = \\
\frac{2+\alpha}{4} 
\sum_{k=0}^{+\infty}
\left[
\binomial{-\alpha/2}{k}^2 \frac{1}{k + \frac{2+\alpha}{4}} 
\left( \frac{(\alpha/2+k)^2}{(1+k)^2} + 1 \right)
\right]  -
\frac{2+\alpha}{2-\alpha} = \\
\frac{\alpha^2}{4} + 1 + 
\frac{2+\alpha}{4} 
\sum_{k=1}^{+\infty}
\left[
\binomial{-\alpha/2}{k}^2 \frac{1}{k + \frac{2+\alpha}{4}} 
\left( \frac{(\alpha/2+k)^2}{(1+k)^2} + 1 \right)
\right]  -
\frac{2+\alpha}{2-\alpha}.
\end{split}
\end{equation}

\begin{nr}
\label{aa:ineq}
If $x\in (0,1)$, then  for every $k\geq 1$, 
\[
|\binomial{-x}{k}|   \leq 
x \left( \frac{k}{2} \right)^{x-1}.
\]
\end{nr}
\begin{proof}
If $k=1$, then it is true since  it reduces to $x\leq x$. 
Otherwise, if $k\geq 2$,
\begin{align*}
\binomial{-x}{k} &= (-1)^k \prod_{j=1}^k(1 + \frac{x-1}{j}),
\end{align*}
hence 
\begin{equation*}
\begin{split}
\left| \binomial{-x}{k} \right| = 
x \prod_{j=2}^k(1 + \frac{x-1}{j}) =
x e^{\sum_{j=2}^k \log(1+\frac{x-1}{j})} \leq \\
x e^{(x-1)\sum_{j=2}^k \frac{1}{j}} \leq 
x e^{(x-1) \log\frac{k}{2}} = x (\frac{k}{2})^{x-1}
\end{split}
\end{equation*}
as claimed.
\end{proof}

\begin{nr}
\label{aa:estim}
\begin{equation*}
\sum_{k=1}^{+\infty}
\left[
\binomial{-\alpha/2}{k}^2 \frac{1}{k + \frac{2+\alpha}{4}} 
\left( \frac{(\alpha/2+k)^2}{(1+k)^2} + 1 \right)
\right]   
<
2^{1-\alpha} {\alpha^2}  \frac{3-\alpha}{2-\alpha}.
\end{equation*}
\end{nr}
\begin{proof}
By $~\ref{aa:ineq}$ it follows that 
\begin{equation*}
\binomial{-\alpha/2}{k}^2 \frac{1}{k + \frac{2+\alpha}{4}} 
\left( \frac{(\alpha/2+k)^2}{(1+k)^2} + 1 \right)
< \frac{\alpha^2}{4} \left( \frac{k}{2} \right)^2 \frac{2}{k}
=
\frac{\alpha^2}{4} (\frac{k}{2})^{\alpha -3},
\end{equation*}
and hence 
\begin{equation*}
\begin{split}
\sum_{k=1}^{+\infty}
\left[
\binomial{-\alpha/2}{k}^2 \frac{1}{k + \frac{2+\alpha}{4}} 
\left( \frac{(\alpha/2+k)^2}{(1+k)^2} + 1 \right)
\right]  <  \\
2^{1-\alpha} \alpha^2  \sum_{k=1} k^{\alpha-3} =
2^{1-\alpha} {\alpha^2}  \zeta(3-\alpha).
\end{split}
\end{equation*}
Now, since for positive $s$
\[
\zeta(s) \leq 1 + \frac{1}{s-1},
\]
it follows that 
\[
\zeta(3-\alpha) \leq 1 + \frac{1}{2-\alpha},
\]
and therefore that 
\begin{equation*}
\sum_{k=1}^{+\infty}
\left[
\binomial{-\alpha/2}{k}^2 \frac{1}{k + \frac{2+\alpha}{4}} 
\left( \frac{(\alpha/2+k)^2}{(1+k)^2} + 1 \right)
\right]  < 
2^{1-\alpha} {\alpha^2}  \frac{3-\alpha}{2-\alpha}.
\end{equation*}
\end{proof}

\begin{proof}[Proof of theorem ~\ref{theo:aa}]
By ~\ref{aa:nr1}, it is suffices to prove theorem ~\ref{theo:aa}
in the case
$\xi$ belongs to $\sphere$. By ~\ref{aa:homogen} we can assume $|\xi|=1$.
By a change of coordinates we can assume that $\xi=1\in \C$,
where $\C=\R^2 \subset \R^3$ is an embedded plane.
Now, by equation ~\ref{aa:hence}
\begin{equation}\begin{split}
\ts(\xi,\sphere) = 
\ts(1,\sphere) = 
\frac{1}{2\pi} \int_0^{+\infty} \int_0^{2\pi} \left[
|t^{2/(2+\alpha)} + e^{i\theta}|^{-\alpha} - t^{-2\alpha/(2+\alpha)}
\right] d\theta dt  =  \\
\frac{\alpha^2}{4} + 1 + 
\frac{2+\alpha}{4} 
\sum_{k=1}^{+\infty}
\left[
\binomial{-\alpha/2}{k}^2 \frac{1}{k + \frac{2+\alpha}{4}} 
\left( \frac{(\alpha/2+k)^2}{(1+k)^2} + 1 \right)
\right]  -
\frac{2+\alpha}{2-\alpha}.
\end{split}
\end{equation}
Therefore, by applying inequality ~\ref{aa:estim},
\begin{equation}\begin{split}
\ts(\xi,\sphere) <
\frac{\alpha^2}{4} + 1 + 
\frac{2+\alpha}{4} 
\left( 
2^{1-\alpha} {\alpha^2}  \frac{3-\alpha}{2-\alpha}
\right)
-\frac{2+\alpha}{2-\alpha}  = \\ 
-\frac{a}{4} \frac{8-2\alpha + \alpha^2}{2-\alpha}
+ 
\frac{2+\alpha}{2} 
{\alpha^2}  \frac{3-\alpha}{2-\alpha}
2^{-\alpha}< 0,  
\end{split}
\end{equation}
since for $a\in [0,2]$
\[
\begin{split}
2^{-\alpha} <  \frac{7}{8}(3-\alpha) <
\frac{(8 - 2\alpha + \alpha^2)(3-\alpha)}{2(2+\alpha)}.
\end{split}
\]
The proof is complete.
\end{proof}

\begin{remark}
The sums in ~\ref{aa:sum1} and ~\ref{aa:sum2} can be written
in terms of hypergeometric functions as follows
\begin{equation*}\begin{split}
\frac{1}{2\pi} \int_0^1 \int_0^{2\pi} \left[
|t^{2/(2+\alpha)} + e^{i\theta}|^{-\alpha} - t^{-2\alpha/(2+\alpha)}
\right] d\theta dt = \\
\hypergeom{3}{2}{\alpha/2}{\alpha/2}{(2+\alpha)/4}{1}{(6+\alpha)/4}{1}
- 
\frac{2+\alpha}{2-\alpha},
\end{split}
\end{equation*}
\begin{equation*}\begin{split}
\frac{1}{2\pi} \int_1^{+\infty} \int_0^{2\pi} \left[
|t^{2/(2+\alpha)} + e^{i\theta}|^{-\alpha} - t^{-2\alpha/(2+\alpha)}
\right] d\theta dt =
\\
\frac{2+\alpha}{2-\alpha} \left(
1 -
\hypergeom{3}{2}{\alpha/2}{\alpha/2}{(\alpha-2)/4}{1}{(2+\alpha)/4}{1}
\right).\end{split}
\end{equation*}
They are nearly-poised (of the second kind) hypergeometric
functions evaluated in $1$. They are balanced (i.e.\ Saalsch\"utzian) 
if and only if $\alpha=1$.
\end{remark}

\section{The standard variation}
\label{section:st}

Assume that  $\bar q$ is a parabolic collision solution
of the particles in $\kk \subset \n$, as defined
in ~\ref{eq:barq} for every $t\geq 0$. 
Up to a change of the time scale
we can assume that $\kappa=1$, so that 
$\bar q = t^{2/(2+\alpha)}\bar s$. 
Let $\delta \in \R^{dk} = {(\R^d)}^k$ be a vector
of norm $|\delta| = \sum_{i\in \kk} \delta_i^2$
sufficiently small and $T>0$ a real number.
In this section we will define the \emph{standard variation}
associated to $\delta$, and show how to use
the averaging estimate ~\ref{theo:aa} of the previous section
in our equivariant context. It will be of crucial
importance to introduce the notions of definitions
~\ref{defi:rotating} and ~\ref{defi:rotatingfor},
since they are the building blocks of the \emph{rotating circle property}
~\ref{propertyK} that will be introduced in the next section.

\begin{defi}
\label{defi:standard_variation}
The \emph{standard variation} associated to $\delta$ and $T$
is defined as follows:
\[
v^\delta(t) = 
\begin{cases}
\delta & \mbox{ if $0 \leq t \leq T-|\delta|$}\\
(T-t) \dfrac{\delta}{|\delta|} & \mbox{ if $T-|\delta|\leq t \leq T$ } \\
0  & \mbox{if $t\geq T$}
\end{cases}
\]
\end{defi}

Let $\pot$ and $\kin$ denote the operators 
\[
\pot(q)(t) = \sum_{i<j} \dfrac{m_im_j}{| q_i(t) - q_j(t) |^\alpha}
\]
and 
\[
\kin(q)(t) = \sum_{i\in \kk} \dfrac{1}{2} m_i \dot q_i^2(t).
\]
\begin{nr}
\label{st:1}
Let $\Delta \action$ denote  the difference
\(
\action(q+v^\delta) - \action(q)
\) (where $\action$ is meant in $[0,T]$). 
Then for $\delta \to 0$ 
\[
\Delta \action = 
|\delta|^{1-\alpha/2}
\sum_{i<j}  
m_i m_j S( \bar s_i -\bar s_j, \dfrac{\delta_i - \delta_j}{|\delta|}) + O(|\delta|)
\]
\end{nr}
\begin{proof}
Since $\dot v^\delta$ and $\dot {\bar q}$  are bounded
in a neighborhood of $T$,
\[
2  \int_0^T \left( \kin(\bar q+v^\delta) - \kin(\bar q) \right) dt=
\int_{T-|\delta|}^T 
\sum_{i\in\kk}  m_i  \dot v^\delta_i( \dot v^\delta_i - 2 \bar {\dot q_i})
dt
\]
and hence there exists a constant $c_1>0$ such that for every $\delta$
sufficiently small
\[
\left| \int_0^T \left( \kin(\bar q+v^\delta) - \kin(\bar q) \right) dt \right|
\leq 
c_1 |\delta|.
\]
Furthermore, 
for every $i>j$, $i,j \in \kk$,
\[
\begin{split}
\int_0^T 
\left( 
\dfrac{1}{ |\bar q_i - \bar q_j + v_i^\delta - v_j^\delta |^\alpha} 
-
\dfrac{1}{ |\bar q_i - \bar q_j |^\alpha }
\right)
dt = \\
\int_0^{+\infty}
\left( 
\dfrac{1}{ |\bar q_i - \bar q_j + \delta_i - \delta_j |^\alpha} 
-
\dfrac{1}{ |\bar q_i - \bar q_j |^\alpha }
\right)
dt 
+ r(\delta),
\end{split}
\]
where the remainder is
\[\begin{split}
r(\delta) = 
\int_{T-|\delta|}^{T}  
\left( 
\dfrac{1}{ |\bar q_i - \bar q_j + v_i^\delta - v_j^\delta |^\alpha} 
-
\dfrac{1}{ |\bar q_i - \bar q_j |^\alpha }
\right)
dt  
- \\
\int_{T-|\delta|}^{+\infty}
\left( 
\dfrac{1}{ |\bar q_i - \bar q_j + \delta_i - \delta_j |^\alpha} 
-
\dfrac{1}{ |\bar q_i - \bar q_j |^\alpha }
\right)
dt. 
\end{split}
\]
Now,  the function 
\(
\left( 
\dfrac{1}{ |\bar q_i - \bar q_j + v_i^\delta - v_j^\delta |^\alpha} 
-
\dfrac{1}{ |\bar q_i - \bar q_j |^\alpha }
\right)
\)
is bounded in a neighborhood of $T$, hence  there is a constant 
$c_2$ such that 
\[
\left| 
\int_{T-|\delta|}^{T}  
\left( 
\dfrac{1}{ |\bar q_i - \bar q_j + v_i^\delta - v_j^\delta |^\alpha} 
-
\dfrac{1}{ |\bar q_i - \bar q_j |^\alpha }
\right)
dt  
\right| \leq c_2|\delta|.
\]
Furthermore, since there is a constant $c_3>0$ such that for every $\delta$ 
sufficiently small and for every $t\geq T$
\[
\left| \dfrac{1}{ |(\bar s_i - \bar s_j) t^{2/(2+\alpha)}
 + \delta_i - \delta_j |^\alpha} 
-
\dfrac{1}{ |(\bar s_i - \bar s_j) t^{2/(2+\alpha)} |^\alpha }
\right|
\leq  c_3  |\delta| t^{2(\alpha+1)/(2+\alpha)},
\]
the inequality 
\[
\left| \int_{T-|\delta|}^{+\infty}
\left( 
\dfrac{1}{ |\bar q_i - \bar q_j + \delta_i - \delta_j |^\alpha} 
-
\dfrac{1}{ |\bar q_i - \bar q_j |^\alpha }
\right)
dt 
\right| \leq c_4  |\delta|
\]
holds for some constant $c_4>0$ and every $\delta \to 0$.
Therefore 
\[
\Delta \action = \int_0^T (\Delta \kin + \Delta \pot) dt  = 
\sum_{i<j}  
m_i m_j S( \bar s_i -\bar s_j, \delta_i - \delta_j) + r(\delta)
\]
where $|r(\delta)| \leq c|\delta|$ for some $c>0$ and every $\delta\to 0$;
$S$ is defined in ~\ref{eq:S}.
The conclusion follows at once by applying equation ~\ref{aa:homogen}.
\end{proof}

\begin{defi}
\label{defi:rotating}
For a group $H$ acting orthogonally on $V$,
a circle $\sphere \subset V$ (with center in $0\in V$)
is termed \emph{rotating under $H$}
if $\sphere$ is invariant under $H$ (that is, for every $g\in H$ 
$g\sphere = \sphere$)
 and for every $g\in H$ the restriction
$g|\sphere \from \sphere \to \sphere$ of 
the orthogonal motion $g\from V\to V$ 
is  a rotation (where the identity is meant as a rotation
of angle $0$). 
\end{defi}

\begin{defi}
\label{defi:rotatingfor}
Let $i\in \n$ be an index and $H\subset G$ a subgroup. 
A circle $\sphere \subset V$ (with center in $0\in V$)
is termed  \emph{rotating for $i$ under $H$}
if $\sphere$ is rotating
under $H$ and 
\[
\sphere \subset  V^{H_i} \subset V,
\]
where $H_i\subset H$ denotes the isotropy subgroup of the index $i$ in $H$
relative to the action of $H$ on the index set 
$\n$ induced by restriction (that is, the 
isotropy $H_i = \{ g\in H \st gi =i \}$).
\end{defi}

\begin{remark}
It is not difficult to show that if  $\sphere$ is 
rotating for $i$ under $H$, then $h\sphere$ is rotating
for $hi$ under $H$ for every $h \in H$. In fact
$h V^{H_i} = V^{hH_ih^{-1}} = V^{H_j}$, and the 
conjugate of a rotation is a rotation.
Moreover,
the motivation of this definition will become clear in the following 
sections.
Actually, the main point will be to move the particle $i$
away from the collision, with the further property
that the configuration at the collision time is 
$H$-equivariant.  Thus, moving away $i$ will automatically
(by equivariance) need to  move away all the images of $i$ in $Hi$ (which 
is automatically $H$-homogeneous),
and of course $i$ can be moved only in a direction
fixed by its isotropy subgroup $H_i$.
Hence, if 
a circle $\sphere \subset V$ is rotating 
for the index $i$ under $H$,
then $i$ can be moved away from $0$ in all the directions
of $\sphere$, since by hypothesis $\sphere \subset V^{H_i}$,
and at the same time the corresponding bodies in $H_i$ 
will be moved by equivariance in the same circle $\sphere$ (since by 
hypothesis $H\sphere = \sphere$); not only, the total
collision of the $|Hi|$ bodies is so replaced by
a regular polygon (given by the rotation of the variation of $i$
in $\sphere$).
\end{remark}

\begin{defi}
A subset $\kk \subset \n$ is termed \emph{$H$-homogeneous} 
for a subgroup $H\subset G$ if
$H$ acts transitively on $\kk$, that is, 
for every $i,j \in \kk$ there exists a $h\in H$
such that $hi=j$. This implies that $m_i = m_j$ 
by ~\ref{eq:property}.
\end{defi}

If $\kk\subset \n$, then 
let $\X_\kk\subset V^n$ denote the space consisting
in all configurations such that $i\not\in \kk \implies
x_i=0$. It is isomorphic to the (not centered) 
configuration space in $V$ of 
the $k$ bodies in $\kk$.

\begin{nr}
\label{st:2}
Let $H\subset G$ be a subgroup.
If  $\kk\subset \n$ is $H$-homogeneous and
$i\in \kk$, then there
is an isomorphism  (and isometry)
\[
\iota_i\from V^{H_i} {\cong}   \X_\kk^H
\]
defined by 
\[
\forall p \in V^{H_i}, \forall j\in \kk : 
\left( \iota_i(p) \right)_j  = h_{i,j}p 
\]
where $h_{i,j}$ is any element in $H$ such that 
$h_{i,j}i=j$. 
\end{nr}
\begin{proof}
The homomorphism (of vector spaces over $\R$) 
$\iota_i$  does not depend
on the choice of the the elements $h_{i,j}$ in $H$: if $h_1i=h_2i$ 
then $h_2^{-1}h_1 \in H_i$, so that 
$h_1p = h_2p$. Furthermore, if for $h\in H$ 
 $l=h^{-1}j\in \n$,
then $l = h^{-1} h_{i,j}i$ 
for any $h_{i,j}$ such that $h_{i,j}i=j$, and hence 
by ~\ref{eq:Gaction} 
\[
\left( h \iota_i(p) \right)_j = 
h \left(\iota_i(p)\right)_{h^{-1}j} = 
h h^{-1} h_{i,j} p = \left( \iota_i(p) \right)_j
\]
that is, $H\iota_i(p)=\iota_i(p)$.
Consider the homomorphism
$r_i\from \X_\kk^H \to V^{H_i}$, defined by the projection 
$r_i( q ) = q_i$ for every $q \in \X_\kk^H$.
It is easy to show, by the homogeneity of $\kk$,
that $r_i$ is the inverse of $\iota_i$. 
The fact that $\iota_i$ is an isometry is easy to prove.
\end{proof}

\begin{nr}
\label{st:3}
Consider a (right) blow-up $\bar q$ as in ~\ref{eq:barq}
of type $\kk \subset \n$ and a subgroup $H\subset G$.
If there exists an index $i\in \kk$ and 
a  circle $\sphere\subset V$ which 
is rotating under 
$H$   for the index $i$,
then 
for every $T>0$
the average  in $\sphere$ 
of the variation of the Lagrangian action 
(in the interval  $[0,T]$)
\[
\int_{\sphere} \left( 
\action(\bar q + v^{\iota_i (p)}) - \action(\bar q) \right) dp <0
\]
is negative. 
\end{nr}
\begin{proof}
First note that 
for $p \in \sphere$ the image $\delta = \iota_i(p)$
belongs to ${\X_\kk^H}$, and 
$v^{\iota_i(p)}$ denotes the standard variation 
~\ref{defi:standard_variation}, defined for $t\geq 0$ 
and with support in $[0,T]$. 
By definition $v^{\iota_i(p)}(t) \in \X_\kk^H$ for every $t$ and 
$\iota_i | \sphere$ 
is an embedding of 
$\sphere$ in $\X_{Hi}^H \subset \X_\kk^H$. 
Without loss of generality we can assume that $i=1$.
By ~\ref{st:1}, $\int_\sphere \Delta \action dp <0$ 
if and only if 
\begin{equation}\label{st:aa}
\sum_{i<j}  
m_i m_j S( \bar s_i -\bar s_j, 
\dfrac{\delta_i - \delta_j}{|\delta|}) < 0
\end{equation}
where $\delta_i = \left( \iota_1(p) \right)_i = h_{1,i}p$
for any $h_{1,i} \in H$ such that 
$h_{1,i}1=i$. 
This implies that $|\delta|$ is constant in $\iota_1(\sphere)$ 
and that as $p$ ranges in $\sphere$,
for every $i,j \in \kk$ the difference
$\dfrac{\delta_i - \delta_j}{|\delta|}$
ranges in a circle $\sphere'$ in 
$V$:
if $i,j \in Hi \subset \kk$ then
\[
\dfrac{\delta_i - \delta_j}{|\delta|} =
\dfrac{(h_{1,i} - h_{1,j})p}{|\delta|} 
\]
which describes a circle of radius $2(1-\cos\theta)$,
where $\theta$ is the angle of the rotation in $\sphere$
given by the composition of the rotations $h_{1,i}^{-1}h_{i,j}$.
On the other hand, if one considers indexes 
$j\not\in Hi$,
then  $\delta_j=0$ and 
\[
\dfrac{\delta_i - \delta_j}{|\delta|} =
\dfrac{h_{1,i}p}{|\delta|} 
\]
which describes in a circle of radius $1$.
Hence we can apply theorem ~\ref{theo:aa} for every $i,j$ 
in the sum in ~\ref{st:aa} to obtain the claimed statement.
\end{proof}

\begin{remark}
The idea of the proof of \ref{st:3} can be also
sketched as follows:
if there exists 
a  circle $\sphere\subset V$ which 
is rotating under the subgroup
$H$   for the index $i$,
then one can replace the colliding particle $i$ 
with a circle in the rotating circle $\sphere$ 
under $H$ (which exists
by hypothesis). The isotropy $H_i$ needs to fix $\sphere$,
and $H$ acts on $\sphere$ by rotation, hence by equivariance 
all the particles in $Hi$ are replaced by circles 
$h\sphere$ (i.e. rotated copies of $\sphere$). The main
point is that the interaction of particles in $Hi$ 
with the other particles and the interaction of 
particles within $Hi$ both yield the same type
of integral (namely, the integral which appears in 
the inequality of theorem \ref{theo:aa}), which lower 
the value of the action functional.
\end{remark}

\section{The rotating circle property and the main theorems}
\label{section:mt}

Once analyzed the standard variation in the previous section,
we can 
can now finally introduce 
the definition of \emph{rotating circle
property} and so
state and  prove the main theorems ~\ref{mt:1} and ~\ref{mt:2}. 
Property ~\ref{propertyK} depends only on the group action,
thus it is computable in terms of $\rho$, $\tau$ and 
$\sigma$. We will see in section ~\ref{section:examples}
that actions often have this property,
so that the results ~\ref{mt:1} and ~\ref{mt:2}
can be applied to wide classes of group actions,
together with proposition ~\ref{propo:coercive}.
The proofs of ~\ref{mt:1} and ~\ref{mt:2} basically rely only on  
the averaging estimate ~\ref{theo:aa} and 
the properties of the standard variation.

The following definition is motivated 
by ~\ref{st:3}, 
where the existence of a rotating circle under $H$ for $i$ 
is exploited to move away from the collision the $i$-th
particle, while keeping the $H$-equivariance of the final
trajectory.

\begin{defi}
\label{propertyK}
We say that a group $G$ acts (on $\T$, $\n$ and $V$) with the 
\emph{rotating circle property} (or, equivalently,
that $G$ \emph{has} the rotating circle property,
once the $G$-action is chosen)
if for every $\T$-isotropy subgroup $G_t\subset G$ 
and for at least $n-1$ indexes $i\in \n$ 
there exists in $V$ a rotating circle $\sphere$ under $G_t$
for $i$.
(See also in ~\ref{defi:Tisotropy} the definition of $\T$-isotropy
and in ~\ref{defi:rotatingfor} the definition of rotating circle).
\end{defi}

The trivial group has 
the rotating circle property.
If a group $G$ has the rotating circle property, then it is 
easy to show that any subgroup $K\subset G$ has it. 
In fact,  
a group has the rotating circle property
if and only if 
all its maximal $\T$-isotropy subgroups
have the property.
We will see in the examples in the last section 
that 
to determine whether a group acts with  the rotating 
circle property
 is usually a straightforward
task. 
\begin{remark}
\label{rem:new23}
The reason that in definition \ref{propertyK} it is required 
the existence in $V$ of a rotating circle $\sphere$ under the isotropy 
$G_t$ for
\emph{at least $(n-1)$ indexes in $\n$} is simply that,
in order to apply the local averaging variation to all possible
collisions, one has to be sure that  there exists
at least one particle $i$ 
that can be moved away, for each choice of a cluster
$\kk\subset \n$. Since in a collision of type $\kk$  there are 
necessarily at least $2$ bodies and $(n-1)$ of the $n$
bodies can be always moved, 
it follows that surely at least one of the colliding particles
can be moved (see \ref{st:3} and also the proof of theorem \ref{mt:1} below).
More generally, if $t\in \T$ is a collision time 
and $G_t$ its isotropy, 
then one can define the subset $\kk_t\subset \n$ 
of all those indexes $i\in \n$ for which 
there is a rotating circle under $G_t$ for $i$ (thus, all those particles
that can be moved away from a collision).
The proof of \ref{mt:1} will show that  in a local minimizer 
all the particles in $\kk_t$ are not colliding. 
Hence if $\kk_t$ has at least $n-1$ elements,
no particle can collide. More generally, if every $\kk_t$ 
has at least $n-k$ elements, then a colliding 
cluster in a local minimizer needs to consist of at most 
$k$ particles. 
\end{remark}

Before stating the next theorem, we would like to recall
that a minimizer for the fixed--ends problem (also known as Bolza problem) 
is a minimizer of the Lagrangian action in the space
$\Lambda_{x,y} H^1([0,1],\X)$ of paths (defined in the time interval $[0,1]$)
starting from the configuration $x\in \X$ and ending in the 
configuration $y\in \X$. If $K$ is a group acting
on $\Lambda$ as in ~\ref{eq:Gaction2}, in particular
it acts on $\X$ and therefore 
on $\Lambda_{x,y}$. 
If $x,y\in \X^K$, 
the $K$-equivariant 
Bolza problem consists in finding 
critical points (in our case, minimizers -- see definition 
~\ref{defi:minimizer}) 
of the the restriction of the Lagrangian functional 
to the fixed subspace $\Lambda_{x,y}^K \subset H^1([0,1],\X^K)$.
Using the defining property of the fundamental domain $\I$, we will apply
this result later to the case $K=\ker \tau$ (Theorem ~\ref{coro:mt:1}).

\begin{theo}\label{mt:1}
Consider a finite
group $K$ acting on $\Lambda$
with the rotating circle property.
Then 
a minimizer of the $K$-equivariant fixed--ends (Bolza) problem 
is free of collisions.
\end{theo}
\begin{proof}
Let $x$ be a minimizer and let  $\I=(T_0,T_1)$ denote the interior 
of its time domain.
By ~\ref{propo:gen_sol},  $x$ is a generalized solution in $\I$.
If  there is an interior collision, then 
by ~\ref{propo:isolated} there is an interior isolated collision 
at a time $t_0\in \I$.
We can assume that $t_0=0$.
Let $\kk\subset \n$ be a colliding cluster for such a collision
solution.
If $T>0$ is small enough, then the  interval
$[-T,T]$ 
is contained in the interior of $\I$ and does not contain collision times
for $t\neq 0$.
Thus there exists a (right and left) blow-up
$\bar q(t)$ 
defined as  in ~\ref{eq:barq} by
\begin{equation*}
\bar q_i(t) = 
\begin{cases}
 t^{2/(2+\alpha)} \xi_i & \mbox{ if $t\geq 0$} \\
 (-t)^{2/(2+\alpha)} \xi_i' & \mbox{ if $t<0$} 
\end{cases}
\end{equation*}
for every $i\in \kk$, 
where $\xi$ and $\xi'$ are suitable central configurations.

By hypothesis 
$K$ has the rotating circle property ~\ref{propertyK}.
Furthermore, since $x(t) \in \X^{K}$ for every $t$,
the centered cluster trajectory
$q(t) \in \X^H_\kk$, where $H\subset K$ is the subgroup
consisting of all the elements of $g\in \ker \tau$ such that 
$g\kk = \kk$.
Therefore there exists $i\in \kk$ and a circle $\sphere \subset V$ 
rotating under $H$ for $i$, and hence 
by applying ~\ref{propo:blowup}  to both sides of 
$[-T,0]$ and $[0,T]$ and adding the results, 
the average in $\sphere$ of the variation of $\action$ 
\[
\int_{\sphere} \left(   
\action(\bar q + v^{\iota_i (p)}) - \action(\bar q) \right) dp <0     
\]
is negative.
Thus there is $p\in \sphere$ such that if $\delta = \iota_i(p)$,
\begin{equation}\label{eq:postman}
\action(\bar q + v^{\delta}) - \action(\bar q) <0. 
\end{equation}
Now, by   ~\ref{st:2}, such $\delta \in \X_\kk^H$.
The subset $K\kk  \subset \n$ 
is the disjoint union of $h$ images of $\kk$, 
where $h$ is the index of $H$ in $K$, 
and $\delta$ yields  in a unique way  an element
$\delta \in \X^H$ (that we will denote still by $\delta$,
with a abuse of terminology), 
by setting $\delta_j = g \delta_i$ if $gi=j$ 
for $g\in K$ and 
with $\delta_j=0$ otherwise.
Hence 
one can apply 
~\ref{eq:postman} and ~\ref{propo:blowup}
$h$ times 
(for both sides -- and projecting
on the centered variations if necessary),
to show that there exists a sequence 
\[
\lim_{n\to \infty}
\int_0^T \left[
\lag(x^{\lambda_n} + v^\delta +\psi_n) - \lag(x^{\lambda_n})
\right] dt
= h
\int_0^T
\left[
\lag(\bar q + v^\delta) - \lag(\bar q)
\right] dt < 0
\]
where $v^\delta(t) + \psi_n(t) \in  \X^{\ker\tau}$ for every 
$t\in [-T,T]$.
But this means that 
$x^{\lambda_n}$ is not a local minimizer for $n\gg 0$,
and by ~\ref{bl:4} that $x$ is not a local minimizer.
This contradicts the assumption, hence there cannot be interior collisions.
\end{proof}

\begin{remark}
As explained in remark \ref{rem:new23}, 
one could easily rephrase and generalize theorem 
\ref{mt:1} as:
\emph{Consider a finite
group $K$ acting on $\Lambda$.
For every $t\in \T$ let 
$K_t$ denote its $\T$-isotropy 
and let $\kk_t$ be the subset $\kk_t\subset \n$ 
of all those indexes $i\in \n$ for which 
there is a rotating circle under $K_t$ for $i$.
If for every $t\in \T$ the order $|\kk_t|$ is at least $n- k$,
then 
a minimizer of the $K$-equivariant fixed--ends (Bolza) problem 
is free of collisions  of type $\kk$, with $|\kk|> k$.
}
\end{remark}

\begin{coro}\label{coro:bolza}
For every $\alpha>0$, minimizers
of the fixed-ends (Bolza) problem  are free of interior
collisions.
\end{coro}
\begin{proof}
If $\alpha\geq 2$ this is well-known (see e.g. \cite{poincare,gordon}). 
For $\alpha\in(0,2)$,
it follows from ~\ref{mt:1} by considering a trivial $K$.
\end{proof}

\begin{theo}\label{coro:mt:1}
Let $G$ be a finite
group acting on $\Lambda$.
If 
$\ker\tau$ has the rotating circle property 
then any local minimizer of $\action^G$  in $\Lambda^G$
does not have interior collisions.
\end{theo}
\begin{proof}
Let $x$ be a local minimum and consider its restriction
$x|\I\from \I \to \X^{\ker\tau}$ 
to a fundamental domain  $\I$ as defined in  ~\ref{defi:fund_dom}.
It needs be a $K$-equivariant  minimizer for the fixed--ends problem,
where $K=\ker \tau$,  
and by ~\ref{mt:1} it does not have interior collisions.
\end{proof}

\begin{coro}
\label{coro:cyclic_type}
If the action of $G$ on $\Lambda$  
is of cyclic type and 
$\ker\tau$ has the rotating circle property 
then any local minimizer of $\action^G$ in $\Lambda^G$ 
is collisionless.
\end{coro}
\begin{proof}
If the action is of cyclic type, then there are no 
$\T$-isotropy groups other than $\ker \tau$, 
and hence \emph{a priori} no boundary collisions.
\end{proof}

\begin{coro}
\label{coro:cyclic_type2}
If the action of $G$ on $\Lambda$  
is of cyclic type and 
$\ker\tau=1$ is trivial 
then any local minimizer of $\action^G$ in $\Lambda^G$ 
is collisionless.
\end{coro}
\begin{proof}
If $\ker \tau = 1$ then $G$ has the rotating circle property, 
and hence the conclusion follows from corollary ~\ref{coro:cyclic_type2}.
\end{proof}

\begin{theo}\label{mt:2}
Consider a finite
group $G$ acting on $\Lambda$ 
so that 
every maximal $\T$-isotropy subgroup of $G$ either 
has the rotating circle property 
or acts trivially on the index set $\n$.
Then any local minimizer of $\action^G$ yields a
collision-free periodic solution of the Newton equations
~\ref{eq:newton} for the $n$-body  problem  in $\R^d$.
\end{theo}
\begin{proof}
Let $x$ be a local minimizer of $\action^G$ in $\Lambda^G$.
We have two cases: either one of the $\T$-isotropy
subgroups acts trivially on $\n$ or not. If yes,
then $\ker \tau$ necessarily acts trivially on $\n$,
and therefore $\ker \tau=1$, 
because otherwise the action would be  reducible (see
equation ~\ref{eq:assume2}). So in both cases $\ker \tau$
acts with the rotating circle property  and Theorem
~\ref{coro:mt:1} can be applied,
and hence the minimizer $x$ does not have interior collisions. 

Assume that at time $t_0\in \T$ a (boundary) collision  occurs.
Let $\kk \subset \n$ be a colliding cluster 
and $\bar q$ a corresponding (right) blow-up. 
First assume that the $\T$-isotropy $H_0$ of  $t_0$
has rotating circle property ~\ref{propertyK}.
Let $H$ be the subgroup of $H_0$
defined by $H=\{g\in G\st gt_0 = t_0 \mbox{ and }
g\kk = \kk\}$.
As in the proof of ~\ref{mt:1}, 
one just needs to show that there is $\delta \in \X_\kk^{H}$
such that $\action(\bar q + v^\delta) < \action(\bar q)$ (since
in this case the variation $v^\delta$ for the right blow-up
can be extended to give rise to  an equivariant variation 
in $\T$). But this follows from the fact that,  since
a maximal $\T$-isotropy has the rotating circle property ~\ref{propertyK},
so $H$  does.
On the other hand, if $H_0$ acts trivially on 
$\n$, then $H=H_0$ and $\dim V^H > 0 $ (otherwise $x$ would be
\bdc). As above, let $\kk \subset \n$ a colliding 
cluster and $\bar q$ a blow-up, with 
$\bar q_i(t) = t^{2/(2+\alpha)} \xi_i$
for some $\xi_i \in V$, for $i\in \kk$.
Let $\pi\from V \to V^H$ denote the projection 
$\pi (p) = |H|^{-1}\sum_{g\in H} gp$. 
Consider an index $i$ such that 
$|\pi(\xi_i)| \geq |\pi(\xi_j)|$ 
for every $j\in \kk$  and define
\[
\delta_j = \begin{cases}
\lambda \pi(\xi_i) & \mbox{ if $j=i$} \\
0 & \mbox{ if $j\neq i$}
\end{cases}
\]
if $\pi(\xi_i)\neq 0$, or 
\[
\delta_j = \begin{cases}
\lambda e & \mbox{ if $j=i$} \\
0 & \mbox{ if $j\neq i$}
\end{cases}
\]
where $e$ is an arbitrary nonzero vector $e \in V^H$ 
if $\pi(\xi_j) = 0 $ for every $j$.
It is not difficult to show that for every $\lambda>0$ 
$S(\xi_i-\xi_j, \delta_i -\delta_j) < 0$,
and hence that $\action(\bar q + v^\delta) < \action(\bar q)$ 
as claimed.
\end{proof}

\begin{remark}
The existence of the variations in the proof of ~\ref{mt:2}
shows that under the same hypotheses of ~\ref{mt:2},
paths in $\Lambda^G$ are not \bdc.
\end{remark}

\section{Examples}
\label{section:examples}

Some 
well-known periodic orbits are now shown to exist
as a consequence of the results of section ~\ref{section:mt}. Furthermore, 
some interesting examples of group actions that fulfill
the hypotheses of ~\ref{coro:mt:1} or ~\ref{mt:2} are given.
A complete classification of such group actions
is beyond the scope of the present paper: 
in this section we include only a few examples that 
we consider particularly significative or interesting,
for the sake of illustrating the power and the limitations
of the approach.
Well-known 
examples are the celebrated Chenciner--Montgomery ``eight'' \cite{monchen}, 
Chenciner--Venturelli ``Hip-Hop'' solutions \cite{chenven}, 
Chenciner ``generalized Hip-Hops'' solutions \cite{chenkyoto}, 
Chen's orbit \cite{chen} and Terracini--Venturelli generalized Hip-Hops
\cite{terven}.
One word about the pictures of planar orbits: 
the configurations at the boundary points of the fundamental
domain $\I$ are denoted with an empty circle (starting point $x_i(0)$) and 
a black disc (ending point $x_i(t)$, with $t$ appropriate), 
with a 
label on the starting point describing the index of the particle.
The trajectories of the particles with the times in $\I$ 
are painted as thicker lines (thus it is possible to recover 
the direction of the movement from $x_i(0)$ to $x_i(t)$). 
Unfortunately this feature was not possible  with the three-dimensional
images.

Also, in all the following examples  but \ref{ex:cc} and 
\ref{ex:cc2} existence of the orbits follows directly from 
the results of the paper. The existence of the orbits described in
examples
\ref{ex:cc} and \ref{ex:cc2}, which goes beyond the scope of this 
article, has been recently proved by Chen in \cite{chen2003}. 
Thousands of other suitable actions and the corresponding orbits 
have been found by a special-purpose
computer program based on GAP \cite{GAP4}.

\begin{example}[Choreographies]
\label{ex:choreo}
Consider the cyclic group $G=\ze_n$ of order $n$ acting
trivially on $V$, with a cyclic permutation of order 
$n$ on the index set $\n=\{1,\dots, n\}$ and with a rotation 
of angle $2\pi/n$ on the time circle $\T$.
Since $\X^G=0$, by ~\ref{propo:coercive} the action
functional $\action^G$ is coercive. Moreover, 
since the action of $G$ on $\T$ is of cyclic type and $\ker \tau=1$, 
by ~\ref{coro:cyclic_type} the minimum exists and it has 
no collisions. 
For several numerical results and a description of choreographies
we refer the reader to \cite{cgms}. Recently 
V.~Barrutello
and the second author  have proved that 
in any dimension the minima with just the choreographic
cyclic symmetry are just rotating regular polygons
(see also \cite{chenICM}, 4.2.(i)).
\end{example}

\begin{example}
\label{ex:1}
Let $n$ be odd. Consider the dihedral group $G=D_{2n}$ of 
order $2n$, with the presentation 
$G = < g_1, g_2 | g_1^2=   g_2^n = (g_1g_2)^2 = 1 >$.
Let $\tau$ be the homomorphism defined by
$\tau(g_1) = 
\matrice{0}{1}{1}{0}
$, 
and $\tau(g_2) = 
\rotationmatrix{n}$.
Furthermore,
let the homomorphism $\rho$ be defined by 
$\rho(g_1) = 
\matrice{-1}{0}{0}{-1}
$ and $\rho(g_2)=
\left[
\begin{array}{cc}
1 &  0 \\
0 &  1\\
\end{array}
\right]
$.
Finally, let $G$ act on $\n$ by the homomorphism $\sigma$ defined 
as 
$\sigma(g_1) = (1,n-1)(2,n-2)\dots( (n-1)/2, (n+1)/2 )$,
$\sigma(g_2) = (1,2,\dots, n)$,
where $(i_1,i_2,\dots, i_k)$ means the standard cycle-decomposition
notation for permutation groups.
By the action of $g_2$ it is easy to show that 
all the loops in $\Lambda^G$ are choreographies,
and thus that, since $\X^G=0$, the action functional 
is coercive.
The maximal $\T$-isotropy  subgroups are the subgroups of order $2$
generated by the elements $g_1g_2^i$ with $i=0\dots n-1$.
Since they are all conjugated, it is enough
to show that one of them acts with the rotating circle property.
Thus consider $H=<g_1>\subset G$.
For every index $i\in \{1,2,\dots, n-1\}$ 
the isotropy $H_i \subset H$  
relative to the action of $H$ on $\n$ is trivial, and 
$g_1$ acts by rotation on $V=\R^2$. Therefore 
for every $i\in \{1,2,\dots, n-1\}$ it 
is possible to choose a circle rotating under $H$ for $i$,
since, being $H_i$ trivial (see definition ~\ref{defi:rotatingfor}),
\(V^{H_i} = V\).
The resulting orbits are not homographic (since all
the particles pass through the origin $0$ at some time
of the trajectory and the configurations are centered).
For $n=3$ this is the \emph{eight with less symmetry} of 
\cite{chenICM}. See also \cite{chenciner}. 
Possible trajectories are shown
in figures ~\ref{fig:1} and ~\ref{fig:1bis}.
\begin{figure}
\begin{minipage}{0.5\textwidth}
\begin{center}
\includegraphics*[width=4truecm]{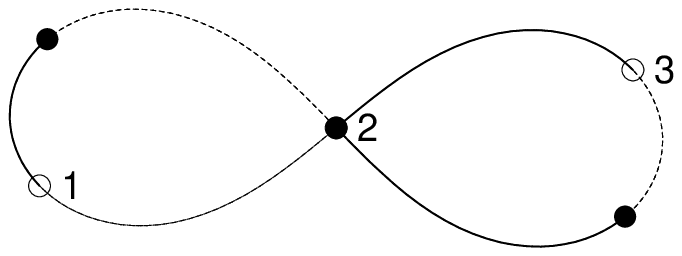}
\caption{The ($D_6$-symmetric) eight for $n=3$}
\label{fig:1}  
\end{center}
\end{minipage}%
\begin{minipage}{0.5\textwidth}
\begin{center}
\includegraphics*[width=4truecm]{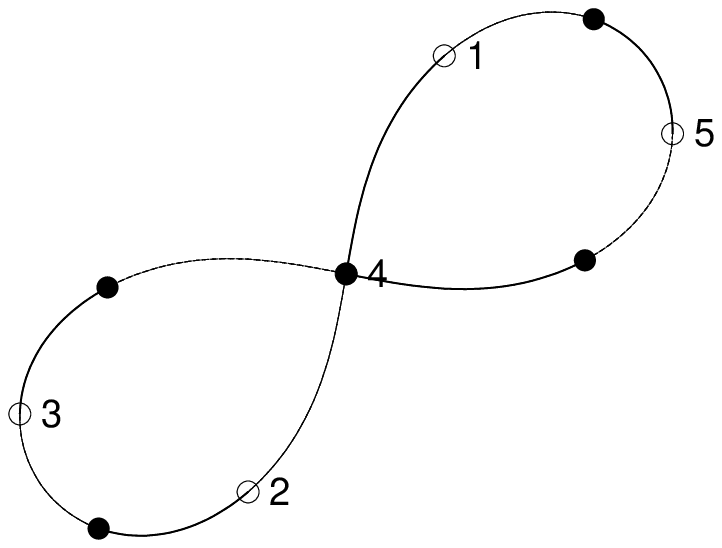}
\caption{The ($D_{10}$-symmetric) eight with $n=5$}
\label{fig:1bis}
\end{center}
\end{minipage}
\end{figure}
\end{example}

\begin{example}
\label{ex:2}
As in the previous example, let $n\geq 3$ be an odd integer.
Let $G=C_{2n}\cong \ze_2+\ze_n$ be the cyclic group of order $2n$,
presented as 
$G = < g_1, g_2 | g_1^2=   g_2^n = g_1g_2g_1^{-1}g_2^{-1} = 1 >$.
The action of $G$ on $\T$ is given by 
$\tau(g_1g_2)=\theta_{2n}$, where $\theta_{2n}$ denotes
the rotation of angle $\pi/{n}$ (hence the action will be 
of cyclic type).
Now, $G$ can act on the plane $V=\R^2$ by the homomorphism
$\rho$ defined by 
$\rho(g_1) = 
\left[
\begin{array}{cc}
-1 &  0 \\
0 &  1\\
\end{array}
\right]
$ and $\rho(g_2)=
\left[
\begin{array}{cc}
1 &  0 \\
0 &  1\\
\end{array}
\right]
$.
Finally, the action of $G$ on $\n=\{1,2,\dots, n\}$
is given by the homomorphism $\sigma\from G \to \Sigma_n$
defined by 
$ \sigma(g_1) = ()$,
$\sigma(g_2) = (1,2,\dots, n)$.
The cyclic subgroup $H_2=<g_2> \subset G$  gives
the symmetry constraints of the choreographies, hence
loops in $\Lambda^G$ are choreographies and the 
functional is coercive.
Furthermore, since the action is of cyclic type,
by ~\ref{coro:cyclic_type} the minimum of the action
functional is collisionless. It is possible 
that such minima coincide with the minima of the previous example:
this would imply that the symmetry group of the minimum contains the
two groups above.
\begin{figure}
\begin{minipage}{0.5\textwidth}
\begin{center}
\includegraphics*[width=4truecm]{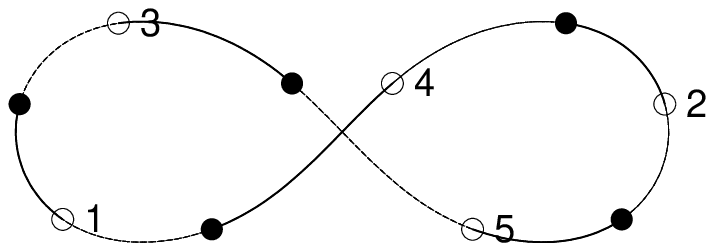}
\caption{Another symmetry constraint for an eight-shaped orbit ($n=5$)}
\label{fig:2}  
\end{center}
\end{minipage}%
\begin{minipage}{0.5\textwidth}
\begin{center}
\includegraphics*[width=4truecm]{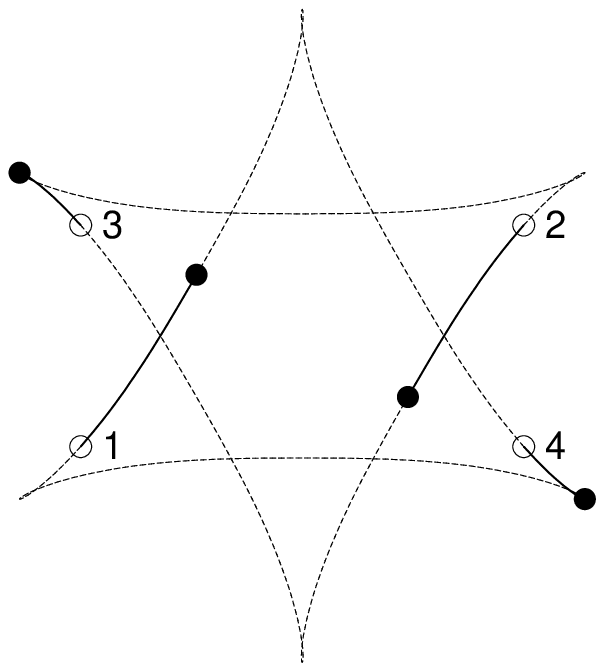}
\caption{The orbit of example ~\ref{ex:cc} with $q=3$}
\label{fig:cc}
\end{center}
\end{minipage}
\end{figure}
\end{example}

\begin{example}%
\label{ex:cc}
Consider four particles with equal masses and an odd integer $q\geq 3$.
Let $G=D_{4q}\times C_2$ be the direct product 
of the  dihedral  group
of order $4q$  with the group $C_2$ of order $2$.
Let $D_{4q}$ be presented by 
\(D_{4q} = < g_1, g_2 | g_1^2 = g_2^{2q} = (g_1g_2)^2 = 1 >\),
and let $c\in C_2$ be the non-trivial element of $C_2$.
Now define the homomorphisms $\rho$, $\tau$ and $\sigma$ 
as follows:
\( \rho(g_1) = \tau(g_1) = 
\matrice{1}{0}{0}{-1} \), 
\( 
\rho(g_2) = \tau(g_2) = 
\rotationmatrix{2q}
\),
\(
\rho(c) =  
\matrice{-1}{0}{0}{-1}
\),
\(
\tau(c) =  
\matrice{1}{0}{0}{1}
\), 
\(
\sigma(g_1) =   (1,2)(3,4) \),
\(
\sigma(g_2) =   (1,3)(2,4) \),
\(
\sigma(c) =   (1,2)(3,4) \).
It is not difficult to show that $\X^G=0$, and thus the action is coercive.
Moreover, $\ker \tau = C_2$, which acts on $\R^2$ with the 
rotation of order $2$, hence $\ker \tau$ acts with the rotating circle
property.  Thus, by ~\ref{coro:mt:1} the minimizer exists and does
not have interior collisions.
To exclude boundary collisions we cannot use ~\ref{mt:2},
since the maximal $\T$-isotropy subgroups do not act with 
the rotating circle property.
A possible graph for such a minimum can be found in figure
~\ref{fig:cc}, for $q=3$ (one needs to prove that the minimum
is not the homographic solution -- with a level estimate -- 
and that there are no boundary collisions -- with an 
argument similar to \cite{chen}). See also \cite{chen2003} 
for an updated and much generalized treatment of such orbits.  
\end{example}

\begin{figure}
\begin{minipage}{0.5\textwidth}
\begin{center}
\includegraphics*[width=4truecm]{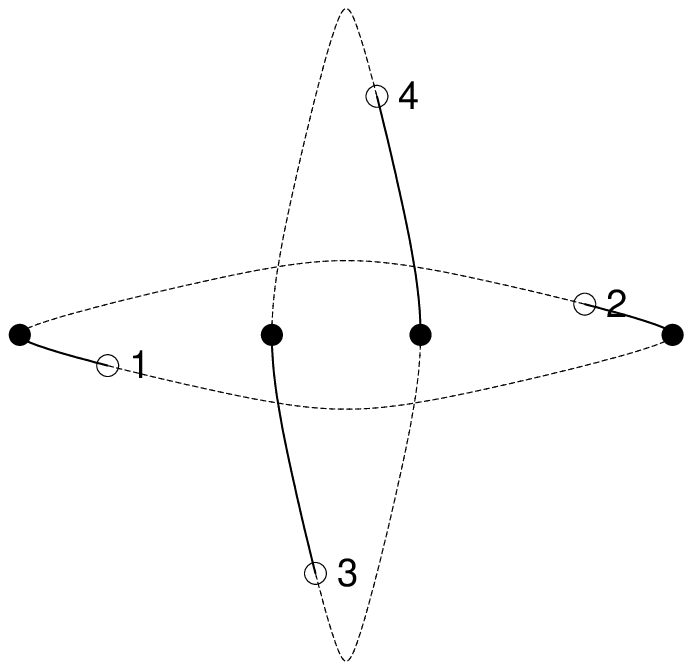}
\caption{A possible minimizer for example ~\ref{ex:cc2}}
\label{fig:cc2}  
\end{center}
\end{minipage}%
\begin{minipage}{0.5\textwidth}
\begin{center}
\includegraphics*[width=6truecm]{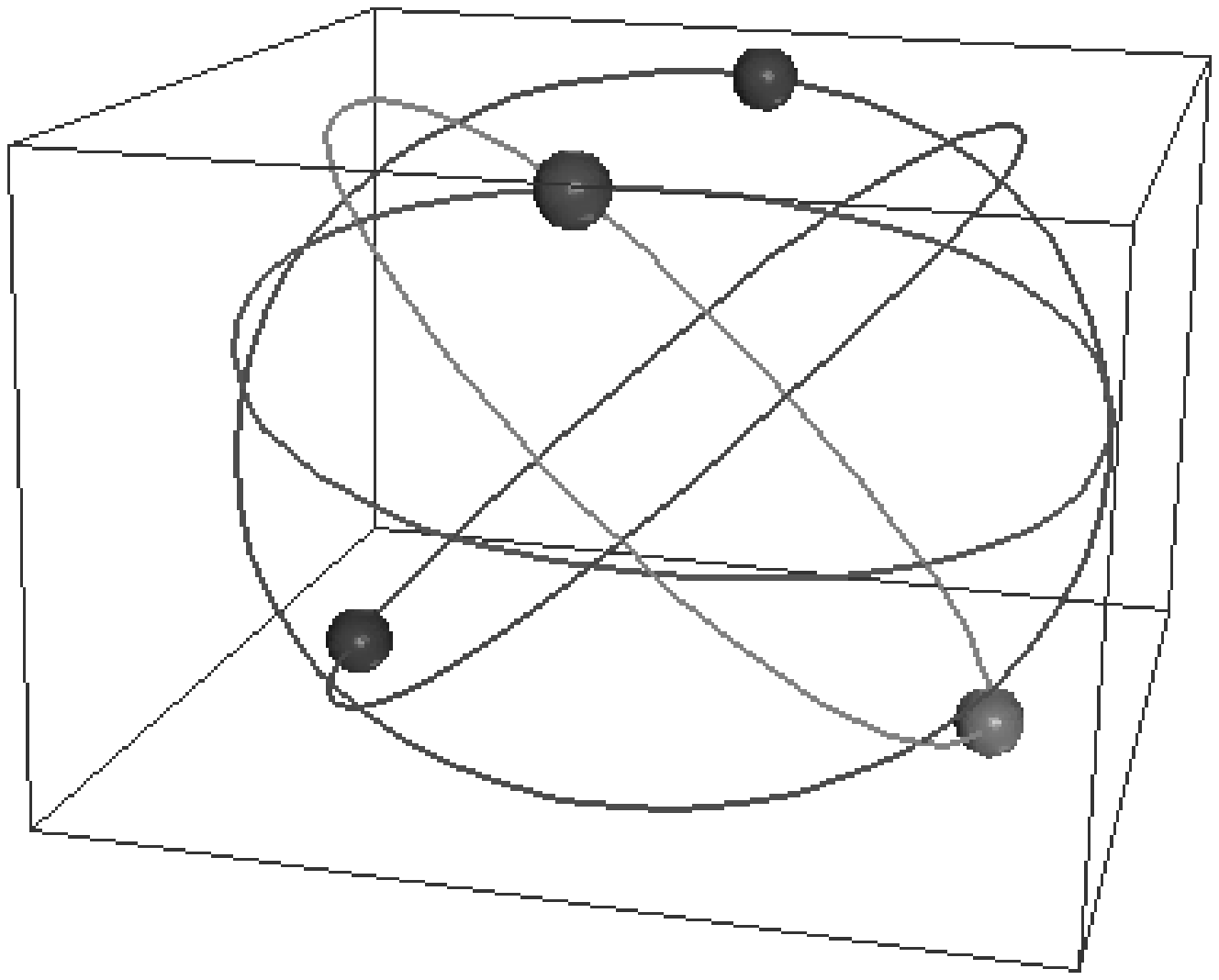}
\caption{The Chenciner--Venturelli Hip-Hop}
\label{fig:5}
\end{center}
\end{minipage}
\end{figure}

\begin{example}
\label{ex:cc2}
Consider four particles with equal masses and an even integer $q\geq 4$.
Let $G=D_{q}\times C_2$ be the direct product 
of the  dihedral  group
of order $2q$  with the group $C_2$ of order $2$.
Let $D_{4q}$ be presented by 
\(D_{4q} = < g_1, g_2 | g_1^2 = g_2^{q} = (g_1g_2)^2 = 1 >\),
and let $c\in C_2$ be the non-trivial element of $C_2$.
As in example ~\ref{ex:cc},
define the homomorphisms $\rho$, $\tau$ and $\sigma$ 
as follows.
\( \rho(g_1) = \tau(g_1) =
\matrice{1}{0}{0}{-1}
\),\(
\rho(g_2) = \tau(g_2) =
\rotationmatrix{q}
\), \(
\rho(c) = 
\matrice{-1}{0}{0}{-1}
\), \(
\tau(c) = 
\matrice{1}{0}{0}{1}
\), \(
\sigma(g_1) =  (1,2)(3,4) \),\(
\sigma(g_2) =  (1,3)(2,4) \), \(
\sigma(c) =  (1,2)(3,4) \)
Again, one can show that a minimizer without 
interior collisions  exists since $\ker\tau=C_2$ 
acts with the rotating circle property
(a possible minimizer is shown in figure ~\ref{fig:cc2}).
This generalizes Chen's orbit \cite{chen}. See also \cite{chen2003}.
\end{example}

\begin{example}[Hip-hops]
\label{ex:hipop}
If 
$G=\ze_2$ is the group of order $2$ acting trivially on $\n$,
acting with the antipodal map on $V=\R^3$  and on the time
circle $\T$, then again $\X^G=0$, so that  ~\ref{propo:coercive} holds.
Furthermore, since the action is of cyclic type proposition 
~\ref{coro:cyclic_type}
assures that minimizers have no  collisions. Such minimizers
were called \emph{generalized Hip-Hops} in \cite{chenICM}. 
See also \cite{chenkyoto}.
A subclass 
of symmetric trajectories leads to 
a generalization of such a Hip-Hop. 
Let $n\geq 4$ an even integer.  Consider $n$ particles
with equal masses, and the group $G=C_{n}\times C_2$
direct product of the cyclic group of order $n$ (with 
generator $g_1$) and the group $C_2$ of order
$2$ (with generator $g_2$).
Let the homomorphisms $\rho$, $\sigma$ and $\tau$ be defined by
\( \rho(g_1) = 
\RotationMatrix{n}
\), \(
\rho(g_2) = 
\Matrice{-1}{0}{0}{0}{-1}{0}{0}{0}{-1} 
\), \( 
\tau(g_1) = 
\matrice%
{1} {0}
{0} {1} 
\), \( 
\tau(g_2) = 
\matrice{-1}{0}{0}{-1}
\), \(
\sigma(g_1) =  (1,2,3,4) \),
\( \sigma(g_2) =  () \)
It is easy to see that $\X^G=0$, and thus 
a minimizer exists.
Since the action is of cyclic type, 
it suffices to exclude interior collisions.
But this follows from the fact that 
$\ker \tau = C_{n}$ has the rotating circle property.
This example is the natural  generalization 
of the Hip-Hop solution of \cite{chenven} to $n\geq 4$
bodies. We can see the trajectories in figure ~\ref{fig:5}.
\end{example}

\begin{figure}
\begin{minipage}{0.5\textwidth}
\begin{center}
\includegraphics*[width=4truecm]{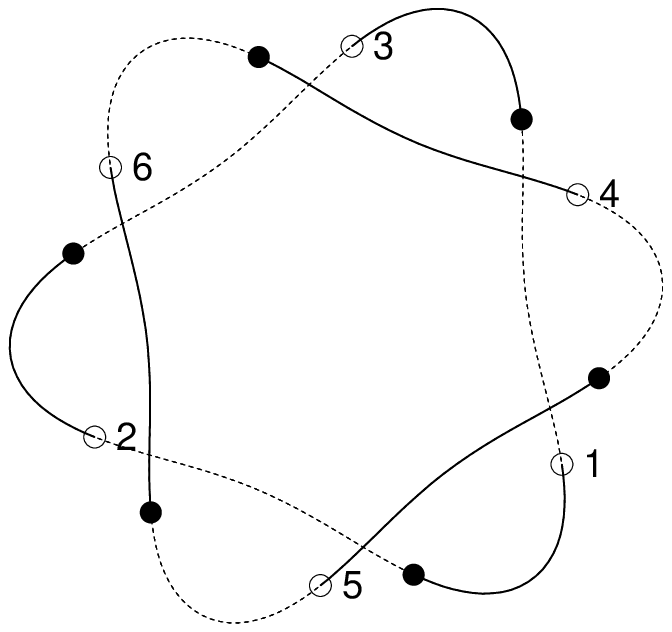}
\caption{The planar equivariant minimizer of example ~\ref{ex:new}}
\label{fig:6}  
\end{center}
\end{minipage}%
\begin{minipage}{0.5\textwidth}
\begin{center}
\includegraphics*[width=6truecm]{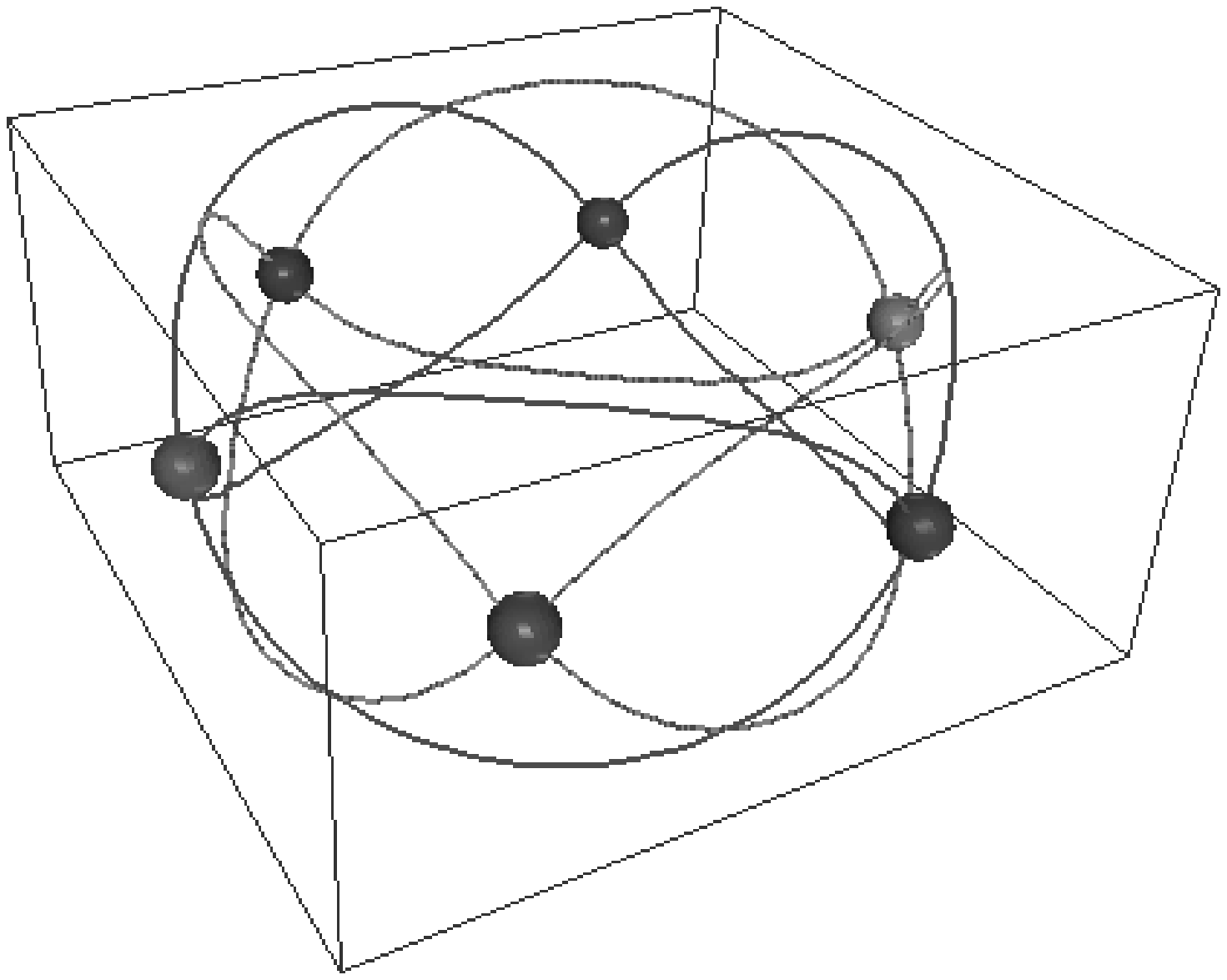}
\caption{The three-dimensional equivariant minimizer
of example ~\ref{ex:new}}
\label{fig:7}
\end{center}
\end{minipage}
\end{figure}

\begin{example}
\label{ex:new}
Consider the direct product $G= D_{6}\times C_3$ 
of the dihedral group
$D_6$ 
(with generators $g_1$ and $g_2$ of order $3$ and 
$2$ respectively) of order $6$ and the cyclic group $C_3$ of order $3$
generated by $c\in C_3$.
Let us consider the planar 
$n$-body problem with $n=6$ with the symmetry
constraints given by the following $G$-action.
\( 
\rho(g_1) = 
\matrice{1}{0}{0}{1}
\),
\(
\rho(g_2) = 
\matrice{-1}{0}{0}{-1}
\), 
\(
\rho(c) = 
\rotationmatrix{3}
\), 
\(
\tau(g_1) = \rotationmatrix{3}
\),
\(
\tau(g_2) = \matrice{0}{1}{1}{0}
\),
\(
\tau(c) = 
\matrice{1}{0}{0}{1}
\), 
\(
\sigma(g_1) =   (1,3,2)(4,5,6)  \),
\(
\sigma(g_2) =   (1,4)(2,5)(3,6) \), 
\(
\sigma(c) =   (1,2,3)(4,5,6)  
\).
By ~\ref{propo:coercive} one  can prove that 
a minimizer exists, and since 
$G$ acts with the rotating circle property (actually,
the elements of the image of $\rho$ are rotations)
on $\T$-maximal isotropy subgroups, the conclusion
of theorem ~\ref{mt:2} holds.  It is not difficult to see 
that configurations in $\X^{\ker \tau}$ are given
by two centered equilateral triangles. Now, to guarantee
that the minimizer is not a homographic solution, of course 
it suffices
to show that there are no homographic solutions
in $\Lambda^G$ (like in the case of example ~\ref{ex:1}).
This follows from the easy observation that 
at some times $t\in \T$ with maximal isotropy it happens
that $x_1 = -x_4$, $x_2=-x_5$ and $x_3=-x_6$, 
while
at some other times it happens that 
$x_1 = -x_5$, $x_2 = -x_6$ and $x_3=-x_4$
or that 
$x_1 = -x_6$, $x_2 = -x_4$ and $x_3 = -x_5$
and this implies that there are no homographic loops
in $\Lambda^G$. 
With no difficulties the same action can be defined for
$n=2k$, where $k$ is any odd integer. We can see 
a possible trajectory in figure ~\ref{fig:6}. 
Also, it is not difficult to consider a similar example in dimension
$3$.
With $n=6$ and the notation of $D_6$ and $C_3$ as above,
consider the group 
$G = D_6 \times C_3 \times C_2$.
Let $g_1$, $g_2$, $c$ be as above,
and let $c_2$ be the generator of $C_2$.
The homomorphisms $\rho$, $\tau$ and $\sigma$ are defined 
in a similar way by
\( 
\rho(g_1) = 
\Matrice{1}{0}{0}{0}{1}{0}{0}{0}{1}
\),
\(
\rho(g_2) = 
\Matrice{-1}{0}{0}{0}{-1}{0}{0}{0}{1}
\), 
\(
\rho(c) = 
\RotationMatrix{3}
\), 
\(
\rho(c_2) = 
\Matrice{1}{0}{0}{0}{1}{0}{0}{0}{-1}
\), 
\(
\tau(g_1) = \rotationmatrix{3}
\),
\(
\tau(g_2) = \matrice{0}{1}{1}{0}
\),
\(
\tau(c) = 
\matrice{1}{0}{0}{1}
\), 
\(
\tau(c_2) = 
\matrice{-1}{0}{0}{-1}
\), 
\(
\sigma(g_1) =   (1,3,2)(4,5,6)  \),
\(
\sigma(g_2) =   (1,4)(2,5)(3,6) \), 
\(
\sigma(c) =   (1,2,3)(4,5,6)  
\),
\(
\sigma(c_2) =   ()  
\).
In the resulting collisionless minimizer (again,
it follows by ~\ref{propo:coercive} and ~\ref{mt:2})
two equilateral triangles rotate in opposite directions and 
have a ``brake'' motion on the third axis.
The likely shape of the trajectories  
can be found in figure ~\ref{fig:7}. 
\end{example}

\begin{figure}
\begin{minipage}{0.5\textwidth}
\begin{center}
\includegraphics*[width=4truecm]{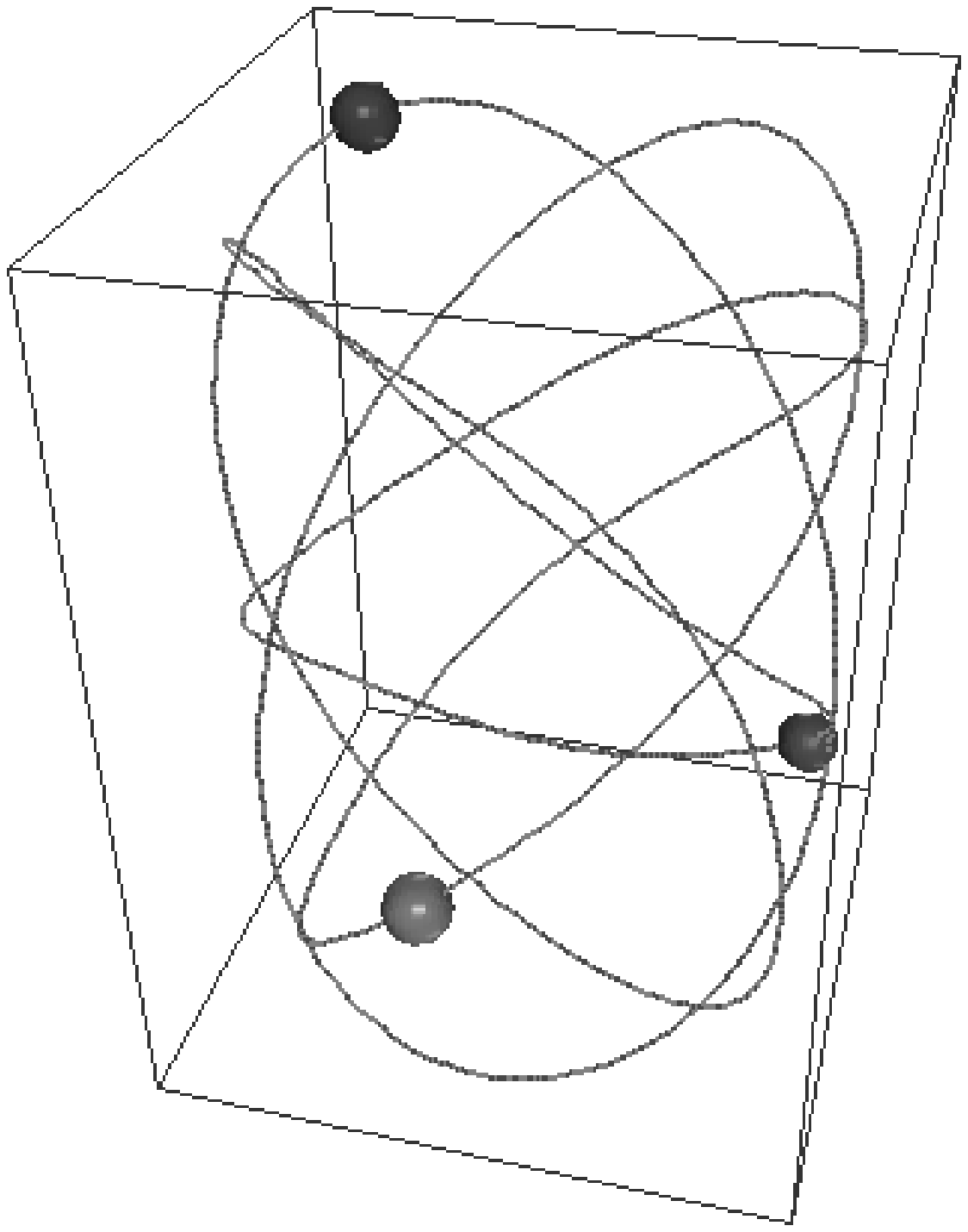}
\caption{The non-planar choreography of ~\ref{ex:matassa} for $k=4$}
\label{fig:matassa}  
\end{center}
\end{minipage}%
\begin{minipage}{0.5\textwidth}
\begin{center}
\includegraphics*[width=6truecm]{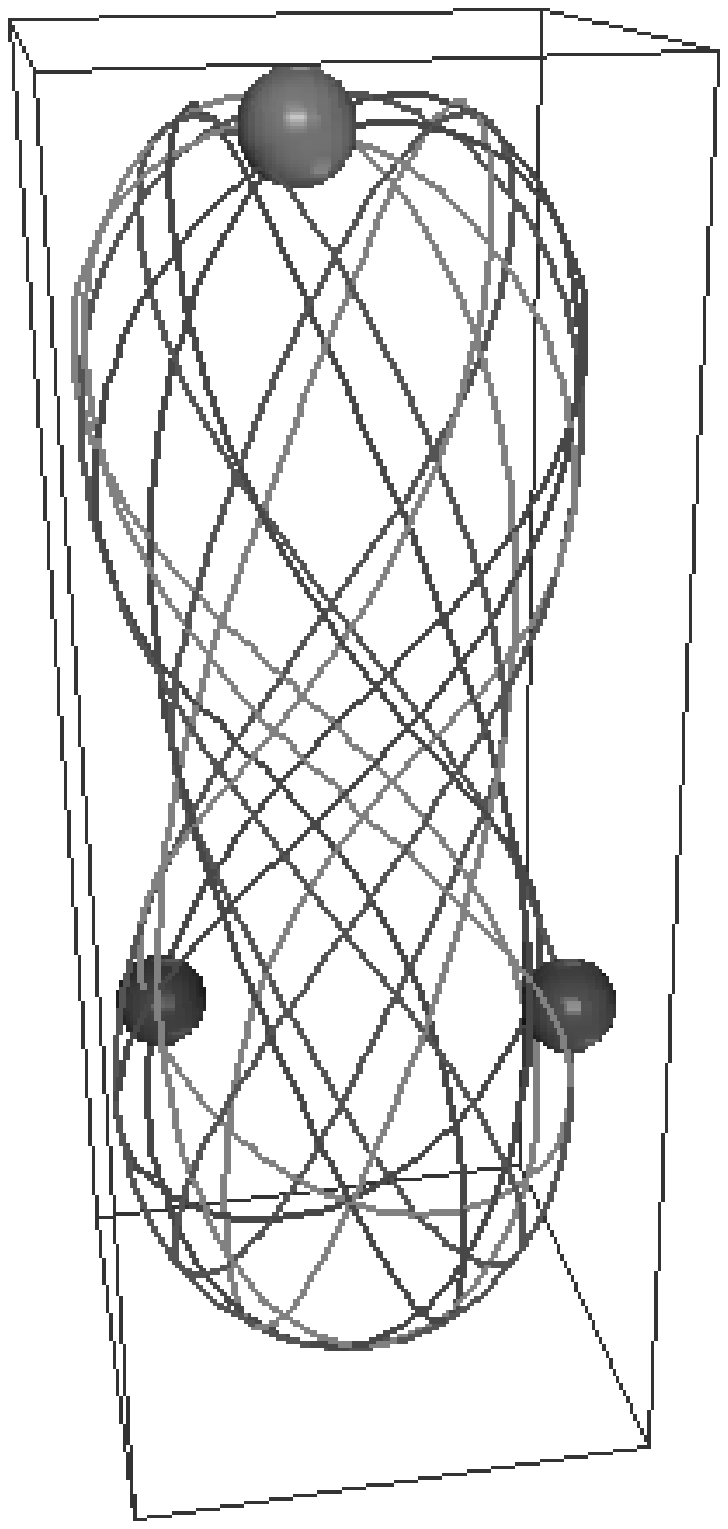}
\caption{A non-planar symmetric orbit of ~\ref{rem:matassa}, with $k=3$ and $p=1$}
\label{fig:9}
\end{center}
\end{minipage}
\end{figure}

\begin{example}
\label{ex:matassa}
Let $k\geq 2$
be  an integer, and consider the cyclic group
$G = C_{6k}$  of order $6k$ generated by the element $c\in G$.
Now consider  orbits for $n=3$ bodies in the space of dimension
$d=3$. With a minimal effort and suitable changes 
the example can be generalized 
for every $n\geq 3$. We leave the details to the reader.
The homomorphisms $\rho$, $\tau$ and 
$\sigma$ are defined by
\(\rho(c) = 
\Matrice%
{\cos \frac{\pi}{k}} {-\sin \frac{\pi}{k}} { 0}
{\sin \frac{\pi}{k}} {\cos \frac{\pi}{k}} {0}
{0} {0} {-1} 
\),
\(
\tau(c) = 
\rotationmatrix{6k}
\),
\(
\sigma(c) = (1,2,3)
\).
Straightforward calculations show that 
$\X^G=0$ and hence proposition ~\ref{propo:coercive} can be applied.
Furthermore, the action is of cyclic type with $\ker \tau=1$,
and hence by ~\ref{coro:cyclic_type} the minimizer does not have
collisions.
It is left to show that this minimum is not a homographic 
motion.
The only homographic motion 
in $\Lambda^G$ is a Lagrange triangle $y(t) = (y_1,y_2,y_3)(t)$,
rotating with 
angular velocity $3-2k$ (assume that the period
is $2\pi$, i.e.\ that $T=|\T|=2\pi$) in the plane
$u_3=0$ (let $u_1,u_2,u_3$ denote the coordinates in $\R^3$).
To be a minimum it needs to be inscribed in the horizontal circle
of radius $(\frac{\alpha 3^{-\alpha/2}}{2(3-2k)^2})^{1/(2+\alpha)}$.
Now, for every  function  $\phi(t)$ defined  on $\T$
such that $\phi(c^3t) =  - \phi(t)$,
the loop given by 
$v_1(t) = (0,0,\phi(t))$, $v_2(t) = (0,0,\phi(c^{-2}t))$
and $v_3(t) = (0,0,\phi(c^2t))$
is $G$-equivariant, and thus belongs to $\Lambda^G$.
If one  computes the value of Hessian of 
the Lagrangian action
$\action$
in $y$ and 
in the direction of the 
loop $v$
one finds that 
\[
D^2_v\action \vert_{y} =  
3 \int_{0}^{2\pi} \dot \phi^2(t) dt 
-
2(3-2k)^2
\int_0^{2\pi} 
(\phi(t) + \phi(ct))^2 dt.
\]
In particular, if we set the 
function $\phi(t) = \sin(kt)$, which has the desired property,
elementary integration yields
\[
D^2_v\action \vert_{y} =  
3\pi ( k^2 - 2(3-2k)^2),
\]
which does not depend on $\alpha$ and 
is negative for every $k\geq 3$.
Thus for every $k\geq 3$ the minimizer is not homographic.
We see a possible trajectory in figure ~\ref{fig:matassa}.
\end{example}

\begin{remark}
\label{rem:matassa}
In the previous example, if $k\not\equiv 0 \mod 3$, the cyclic
group $G$ can be written as the sum $C_3+C_{2k}$.
The generator of $C_3$ acts trivially on $V$,
acts with a rotation of order $3$ on $\T$ and 
with the cyclic permutation $(1,2,3)$ on $\{1,2,3\}$.
This means that for all $k\not\equiv 0 \mod 3$ 
the orbits of example ~\ref{ex:matassa} are non-planar choreographies.
Furthermore,  it is possible to define a cyclic action of the same kind
by setting $\tau$ and $\sigma$ as above and 
\(\rho(c) = 
\Matrice%
{\cos \frac{p}{3k}\pi } {-\sin \frac{p}{3k}\pi} { 0}
{\sin \frac{p}{3k}\pi } {\cos \frac{p}{3k}\pi } {0}
{0} {0} {-1} 
\),
where $p$  is a non-zero integer. If $p=3$ one obtains the same
action as in example ~\ref{ex:matassa}. 
One can perform similar computations and obtain that 
the Lagrange orbit  (with angular velocity $p-2k$, this time) 
is not a minimizer
for all $(p,k)$  such that 
$0<p<3k$ 
and 
$k^2 - 2(p-2k)^2 <0$.
\end{remark}

\begin{remark}
We would like to conclude the article with the observation that,
after ~\ref{propo:coercive} and ~\ref{mt:2}, 
it is interesting to determine a classification
of the $G$-actions on $\Lambda$ (given by $\rho$, $\sigma$ and $\tau$ as in 
~\ref{eq:Gaction2}) with the rotating circle property and 
such that $\X^G=0$.
These conditions  can be tested easily on a computer algebra system 
(we have used GAP \cite{GAP4} to check and find examples).
Some preliminary results on such a classification can be 
found in \cite{or} and this is the topic of a paper in preparation.
Furthermore, we are planning to put on-line on a web page 
some  animations of the 
periodic orbits found (in the order of hundreds, at the moment).
Please contact one of the authors if interested.
\end{remark}

\def\cfudot#1{\ifmmode\setbox7\hbox{$\accent"5E#1$}\else
  \setbox7\hbox{\accent"5E#1}\penalty 10000\relax\fi\raise 1\ht7
  \hbox{\raise.1ex\hbox to 1\wd7{\hss.\hss}}\penalty 10000 \hskip-1\wd7\penalty
  10000\box7} \def\cprime{$'$} \def\cprime{$'$} \def\cprime{$'$}
  \def\cprime{$'$} \def\cprime{$'$} \def\cprime{$'$}

\end{document}